\PassOptionsToPackage{x11names}{xcolor}
\documentclass[12pt,a4paper,twoside]{article}
\usepackage[utf8]{inputenc}
\usepackage[OT4]{fontenc}
\usepackage{geometry}
\newgeometry{tmargin=3cm, bmargin=3cm, lmargin=3cm, rmargin=3cm}
\usepackage{pdflscape}
\usepackage{csvsimple}
\usepackage{longtable}
\usepackage{caption}
\usepackage{booktabs}
\usepackage{subcaption}
\usepackage{pgfplotstable,filecontents}
\usepackage{hyperref}
\counterwithout{figure}{section}
\usepackage{tabularx}
\usepackage{graphicx}
\usepackage{epstopdf}
\usepackage{placeins}

\usepackage[colorinlistoftodos, textwidth=30mm, shadow]{todonotes}
\usepackage{changes}
\usepackage{listings}
\colorlet{cpop}{Chartreuse3}
\colorlet{cmon}{DarkOrchid1}
\colorlet{ckonrad}{cyan}
\colorlet{cjulia}{teal}

\definechangesauthor[name={Konrad}, color=ckonrad]{Konrad}
\definechangesauthor[name={Monika}, color=cmon]{Monika}
\definechangesauthor[name={Julia}, color=cjulia]{Julia}
\definechangesauthor[name={POP}, color=cpop]{POP}

\begin{document}
	
	\pagestyle{myheadings}
	\markright{{\footnotesize Piotrowska ~M.J., Rymuza J., Sakowski K.: Analysis of hospital claims data from Germany}}
	
	\noindent\begin{tabular}{|p{\textwidth}}		
		\Large\bf Analysis of the hospital claims data from Germany (years 2013 -- 2018) and derivation of transfer matrices
		\\\vspace{0.01cm}
		\it  Piotrowska, M.J.$^{\dagger,1, \star}$, Rymuza, J.$^{\dagger,2}$ \& Sakowski, K.$^{\dagger,\clubsuit,3}$\\\vspace{0.02cm}
		\it\small $^\dagger$Institute of Applied Mathematics and Mechanics,\\
		\it\small University of Warsaw, Banacha 2, 02-097 Warsaw, Poland\\\vspace{0.01cm}
		\it\small $^\clubsuit$Institute of High Pressure Physics,\\
		\it\small Polish Academy of Sciences, Sokolowska 29/37, 01-142 Warsaw, Poland\\\vspace{0.01cm}

\small  $^1$\texttt{monika@mimuw.edu.pl}, $^2$\texttt{r.julia98@gmail.com} , $^3$\texttt{konrad@mimuw.edu.pl}  \\
		$^\star$ {\footnotesize alphabetic order}
		\\
		\hline
		ORCID:\\
		\it\small Piotrowska, M.J.:\,\normalfont 0000-0003-0156-8290\\
		\it\small Sakowski, K.:\,\normalfont 0000-0003-1511-074X\\
	\end{tabular}
	\thispagestyle{empty}
	
	\vspace{2em}
	
	
	\textbf{Keywords:}  
	healthcare data analysis, overlapping data, transfer probability matrix
	\begin{abstract}	
		Antibiotic resistant bacteria are a serious threat to public health system. In order to be able to model the spread of this pathogens within healthcare system network, it is necessary to obtain information on patient movements and the structure of the network. To accomplish that we need informations about patient population and their exchange within the network. In this paper, we analysed hospitalisation records which were provided for the EMerGE-Net project to the Martin Luther University Halle-Wittenberg.
		We studied and discussed properties of the data with respect to German states. Moreover, we investigated the patterns of patients movements between different states. Then, we took a closer look at resulting hospital network structure, which can be further used to numerically model spread of pathogens.      
	\end{abstract}

	\section{Introduction}
	
	Antibiotic bacterial resistance is an increasing problem in healthcare systems all over the world.
 	Strains of resistant bacteria are spreading and they affect our ability to treat other diseases and health problems. They impose increased healthcare costs, longer stay at hospitals and increased mortality. Even though new antibiotics are being developed, they are expected to not be effective against most dangerous antibiotic-resistant bacteria~\cite{WHO}. High rates of resistance against antibiotics frequently used to treat common bacterial infections have been observed world-wide, which indicates that we are running out of effective antibiotics~\cite{WHO2}.
	European Centre for Disease Prevention and Control estimates that more than 670\,000 infections occur in the EU/EEA due to bacteria resistant to antibiotics and that number increases the overall healthcare costs by about 1.1 billion euros. Moreover, approximately 33\,000 people die each year in EU/EEA as a direct consequence of these infections~\cite{ECDC}.   
	
	In many studies concerning spread of different types of pathogens, it is essential to derive so-called patients contact networks or network of hospitals which exchange patients (based on e.g. hospital records), cf.~\cite{Donker2010,huang2010quantifying, Donker2012, ciccolini2013infection, Donker2014, Donker2017, Nekkab2017, Lee2012, Belik2016, Lonc2019, Vilches2019, Nekkab2020}. In recent studies \cite{PiotrowskaSIVW2020, Piotrowska2020PlosComp, Xia2021, Vilches2019} the focus was on the role of indirect patient exchange between healthcare facilities i.e. when the patient spent at least one day between following admissions at home. In the approach proposed in \cite{PiotrowskaSIVW2020, Piotrowska2020PlosComp}  authors define network consist of the healthcare facilities and society and also derived so-called transfer matrices describing the probability of the transfer of the patient from one to another healthcare facility or from/to society. To do so, it is necessary to estimate several characteristics such as: size of the healthcare facilities, average length of stay of patients in particular facilities and society. In some cases, for re-sampling and up-scaling techniques of data (see \cite{Xia2021}) or agent based modelling of pathogen spread in a singe hospital (see~\cite{Deeny2013, Sadsad2013, Tahir2021}), detailed information about the patient population structure are required, including sex, age, diagnosis, etc. Thus, to meet these expectations,we conducted an analysis of dataset provided for the EMerGE-Net project to the Martin Luther University Halle-Wittenberg.
	To achieve our goal, we used publicly available code~\cite{emergenetpackage} written by us, which detects and classifies transfers from the anonymous hospital stay records.

	\section{Description of the dataset}
	
	The available data consist of 9 415 914 hospitalisation records for 4 101 647 patients collected by the insurance company from January 2013 till end of August 2018. In particular, the database contains the following information: patient anonymized ID, anonymized healthcare facility ID, federal state of healthcare facility, day of the admission, day of discharge, diagnosis (international ICD-10-GM code), patient’s sex and year of birth.

	Within dataset we found 9 396 578 complete healthcare facility stay records (i.e. with assigned location and diagnosis codes). There are 1 839 hospital facilities among the whole database located in all regions of Germany (for more details see Table~\ref{tab1}). Records without the diagnosis code or region code corresponding to one of the states were omitted from further analysis. 
	States in all tables and images are sorted by their population reported in~\cite{Destatis}. 
		
	\begin{table}[h!]
		\centering
		\caption{Number of admissions and healthcare facilities depending on the state. States are ordered by their population according to~\cite{Destatis}.}
		\label{tab1}
		\begin{filecontents*}{tabels/general_stats.tsv}
State	No. of admissions	No. of facilities
North Rhine-Westphalia	2523945	399
Bavaria	1165249	343
Baden-Württemberg	958275	227
Lower Saxony	808412	205
Hesse	779243	151
Rhineland-Palatinate	433619	94
Saxony	231291	79
Berlin	743739	54
Schleswig-Holstein	373367	74
Brandenburg	204805	54
Saxony-Anhalt	152512	51
Thuringia	146843	44
Hamburg	489709	37
Mecklenburg-West \mbox{Pomerania}	173568	41
Saarland	104963	23
Bremen	107038	14
\end{filecontents*}
\pgfplotstabletypeset[
		col sep=tab, 
		display columns/0/.style={column type={|p{5.5cm}|},string type}, 
		display columns/1/.style={column type={p{3.3cm}|}, int detect, 1000 sep={\;},precision=3},
		display columns/2/.style={column type={p{2.8cm}|},},
		every head row/.style={before row=\hline,after row=\hline},
		every nth row={1}{before row=\hline},
		every last row/.style={after row=\hline},
		]{tabels/general_stats.tsv}
	\end{table}

	\section{Data analysis}

	\subsection{Methods}
	The data was analysed at the server of the Martin Luther University Halle-Wittenberg using secure external access. We used freely available EMERGENERT package developed by Piotrowska~\&~Sakowski, for the documentation, hardware requirements and software see~\cite{emergenetpackage}.
	
	\subsection{Population structure}
\begin{figure}[h!]
		\centering{}
		
		\includegraphics[height=7.5cm]{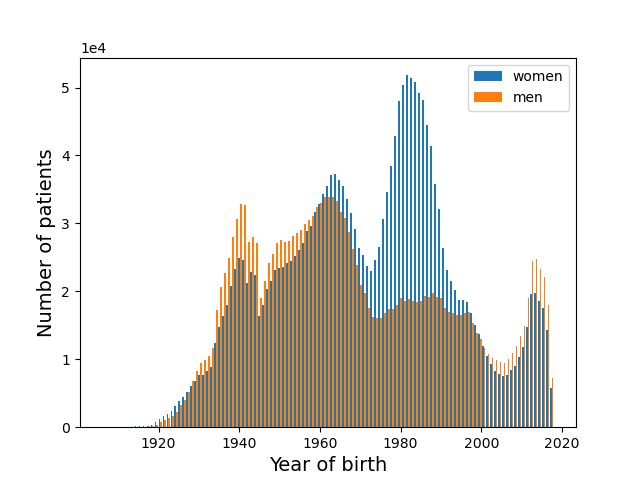}
		\caption{Structure of patient population.}
		\label{fig1}
	\end{figure}

	Within analysed records we found records for 2 205 843 women and 1 893 988 for men. Maximum number of hospitalizations per patient was between 61 (Saxony) and 214 (North Rhine-Westphalia), except for Baden-Württemberg, where one of the patients had 454 hospitalizations (maximal value) within one state. The average number of hospitalization ranged from 1.9 (Brandenburg) to 2.3 (North Rhine-Westphalia and Thuringia). In all the states, average number of hospitalizations was larger for men than for women, see~Table~\ref{tab2}. The median of number of admissions per person was equal to 1 independently of sex and location. 
	In Figure~\ref{fig1} we present the structure of patient population in the database.
	
	\begin{landscape}
		\thispagestyle{empty}
		\begin{table}[h!]
			\centering
			\caption{Hospitalizations statistic for given location.  States are ordered by their population according to~\cite{Destatis}.}
			\label{tab2}
			\begin{filecontents*}{tabels/general_hosp.tsv}
Land	Patients in database: women	men	Hospitalizations per patient min	max	Average hospitalizations per patient	women	men	Median of hospitalizations per patient	women	men
North Rhine-Westphalia	588914	510315	1	214	2.3	2.2	2.4	1	1	1
Bavaria	302142	252417	1	104	2.1	2.0	2.2	1	1	1
Baden-Württemberg	241475	213983	1	454	2.1	2.0	2.2	1	1	1
Lower Saxony	208065	184622	1	200	2.1	2.0	2.1	1	1	1
Hesse	203685	175437	1	181	2.1	2.0	2.1	1	1	1
Rhineland-Palatinate	115049	101514	1	72	2.0	1.9	2.1	1	1	1
Saxony	52767	54074	1	61	2.2	2.0	2.4	1	1	1
Berlin	196820	149516	1	88	2.1	2.0	2.3	1	1	1
Schleswig-Holstein	102692	84452	1	91	2.0	1.9	2.1	1	1	1
Brandenburg	55166	51171	1	169	1.9	1.8	2.0	1	1	1
Saxony-Anhalt	35750	35184	1	63	2.2	2.0	2.3	1	1	1
Thuringia	32059	32609	1	100	2.3	2.1	2.4	1	1	1
Hamburg	137023	109378	1	100	2.0	1.9	2.1	1	1	1
Mecklenburg-West Pomerania	43889	41124	1	94	2.0	1.9	2.2	1	1	1
Saarland	24070	23026	1	65	2.2	2.1	2.3	1	1	1
Bremen	26418	24168	1	96	2.1	2.0	2.2	1	1	1
\end{filecontents*}
\pgfplotstabletypeset[
			col sep=tab, 
			display columns/0/.style={column name= State, column type={|p{5.4cm}|},string type}, 
			display columns/1/.style={column name= women, column type={p{2cm}|},int detect, 1000 sep={\;},precision=3},
			display columns/2/.style={column type={p{1.5cm}|},int detect, 1000 sep={\;},precision=3},
			display columns/3/.style={column name= min,column type={p{1cm}|}},
			display columns/4/.style={column type={p{1cm}|}},
			display columns/5/.style={column name= in general, column type={p{1.8cm}|}},
			display columns/6/.style={column type={p{1.1cm}|}},
			display columns/7/.style={column name= men, column type={p{1.3cm}|}},
			display columns/8/.style={column name= in general, column type={p{1.8cm}|}},
			display columns/9/.style={column name= women,column type={p{1.3cm}|}},
			display columns/10/.style={column name= men,column type={p{1.1cm}|}},
			every head row/.style={
				before row={
					\hline
					& \multicolumn{2}{p{3cm}|}{No. of patients in
						database} & \multicolumn{2}{p{3cm}|}{No. of hospitalizations
						per patient} & \multicolumn{3}{p{4cm}|}{Average no. of
						hospitalizations
						per patient} & \multicolumn{3}{p{4cm}|}{Median of
						hospitalizations
						per patient}\\
					\cline{2-11} 
				},
				after row=\hline,
			},
			every nth row={1}{before row=\hline},
			every last row/.style={after row=\hline},
			]{tabels/general_hosp.tsv}
		\end{table} 
		\vfill
		\raisebox{0ex}{\makebox[\linewidth]{\thepage}}
	\end{landscape}

	In Table~\ref{tab:average:LOS} we present estimated, directly from the dataset, numbers of patients in healthcare facilities and corresponding societies according to each state. It was assumed that patient discharged from a given hospital which is not immediately admitted to another hospital was counted as a member of the society corresponding to the discharging hospital. To classify and count the sizes of the healthcare facilities and corresponding community nodes we used code available here~\cite{emergenetpackage}.   
	In addition, we report the average length of stays in healthcare facilities, which varied between 8.2 (Thuringia) and 9.5 (Bremen). The average length of time spent at home between two consecutive admissions for a single person ranged from 243.1 (Bremen) and 293.4 (Hesse), while the average number of patients in healthcare facilities per day was highest in the state with largest population (North Rhine-Westphalia).
	Both Hamburg and Berlin had large number of patients in hospitals compared to their population.

	\clearpage

	\begin{table}[h!]
		\caption{Estimated from dataset number of patients in healthcare facilities and corresponding societies. Average length of stay in healthcare facilities and societies for given location. States are ordered by their population according to~\cite{Destatis}.
		} \label{tab:average:LOS}
		\begin{filecontents*}{tabels/general_stay.tsv}
Land	Estimated patients: facilities 	ghost-nodes	Hospital stay mean   (standard deviation) [days]	Society stay mean (standard deviation) [days]
North Rhine-Westphalia	10075	179841	8.3 (12.3)	279.5(352.2)
Bavaria	4812	77833	8.6 (14.0)	283.3 (356.8)
Baden-Württemberg	3864	61558	8.4 (13.4)	274.5 (359.5)
Lower Saxony	3320	53872	8.5 (13.6)	287.6 (359.8)
Hesse	3321	52506	8.9 (15.0)	293.4 (363.5)
Rhineland-Palatinate	1726	28598	8.3 (11.7)	288.5 (359.8)
Saxony	957	15291	8.6 (13.0)	271.0 (350.1)
Berlin	2986	48898	8.4 (13.3)	275.6 (354.7)
Schleswig-Holstein	1507	23613	8.4 (13.4)	281.6 (359.1)
Brandenburg	858	12250	8.7 (12.9)	275.1 (358.8)
Saxony-Anhalt	619	10314	8.4 (13.0)	276.2 (354.8)
Thuringia	579	9607	8.2 (12.0)	254.3 (343.3)
Hamburg	2066	31181	8.8 (13.4)	284.6 (357.2)
Mecklenburg-West \mbox{Pomerania}	708	11139	8.5 (13.0)	277.0 (358.9)
Saarland	425	7028	8.5 (12.5)	272.0 (351.4)
Bremen	486	6123	9.5 (15.0)	243.1 (338.1)
\end{filecontents*}
\pgfplotstabletypeset[
		col sep=tab, 
		display columns/0/.style={column name= State, column type={|p{2.4cm}|},string type}, 
		display columns/1/.style={column name= facilities,column type={p{2cm}|}, int detect, 1000 sep={\;},precision=3},
		display columns/2/.style={column name= societies, column type={p{2cm}|}, int detect, 1000 sep={\;},precision=3},
		display columns/3/.style={column type={p{2.8cm}|},string type},
		display columns/4/.style={column type={p{2.8cm}|},string type},
		every head row/.style={
			before row={
				\hline
				& \multicolumn{2}{c|}{Estimated no. of patients in} &  &\\
				\cline{2-3} 
			},
			after row=\hline,
		},
		every nth row={1}{before row=\hline},
		every last row/.style={after row=\hline},
		]{tabels/general_stay.tsv}
	\end{table}

	In Figure~\ref{fig:hosp_social} we present the relationship between sizes of healthcare facilities and corresponding communities sizes. 
	For almost all states we  distinguished two groups: one consists of small hospitals, which had a relatively small community sizes and second (larger one) where facilities had larger corresponding communities. 

	\begin{figure}[h!]
	\centering
	\begin{subfigure}[b]{0.45\textwidth}
		\centering
		\includegraphics[width=\textwidth]{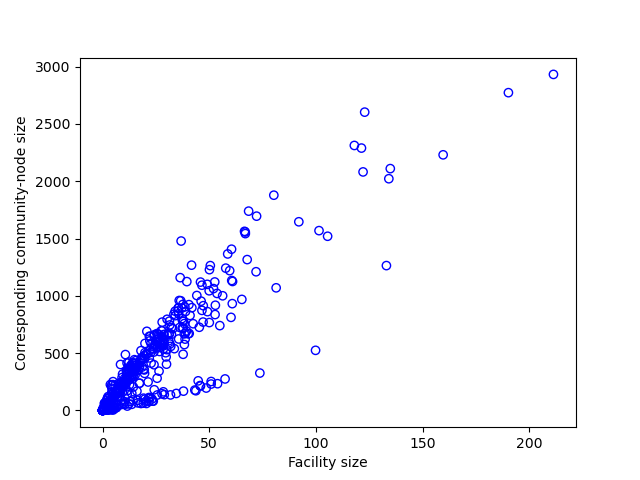}
		\caption{North Rhine-Westphalia}
	\end{subfigure}
	\hfill
	\begin{subfigure}[b]{0.45\textwidth}
		\centering
		\includegraphics[width=\textwidth]{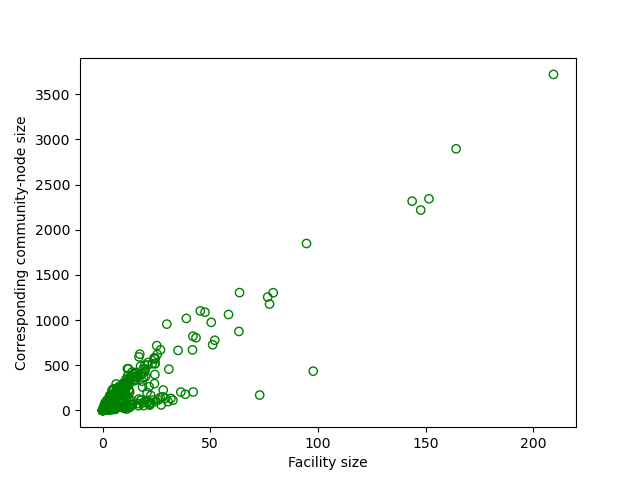}
		\caption{Bavaria}
	\end{subfigure}
	
	\begin{subfigure}[b]{0.45\textwidth}
		\centering
		\includegraphics[width=\textwidth]{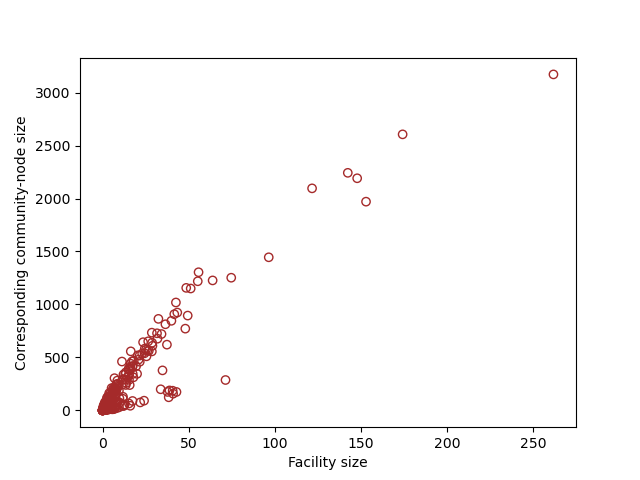}
		\caption{Baden-Württemberg}
	\end{subfigure}
	\hfill
	\begin{subfigure}[b]{0.45\textwidth}
		\centering
		\includegraphics[width=\textwidth]{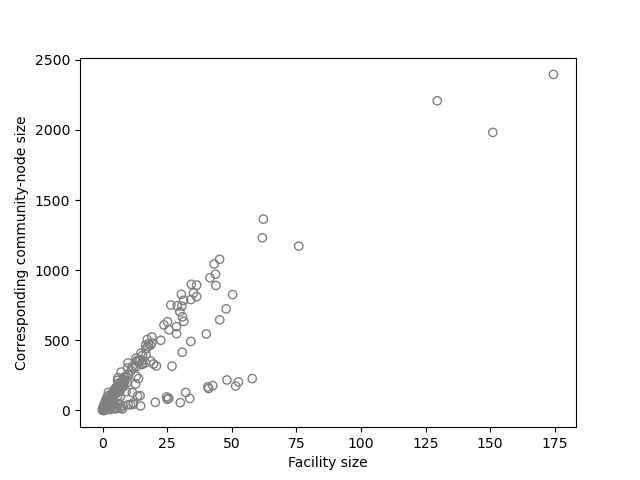}
		\caption{Lower Saxony}
	\end{subfigure}
	
	\begin{subfigure}[b]{0.45\textwidth}
		\centering
		\includegraphics[width=\textwidth]{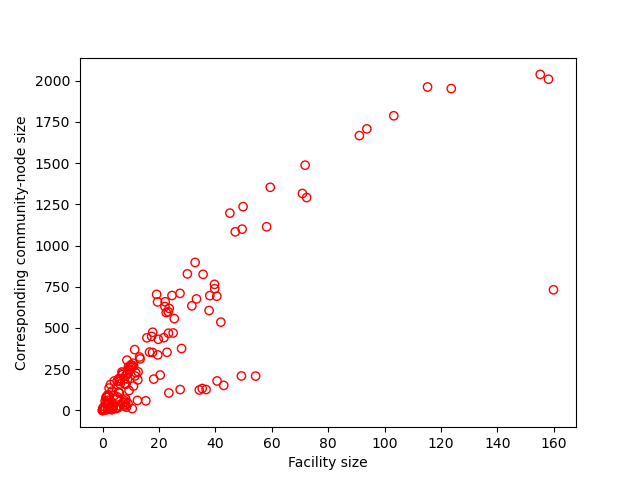}
		\caption{Hesse}
	\end{subfigure}
	\hfill
	\begin{subfigure}[b]{0.45\textwidth}
		\centering
		\includegraphics[width=\textwidth]{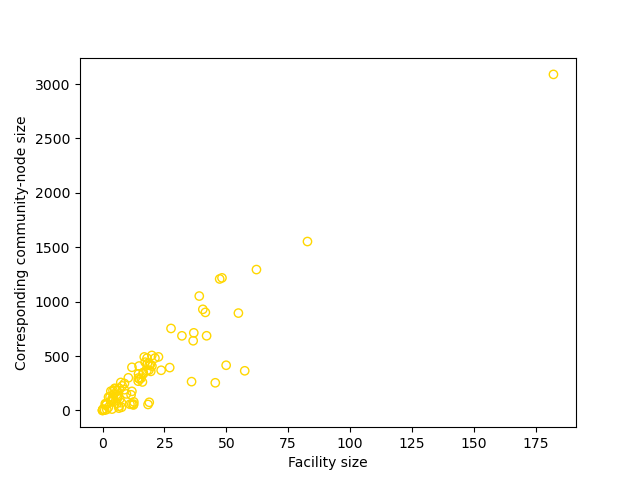}
		\caption{Rhineland-Palatinate}
	\end{subfigure}
\end{figure}
\clearpage   
\begin{figure}[tb]\ContinuedFloat         
	\begin{subfigure}[b]{0.45\textwidth}
		\centering
		\includegraphics[width=\textwidth]{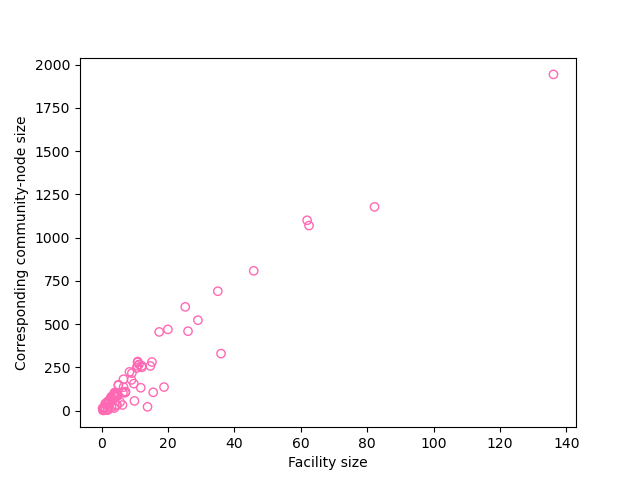}
		\caption{Saxony}
	\end{subfigure}
	\hfill
	\begin{subfigure}[b]{0.45\textwidth}
		\centering
		\includegraphics[width=\textwidth]{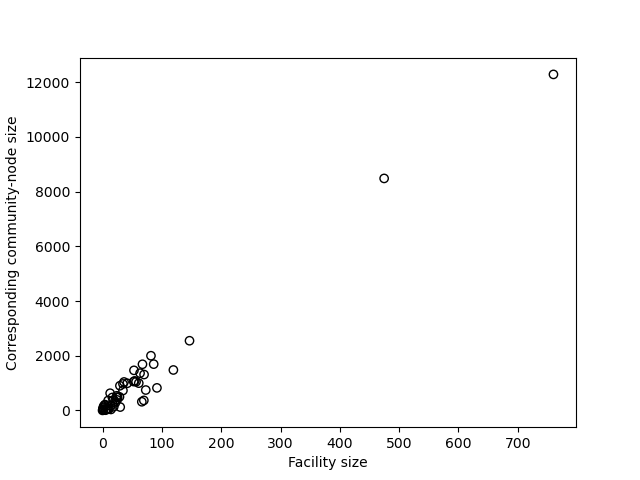}
		\caption{Berlin}
	\end{subfigure}
	
	\begin{subfigure}[b]{0.45\textwidth}
		\centering
		\includegraphics[width=\textwidth]{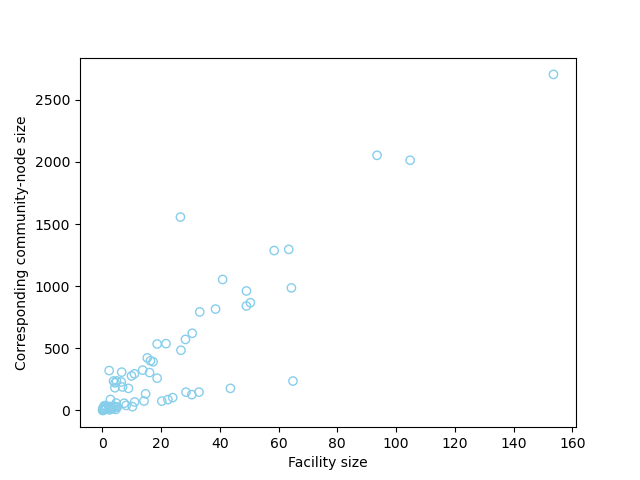}
		\caption{Schleswig-Holstein}
	\end{subfigure}
	\hfill
	\begin{subfigure}[b]{0.45\textwidth}
		\centering
		\includegraphics[width=\textwidth]{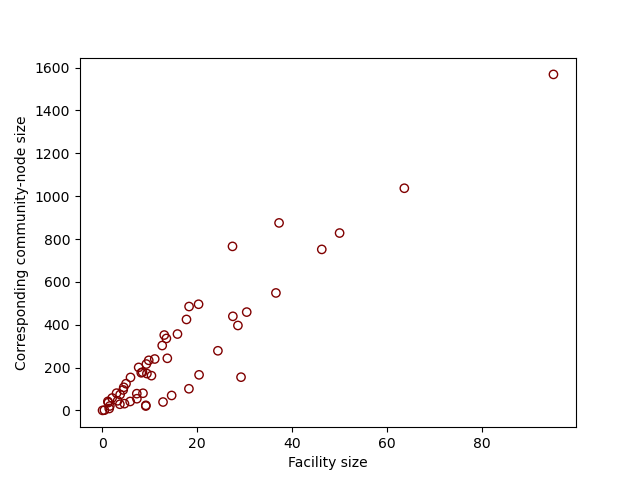}
		\caption{Brandenburg}
	\end{subfigure}   
	
	\begin{subfigure}[b]{0.45\textwidth}
		\centering
		\includegraphics[width=\textwidth]{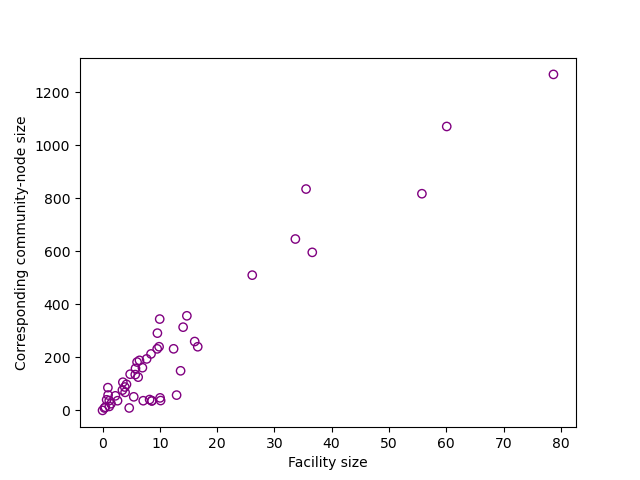}
		\caption{Saxony-Anhalt}
	\end{subfigure}
	\hfill
	\begin{subfigure}[b]{0.45\textwidth}
		\centering
		\includegraphics[width=\textwidth]{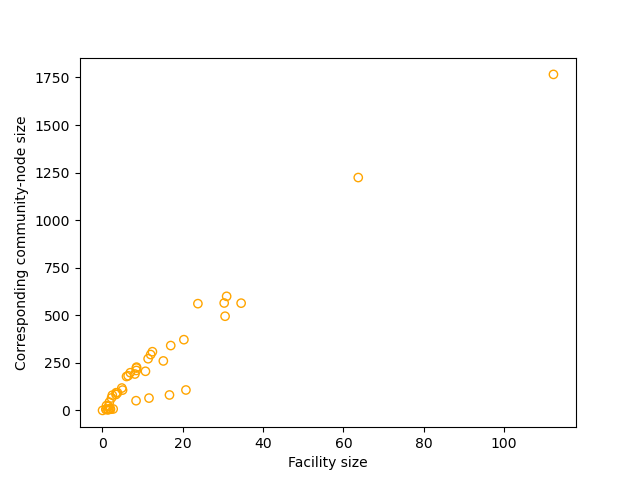}
		\caption{Thuringia}
	\end{subfigure}  
\end{figure}
\clearpage   
\begin{figure}[tbh!]\ContinuedFloat 
	\begin{subfigure}[b]{0.45\textwidth}
		\centering
		\includegraphics[width=\textwidth]{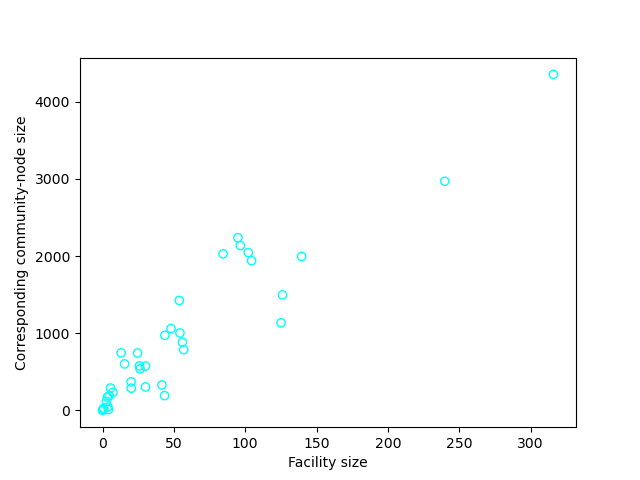}
		\caption{Hamburg}
	\end{subfigure}
	\hfill
	\begin{subfigure}[b]{0.45\textwidth}
		\centering
		\includegraphics[width=\textwidth]{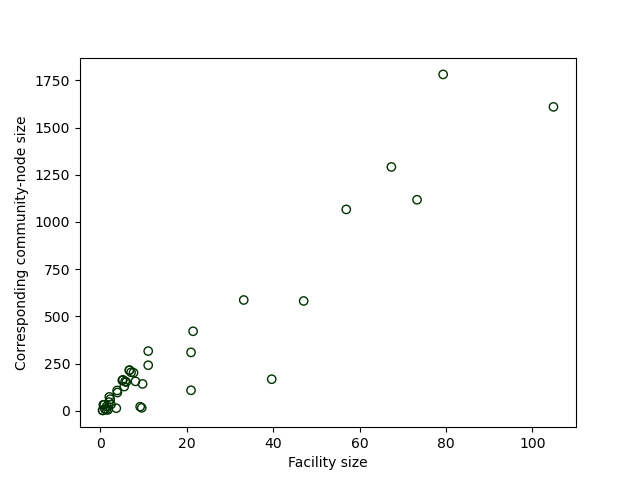}
		\caption{Mecklenburg-West Pomerania}
	\end{subfigure} 
	
	\begin{subfigure}[b]{0.45\textwidth}
		\centering
		\includegraphics[width=\textwidth]{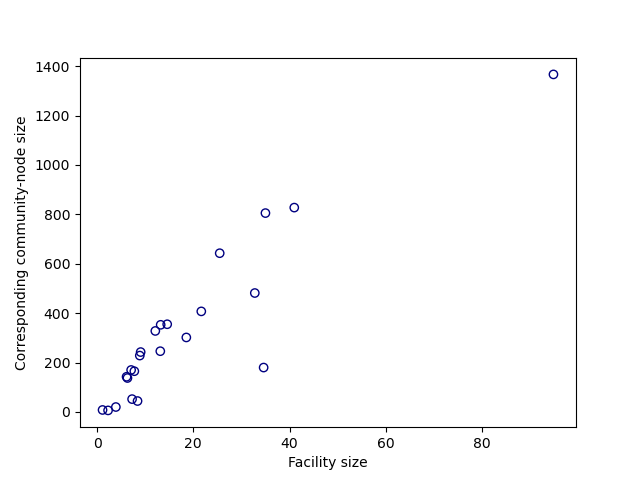}
		\caption{Saarland}
	\end{subfigure}
	\hfill
	\begin{subfigure}[b]{0.45\textwidth}
		\centering
		\includegraphics[width=\textwidth]{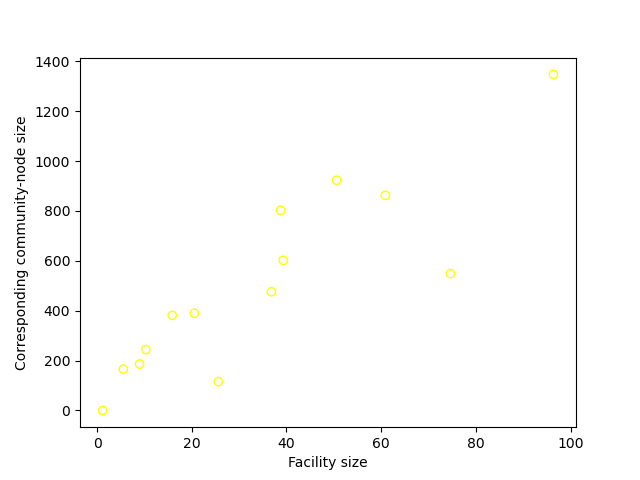}
		\caption{Bremen}
	\end{subfigure} 
	\caption{Dependence of community-node sizes on corresponding healthcare facilities. }
	\label{fig:hosp_social}
\end{figure}

	\newpage
	
	\subsection{Admissions}
	Let focus on the characterisation of the healthcare facilities reported in the database. 
	In Figure~\ref{fig4} we see that in all states most of the healthcare facilities had between 1\,000 and 9\,999 admissions except Hamburg where most of facilities had between 10 000 and 99 999 admissions. Moreover, in all the states hospitals had less than 100 000 admissions apart from Berlin.

\begin{landscape}
	\thispagestyle{empty}
	\begin{figure}[h!]
		\centering
		\includegraphics[height=12cm]{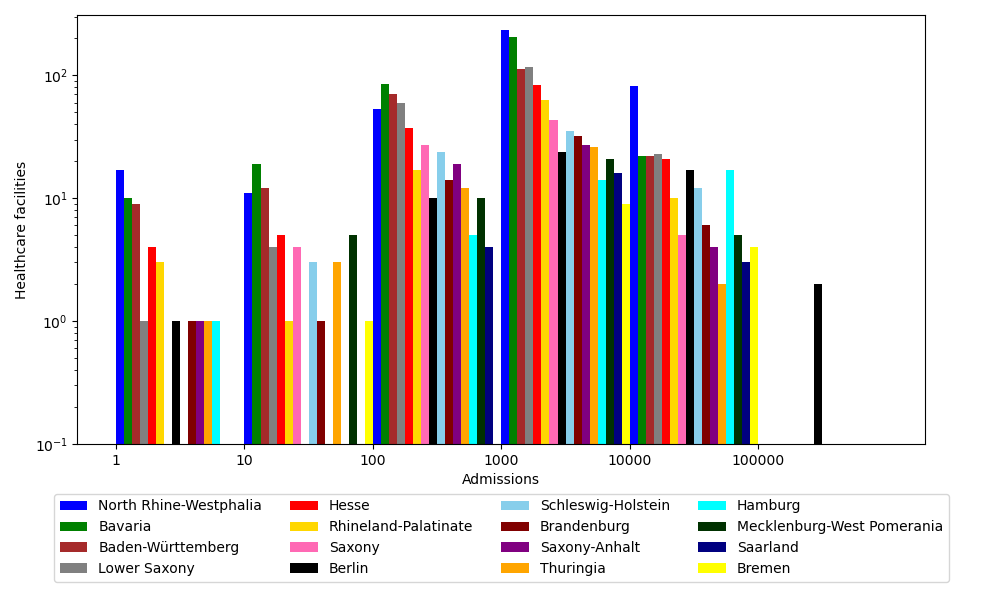}
		\caption{Number of healthcare facilities with given number of admissions.}
		\label{fig4}
	\end{figure}
	\vfill
	\raisebox{0ex}{\makebox[\linewidth]{\thepage}}
\end{landscape}

	\subsection{Number of patients}
	Further more, if we look at  the number of patients admitted to the hospitals (Figure~\ref{fig6}) we see that in almost all states most of the healthcare facilities had between 1 000 and 9 999 patients in the dataset. The only exception was Saxony -- most of facilities had between 100 and 999 patients within considered time period. In addition, we see that Berlin was the only state with hospitals which admitted more then 100 000 patients.

\begin{landscape}
	\thispagestyle{empty}
	\begin{figure}[h!]
		\centering
		\includegraphics[height=12cm]{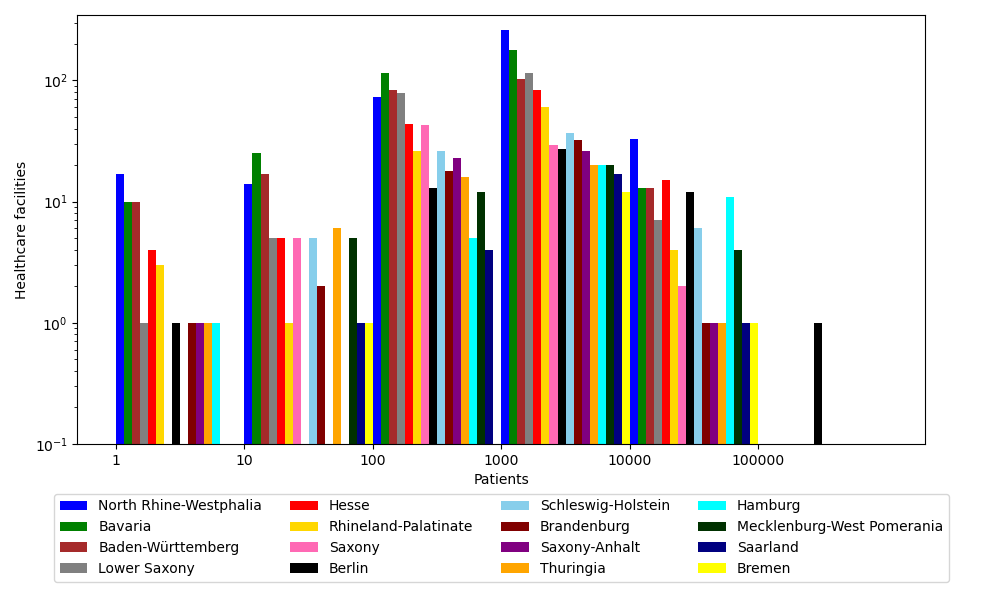}
		\caption{Number of healthcare facilities with a given number of patients depending on the state.}
		\label{fig6}
	\end{figure}
	\vfill
	\raisebox{0ex}{\makebox[\linewidth]{\thepage}}
\end{landscape}

	\newpage
	
	\subsection{Average length of stay}
	We have investigated duration of reported stays of patients in particular healthcare facilities for all German states. In Figure~\ref{fig_hop_stay} we present the histograms of the duration of the hospitalisations for all healthcare facilities in given location.
	Furthermore, in Figure~\ref{fig:society_len} we present the dependence between the duration of the length of stay in society between two admission and a number of such stays in given state.
	In all the states, the majority of the hospitalizations was under ten days, and most of them were 2 or 3 days long. Number of hospitalizations longer then 30 days corresponds to only 4.58\% of all data.
	Only for one record in dataset the duration of time spend in healthcare facility exceeded 1 000 days.
	From Figure~\ref{fig:society_len} we deduce that for all the states the structure of the length of stays in society (i.e. between hospitalizations in given state only -- interstate records were omitted) was similar.
	We observe a significant drop of the number of stays in society shorter than 250 days. For longer stays we still see the decrease in number of stays but it is not that fast. Finally in some states, especially ones with highest population (according to~\cite{Destatis}), we see another sharp drop after 1 500 days.	

	\begin{figure}[h!]
		\centering
		\begin{subfigure}[b]{0.45\textwidth}
			\centering
			\includegraphics[width=\textwidth]{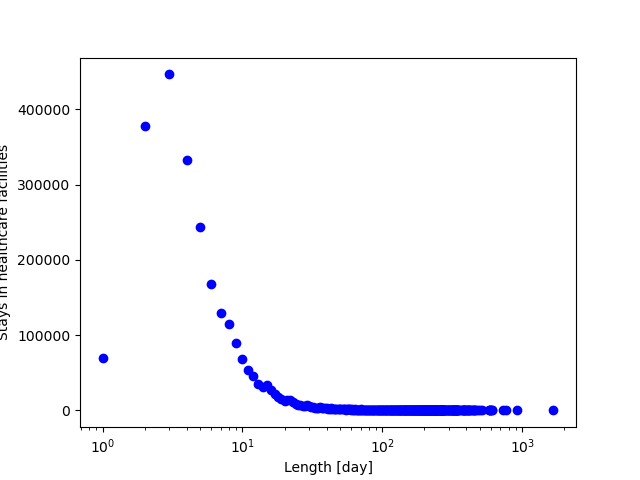}
			\caption{North Rhine-Westphalia}
		\end{subfigure}
		\hfill
		\begin{subfigure}[b]{0.45\textwidth}
			\centering
			\includegraphics[width=\textwidth]{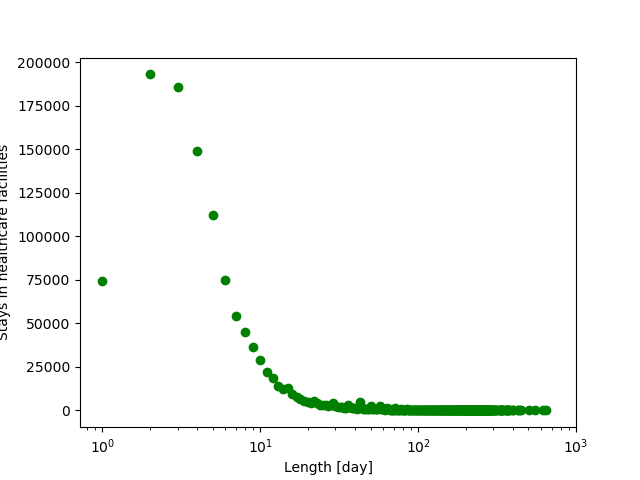}
			\caption{Bavaria}
		\end{subfigure}
		
		\begin{subfigure}[b]{0.45\textwidth}
			\centering
			\includegraphics[width=\textwidth]{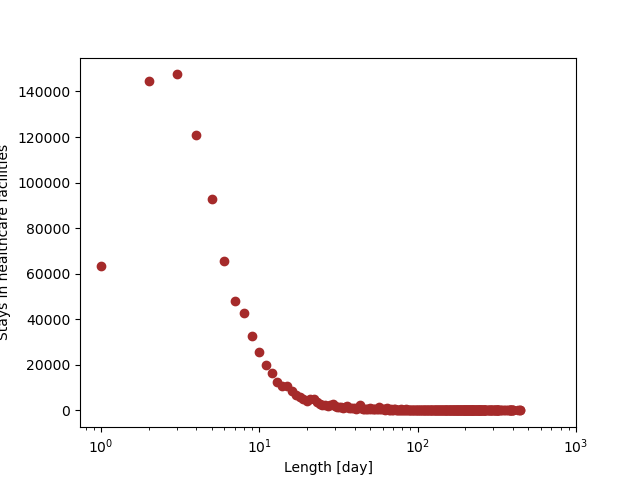}
			\caption{Baden-Württemberg}
		\end{subfigure}
		\hfill
		\begin{subfigure}[b]{0.45\textwidth}
			\centering
			\includegraphics[width=\textwidth]{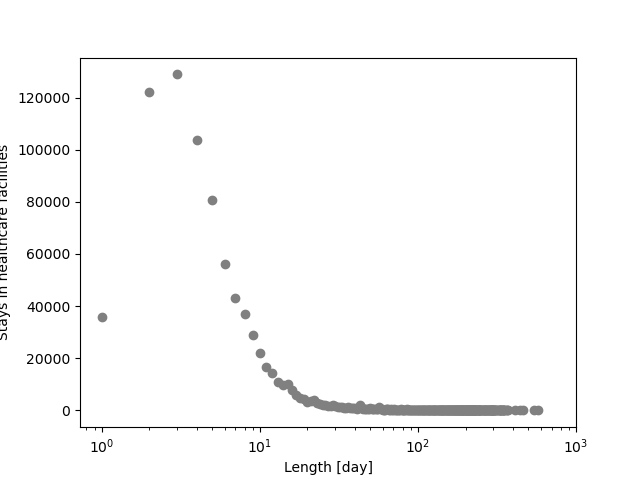}
			\caption{Lower Saxony}
		\end{subfigure}
	\end{figure}
	\clearpage   
	\begin{figure}[tb]\ContinuedFloat 		  		
		\begin{subfigure}[b]{0.45\textwidth}
			\centering
			\includegraphics[width=\textwidth]{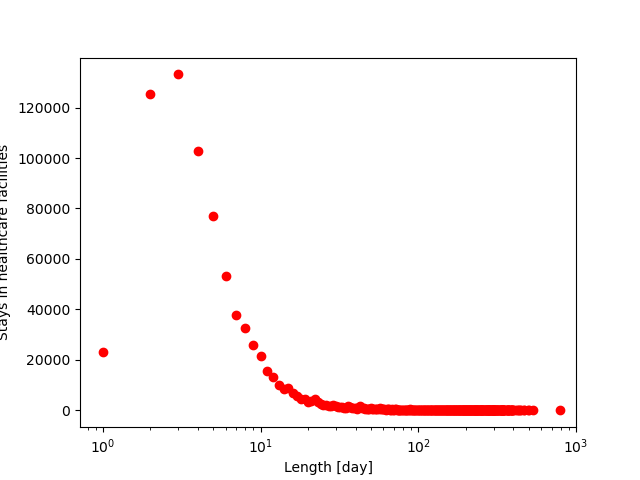}
			\caption{Hesse}
		\end{subfigure}
		\hfill
		\begin{subfigure}[b]{0.45\textwidth}
			\centering
			\includegraphics[width=\textwidth]{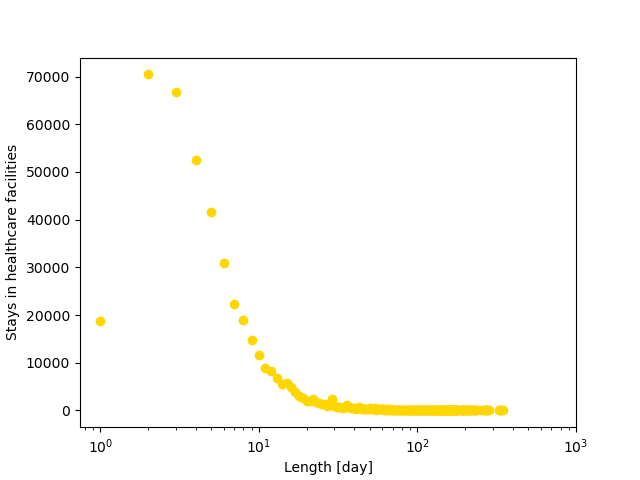}
			\caption{Rhineland-Palatinate}
		\end{subfigure}
       
		\begin{subfigure}[b]{0.45\textwidth}
			\centering
			\includegraphics[width=\textwidth]{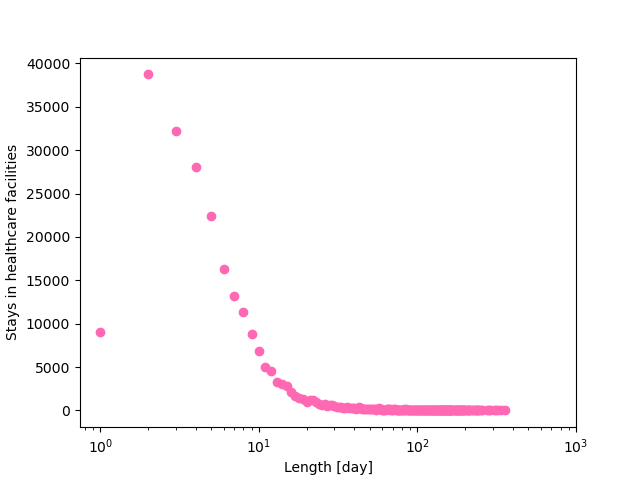}
			\caption{Saxony}
		\end{subfigure}
		\hfill
		\begin{subfigure}[b]{0.45\textwidth}
			\centering
			\includegraphics[width=\textwidth]{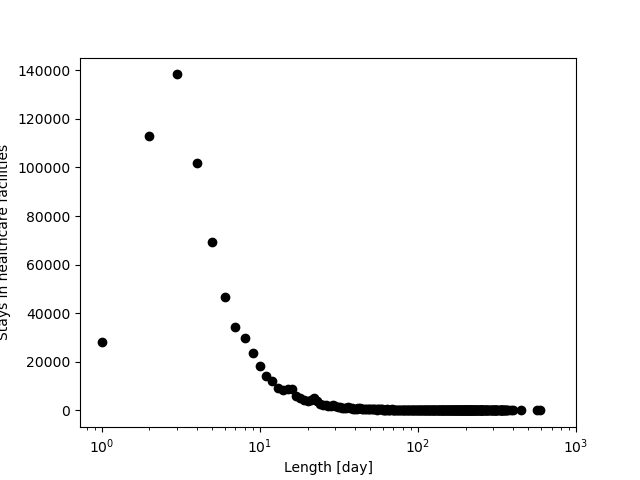}
			\caption{Berlin}
		\end{subfigure}
		
		\begin{subfigure}[b]{0.45\textwidth}
			\centering
			\includegraphics[width=\textwidth]{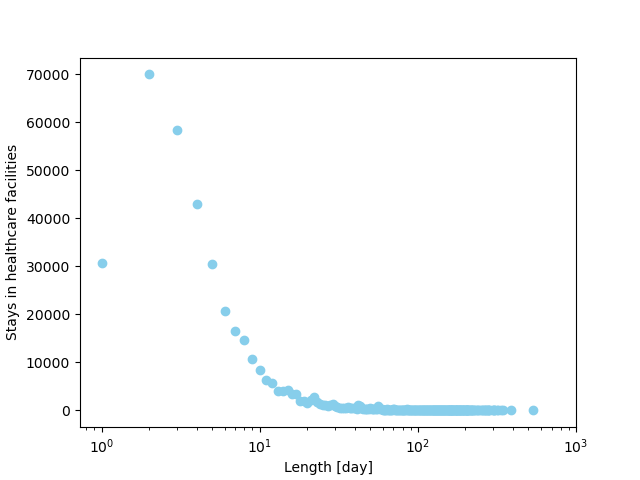}
			\caption{Schleswig-Holstein}
		\end{subfigure}
		\hfill
		\begin{subfigure}[b]{0.45\textwidth}
			\centering
			\includegraphics[width=\textwidth]{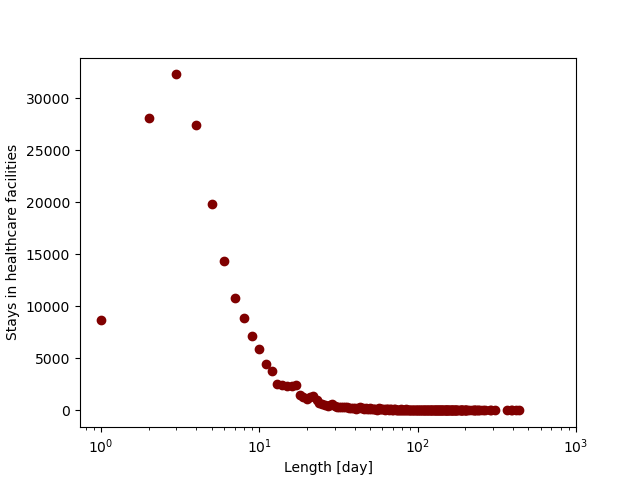}
			\caption{Brandenburg}
		\end{subfigure}   
	\end{figure}
	\clearpage   
	\begin{figure}[tb]\ContinuedFloat 		
		\begin{subfigure}[b]{0.45\textwidth}
			\centering
			\includegraphics[width=\textwidth]{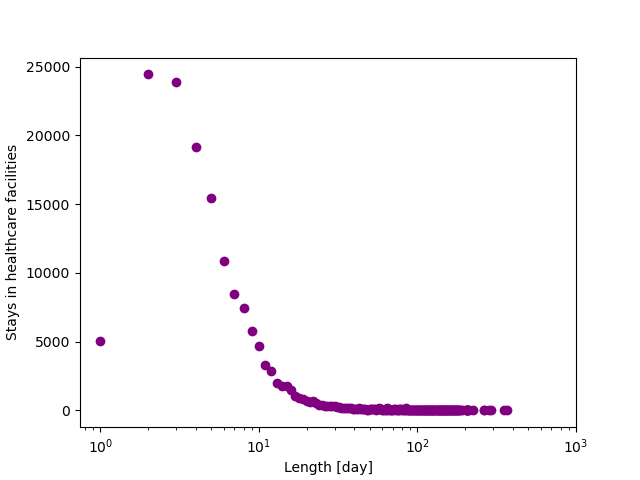}
			\caption{Saxony-Anhalt}
		\end{subfigure}
		\hfill
		\begin{subfigure}[b]{0.45\textwidth}
			\centering
			\includegraphics[width=\textwidth]{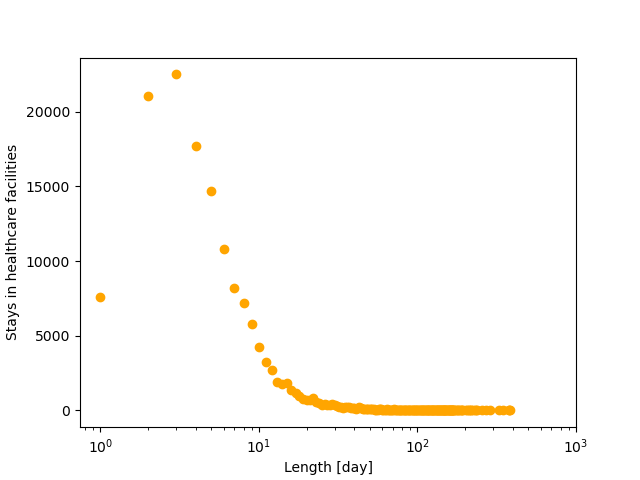}
			\caption{Thuringia}
		\end{subfigure}  

		\begin{subfigure}[b]{0.45\textwidth}
			\centering
			\includegraphics[width=\textwidth]{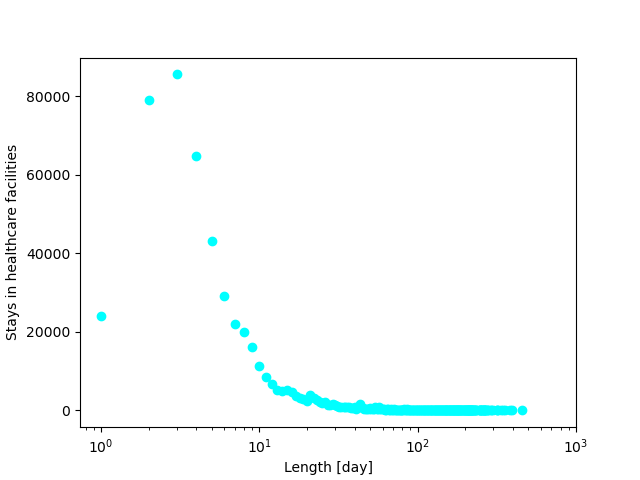}
			\caption{Hamburg}
		\end{subfigure}
		\hfill
		\begin{subfigure}[b]{0.45\textwidth}
			\centering
			\includegraphics[width=\textwidth]{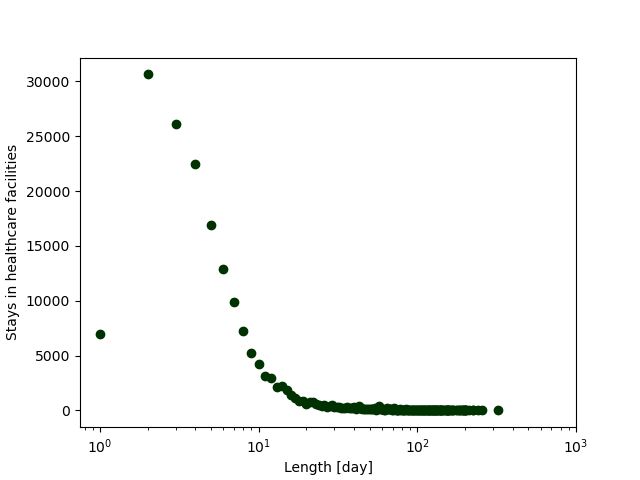}
			\caption{Mecklenburg-West Pomerania}
		\end{subfigure} 
		
		\begin{subfigure}[b]{0.45\textwidth}
			\centering
			\includegraphics[width=\textwidth]{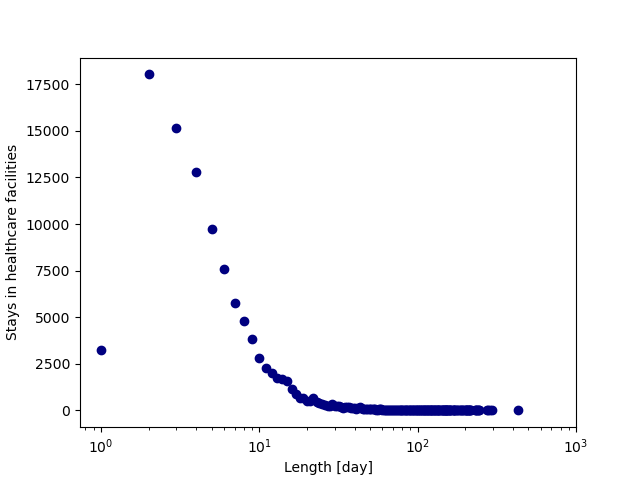}
			\caption{Saarland}
		\end{subfigure}
		\hfill
		\begin{subfigure}[b]{0.45\textwidth}
			\centering
			\includegraphics[width=\textwidth]{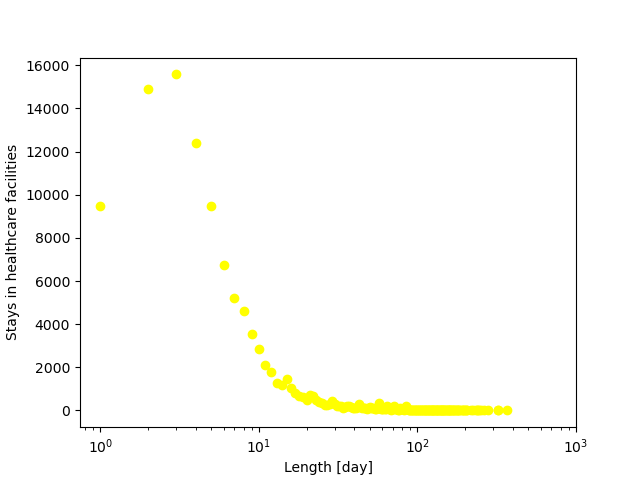}
			\caption{Bremen}
		\end{subfigure} 
		\caption{Duration of patients stays in healthcare facilities located in considered state.}
		\label{fig_hop_stay}
	\end{figure}
	
	\begin{figure}
		\centering
		\begin{subfigure}[b]{0.45\textwidth}
			\centering
			\includegraphics[width=\textwidth]{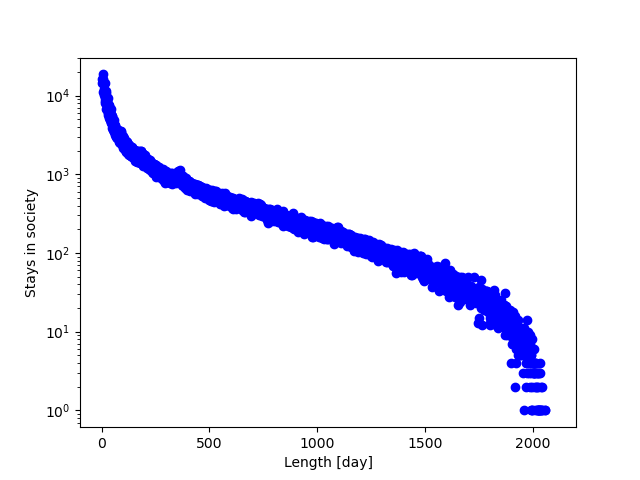}
			\caption{North Rhine-Westphalia}
		\end{subfigure}
		\hfill
		\begin{subfigure}[b]{0.45\textwidth}
			\centering
			\includegraphics[width=\textwidth]{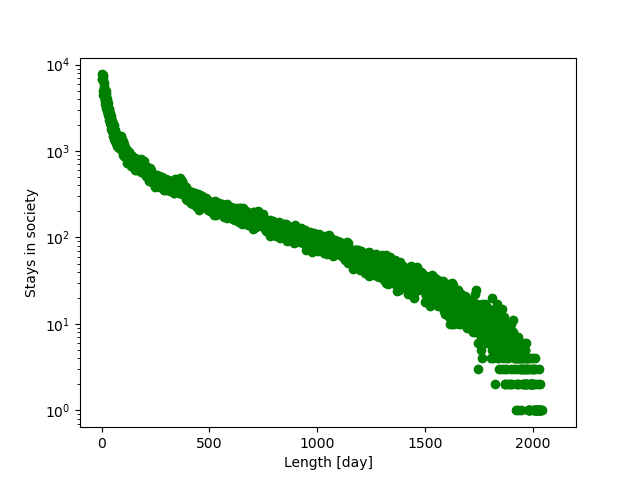}
			\caption{Bavaria}
		\end{subfigure}
		\begin{subfigure}[b]{0.45\textwidth}
			\centering
			\includegraphics[width=\textwidth]{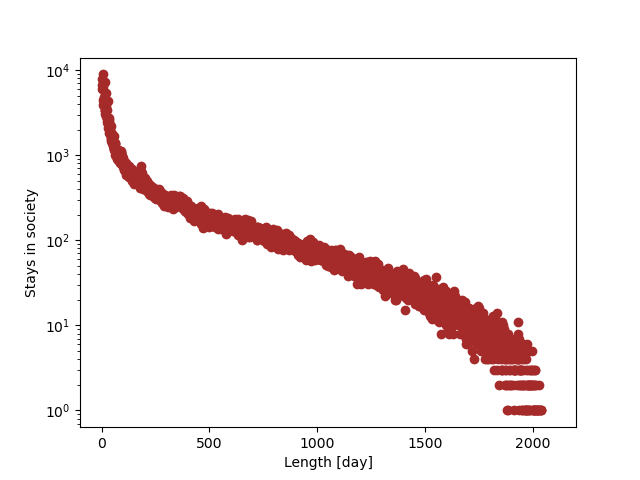}
			\caption{Baden-Württemberg}
		\end{subfigure}
		\hfill
		\begin{subfigure}[b]{0.45\textwidth}
			\centering
			\includegraphics[width=\textwidth]{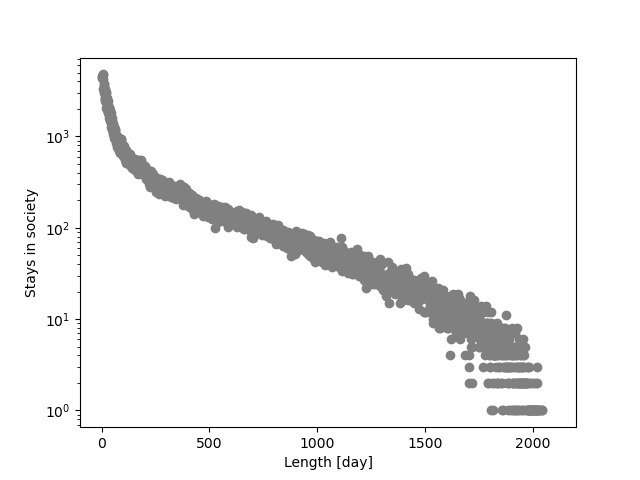}
			\caption{Lower Saxony}
		\end{subfigure}
		
		\begin{subfigure}[b]{0.45\textwidth}
			\centering
			\includegraphics[width=\textwidth]{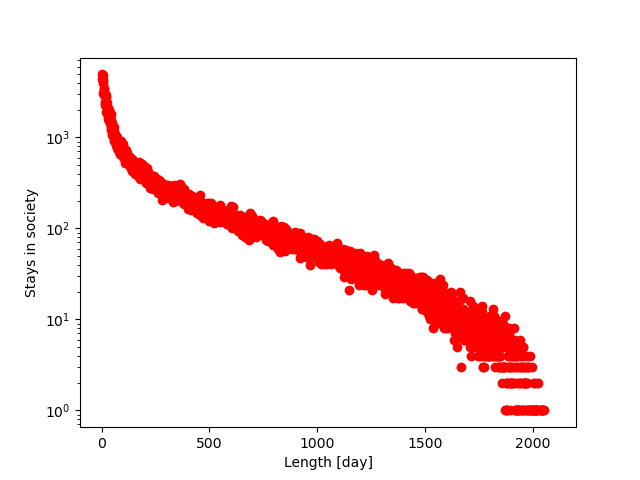}
			\caption{Hesse}
		\end{subfigure}
		\hfill
		\begin{subfigure}[b]{0.45\textwidth}
			\centering
			\includegraphics[width=\textwidth]{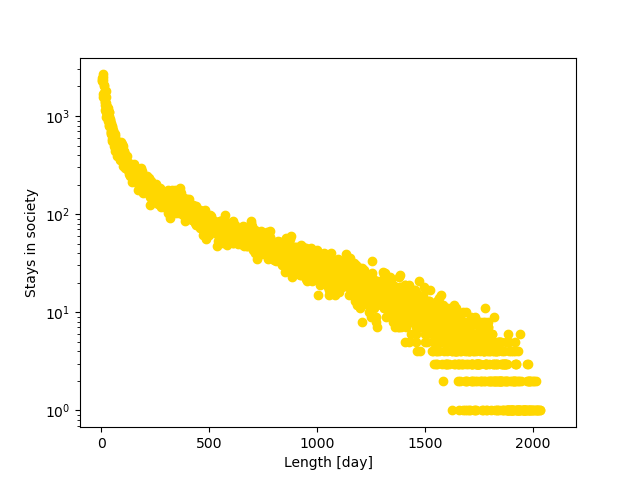}
			\caption{Rhineland-Palatinate}
		\end{subfigure}
	\end{figure}
	\clearpage   
	\begin{figure}[tb]\ContinuedFloat         
		\begin{subfigure}[b]{0.45\textwidth}
			\centering
			\includegraphics[width=\textwidth]{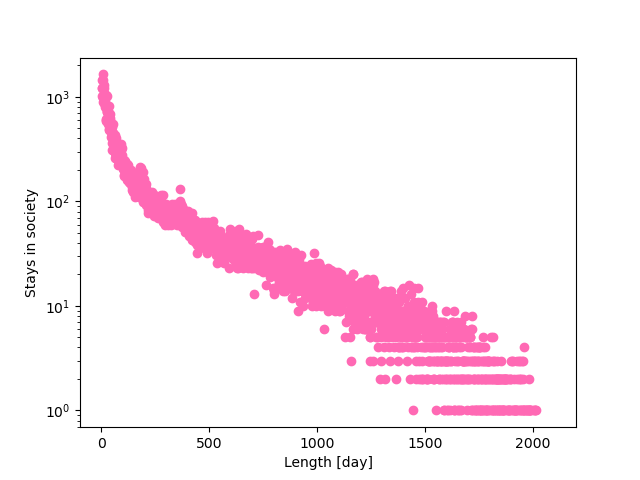}
			\caption{Saxony}
		\end{subfigure}
		\hfill
		\begin{subfigure}[b]{0.45\textwidth}
			\centering
			\includegraphics[width=\textwidth]{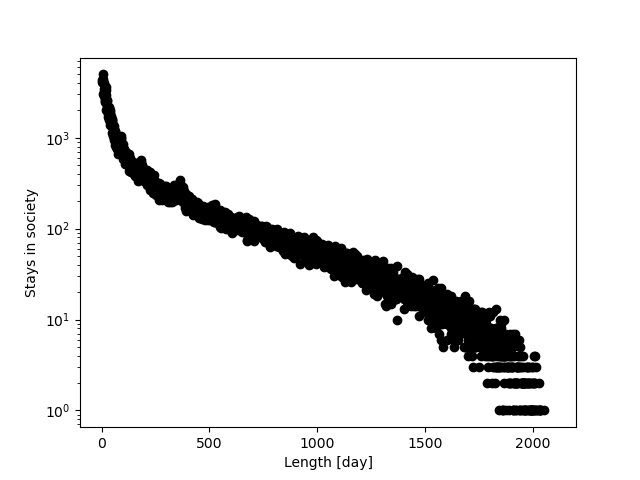}
			\caption{Berlin}
		\end{subfigure}
		
		\begin{subfigure}[b]{0.45\textwidth}
			\centering
			\includegraphics[width=\textwidth]{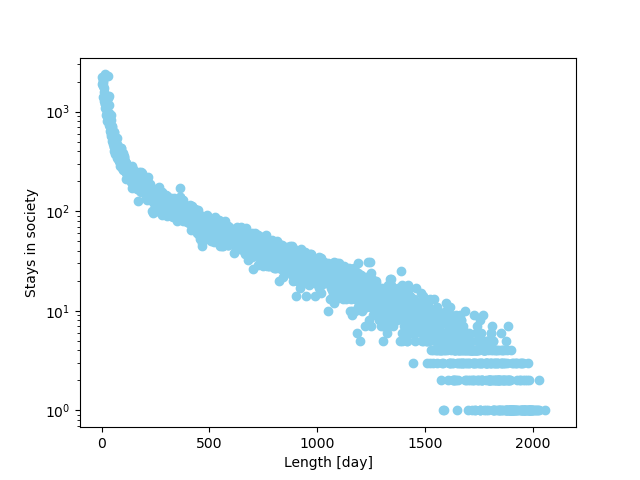}
			\caption{Schleswig-Holstein}
		\end{subfigure}
		\hfill
		\begin{subfigure}[b]{0.45\textwidth}
			\centering
			\includegraphics[width=\textwidth]{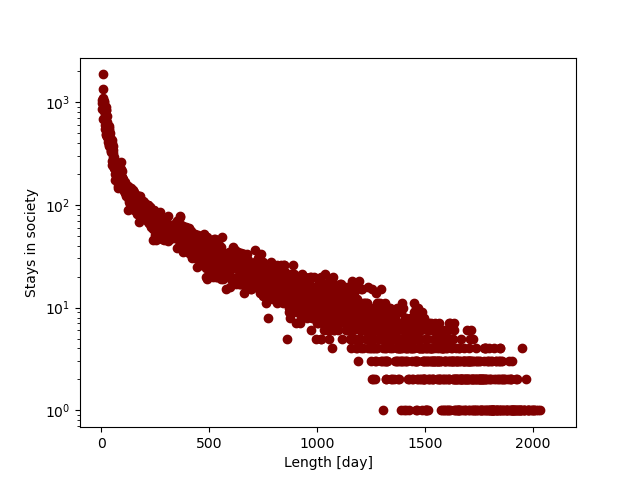}
			\caption{Brandenburg}
		\end{subfigure}   
		
		\begin{subfigure}[b]{0.45\textwidth}
			\centering
			\includegraphics[width=\textwidth]{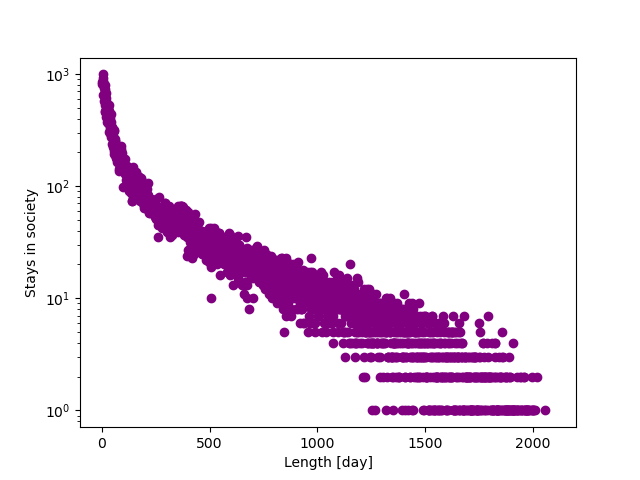}
			\caption{Saxony-Anhalt}
		\end{subfigure}
		\hfill
		\begin{subfigure}[b]{0.45\textwidth}
			\centering
			\includegraphics[width=\textwidth]{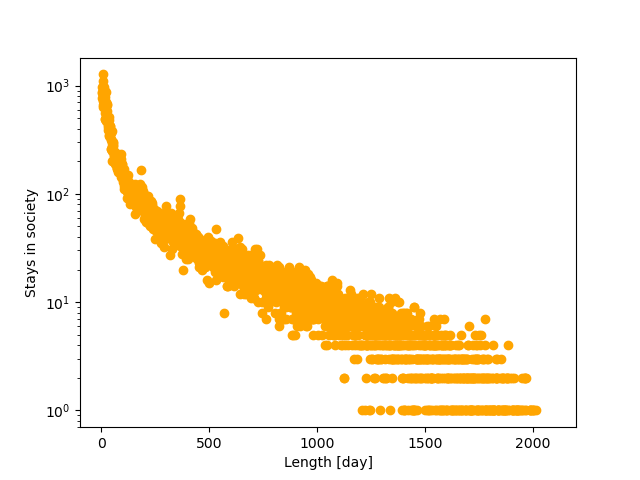}
			\caption{Thuringia}
		\end{subfigure}  
	\end{figure}
	\clearpage   
	\begin{figure}[tb]\ContinuedFloat  
		\begin{subfigure}[b]{0.45\textwidth}
			\centering
			\includegraphics[width=\textwidth]{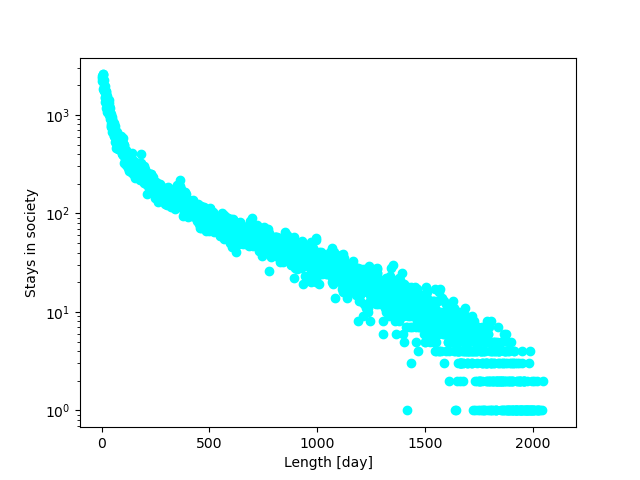}
			\caption{Hamburg}
		\end{subfigure}
		\hfill
		\begin{subfigure}[b]{0.45\textwidth}
			\centering
			\includegraphics[width=\textwidth]{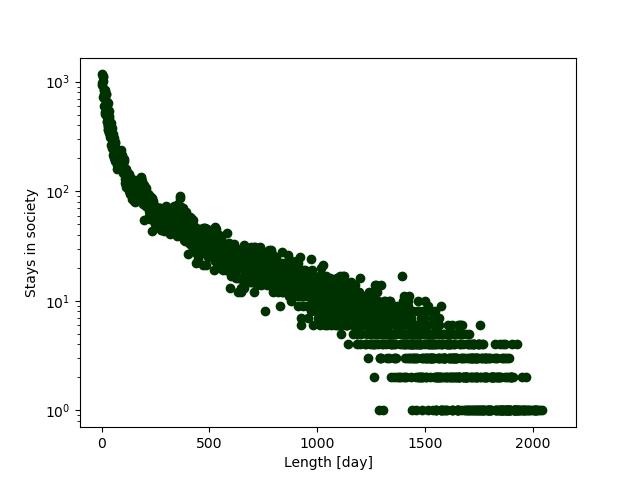}
			\caption{Mecklenburg-West Pomerania}
		\end{subfigure} 
		
		\begin{subfigure}[b]{0.45\textwidth}
			\centering
			\includegraphics[width=\textwidth]{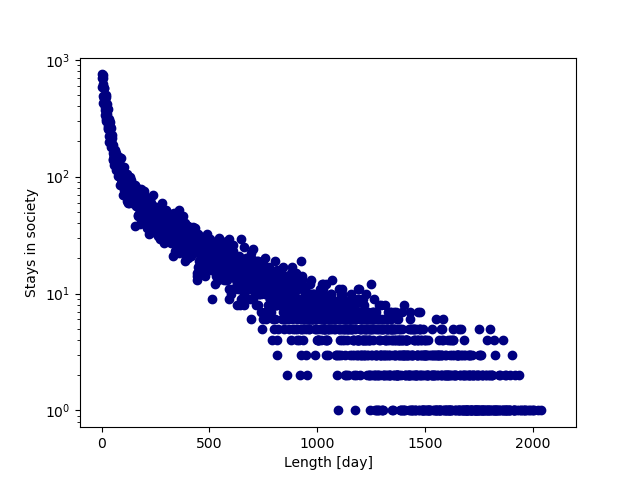}
			\caption{Saarland}
		\end{subfigure}
		\hfill
		\begin{subfigure}[b]{0.45\textwidth}
			\centering
			\includegraphics[width=\textwidth]{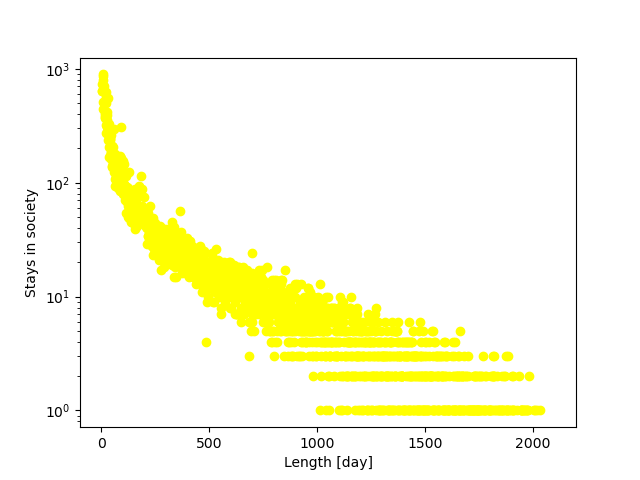}
			\caption{Bremen}
		\end{subfigure} 
		\caption{Duration of patients stays in society for given location.}
		\label{fig:society_len}
	\end{figure}

	\subsection{Patient transfers}
	
	For provided dataset, following requirement of modelling framework proposed in e.g. \cite{PiotrowskaSIVW2020, Piotrowska2020PlosComp}, we analysed patient transfers in detail. In general, we distinguished between two types of them: direct transfers between healthcare facilities (without patient stay at home period) and indirect transfer (otherwise). The indirect transfers were further divided into two types: auto-transfer (readmission to the same hospital) and  different destination (admission to a different hospital). We detected  5 227 973 patient transfers: 4 774 258 within the same state and 453 715 (8.7\% of all transfers) between states. Considerably high number of transfers occurred in Berlin and Hamburg relative to their populations~\cite{Destatis}.
	
	\begin{table}[h!]
		\caption{Number of patient transfers for given state. States are ordered by their population according to~\cite{Destatis}.}
		\label{tab:transfer}
		\begin{filecontents*}{tabels/transfer_bet.tsv}
State	No. of all transfers	direct transfers	indirect transfers	auto-transfers	different destination
North Rhine-Westphalia	  1392899	    86751	  1306148	   720498	   585650
Bavaria	   591704	    34905	   556799	   315343	   241456
Baden-Württemberg	   479669	    28488	   451181	   285102	   166079
Lower Saxony	   393139	    24452	   368687	   225335	   143352
Hesse	   379755	    24425	   355330	   203396	   151934
Rhineland-Palatinate	   202829	    10127	   192702	   121787	    70915
Saxony	   119331	     5774	   113557	    71713	    41844
Berlin	   377911	    23174	   354737	   213118	   141619
Schleswig-Holstein	   174432	    10638	   163794	   107560	    56234
Brandenburg	    90150	     4820	    85330	    62198	    23132
Saxony-Anhalt	    77638	     3333	    74305	    48532	    25773
Thuringia	    78111	     2919	    75192	    55079	    20113
Hamburg	   226446	    13217	   213229	   127535	    85694
Mecklenburg-West \mbox{Pomerania}	    84136	     3516	    80620	    56993	    23627
Saarland	    55040	     3743	    51297	    29370	    21927
Bremen	    51068	     2712	    48356	    33817	    14539
Between	   453715	    37248	   416467	        0	   416467
\end{filecontents*}
\pgfplotstabletypeset[
		col sep=tab, 
		display columns/0/.style={column type={|p{2.5cm}|},string type}, 
		display columns/1/.style={column type={p{2cm}|}, int detect, 1000 sep={\;},precision=3},
		display columns/2/.style={column name= Direct transfers, column type={p{1.8cm}|}, int detect, 1000 sep={\;},precision=3},
		display columns/3/.style={column name= all,column type={p{2cm}|}, int detect, 1000 sep={\;},precision=3},
		display columns/4/.style={column type={p{1.5cm}|}, int detect, 1000 sep={\;},precision=3},
		display columns/5/.style={column type={p{2cm}|}, int detect, 1000 sep={\;},precision=3},
		every head row/.style={
			before row={
				\hline
				& & &\multicolumn{3}{c|}{Indirect transfers}\\
				\cline{4-6} 
			},
			after row=\hline,
		},
		every nth row={1}{before row=\hline},
		every last row/.style={ after row=\hline},
		]{tabels/transfer_bet.tsv}
	\end{table}

	\clearpage
  
\begin{figure}
	\centering
	\begin{subfigure}[b]{0.45\textwidth}
		\centering
		\includegraphics[width=1.2\textwidth]{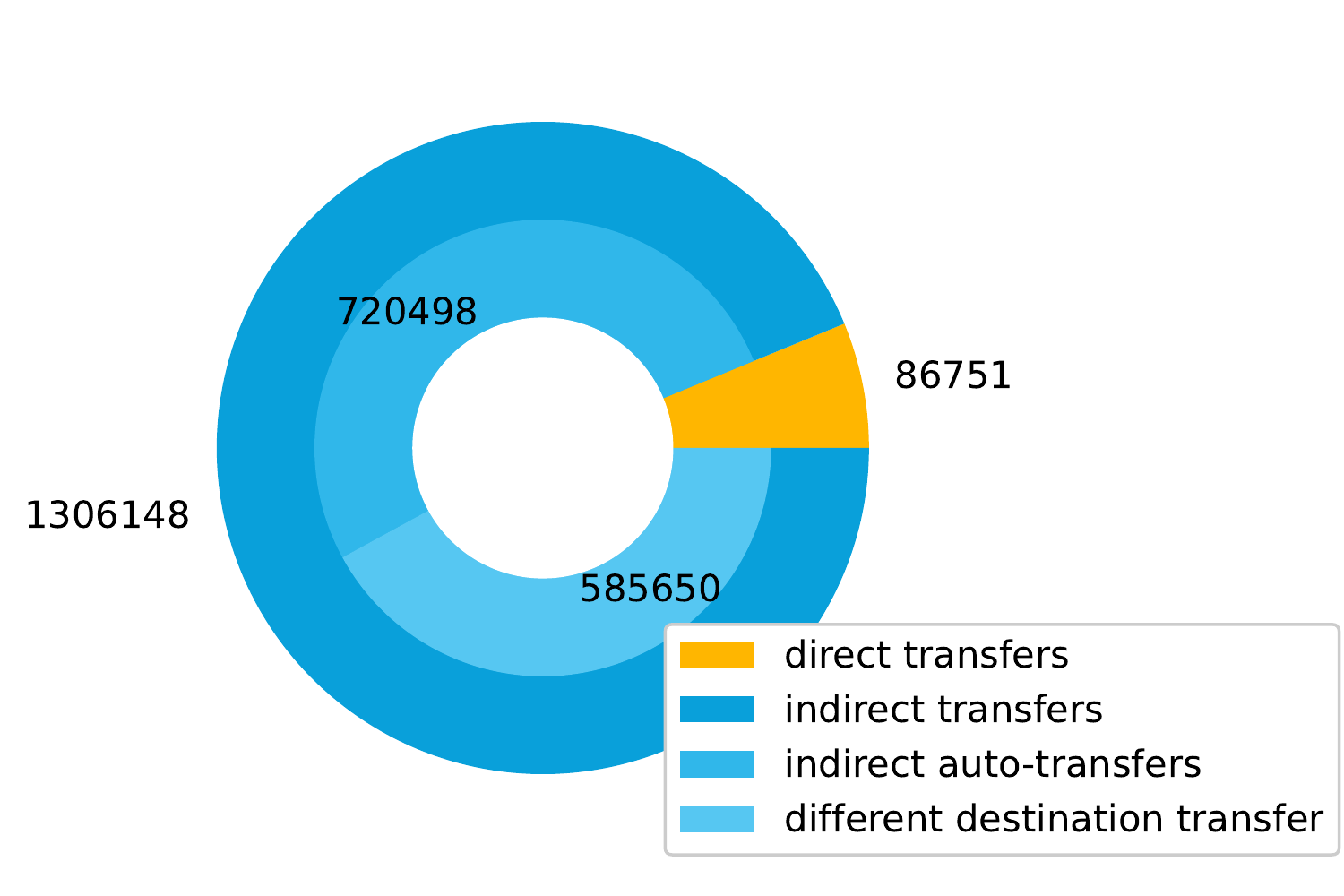}
		\caption{North Rhine-Westphalia}
	\end{subfigure}
	\hfill
	\begin{subfigure}[b]{0.45\textwidth}
		\centering
		\includegraphics[width=1.2\textwidth]{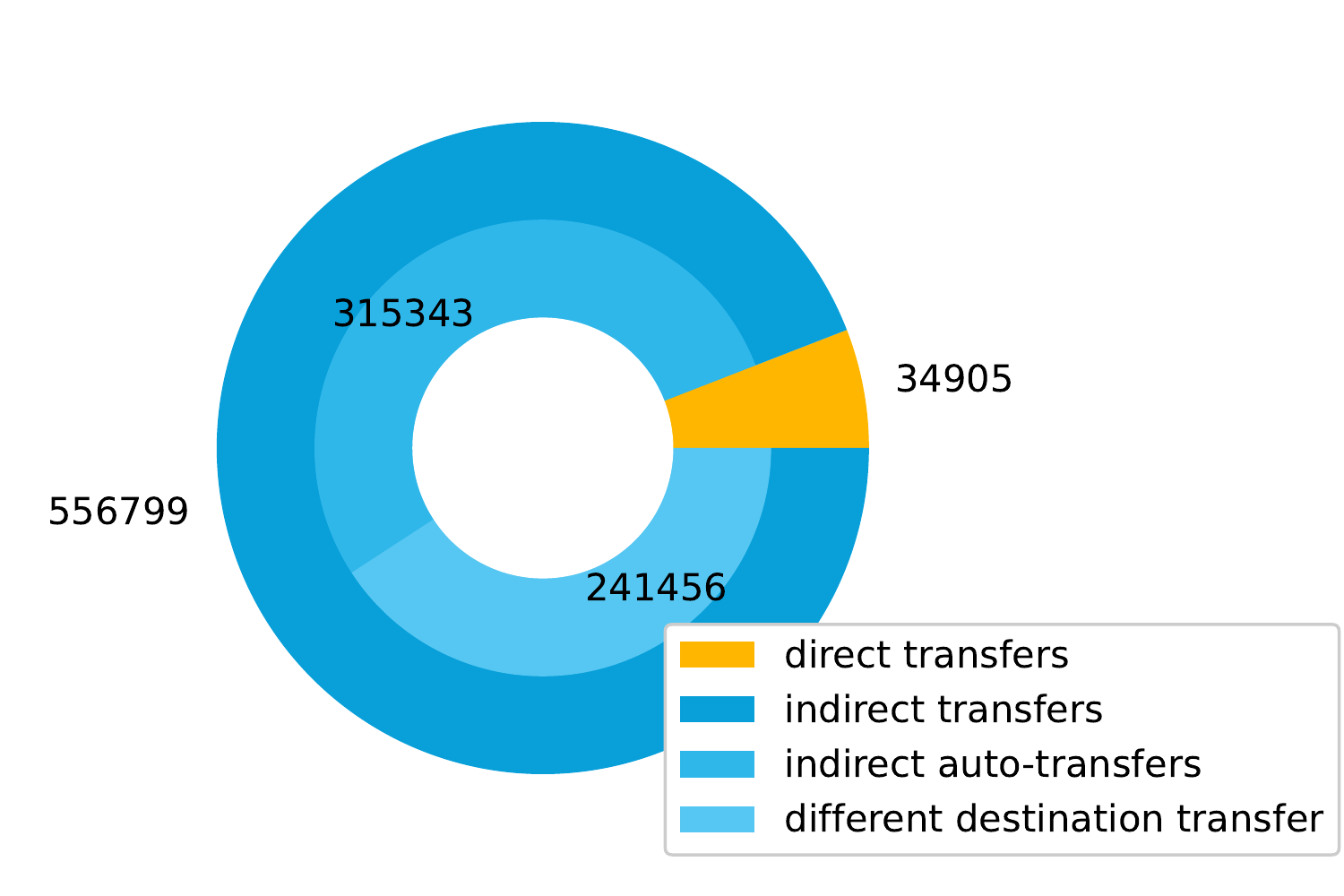}
		\caption{Bavaria}
	\end{subfigure}
	
	\begin{subfigure}[b]{0.45\textwidth}
		\centering
		\includegraphics[width=1.2\textwidth]{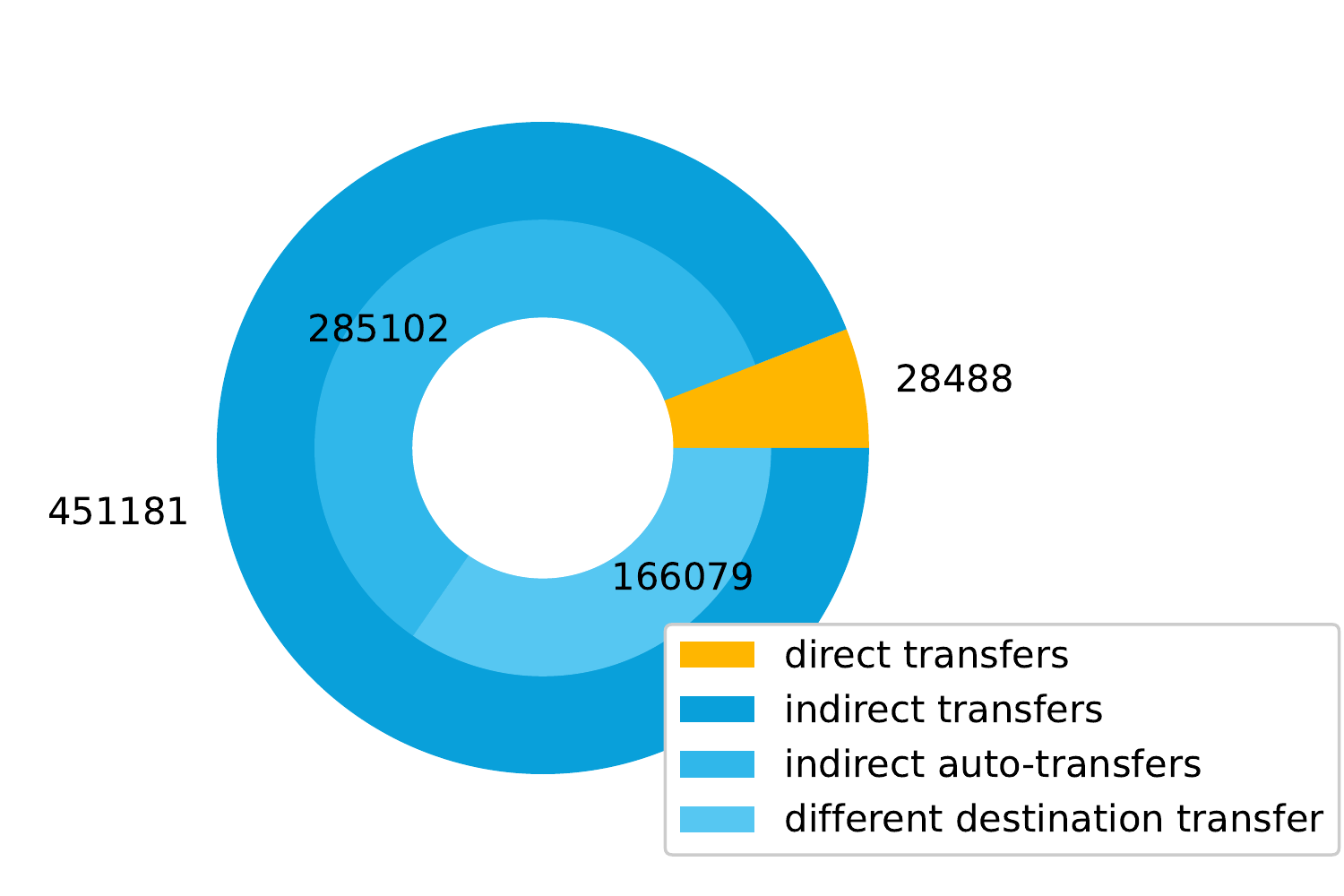}
		\caption{Baden-Württemberg}
	\end{subfigure}
	\hfill
	\begin{subfigure}[b]{0.45\textwidth}
		\centering
		\includegraphics[width=1.2\textwidth]{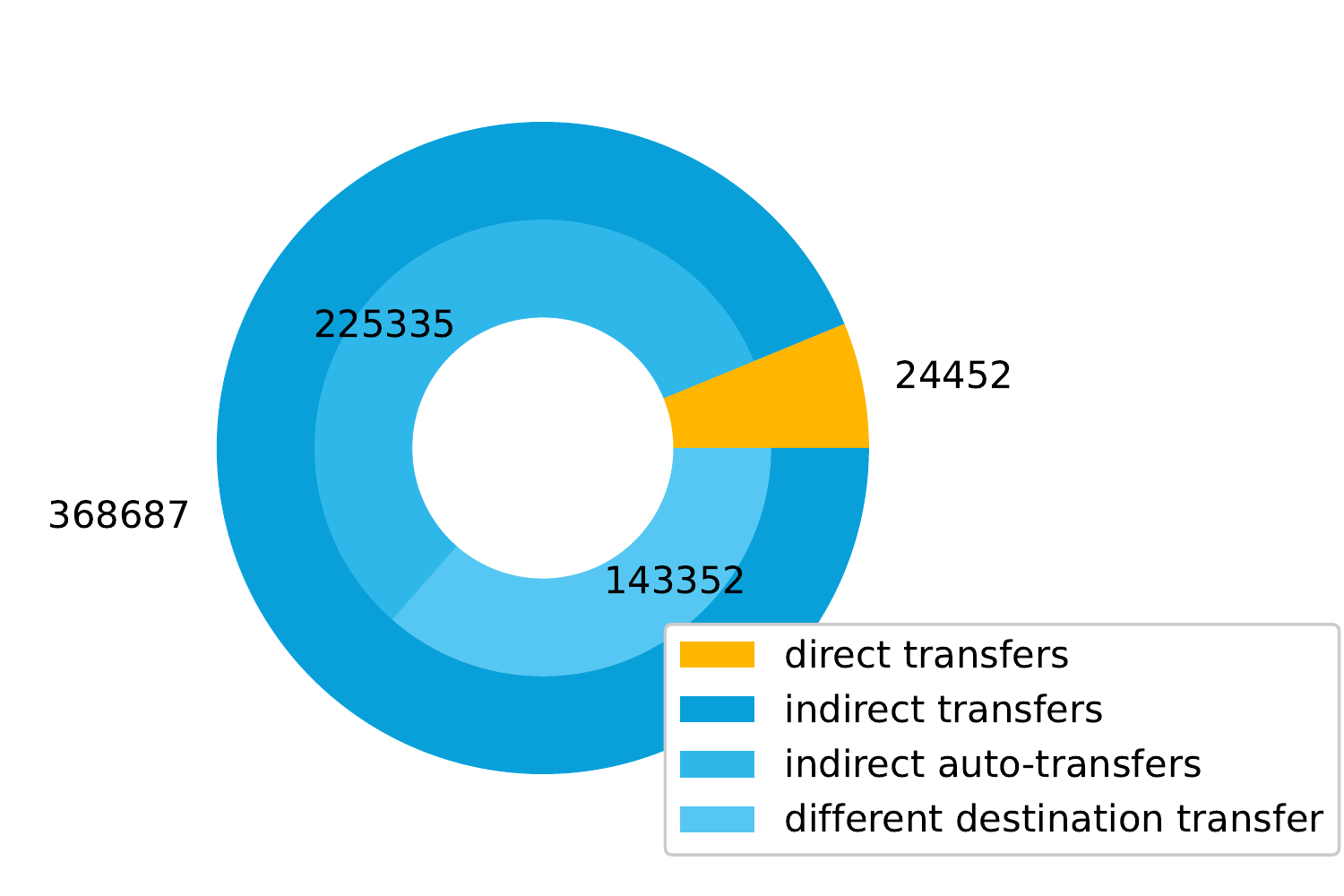}
		\caption{Lower Saxony}
	\end{subfigure}
	
	\begin{subfigure}[b]{0.45\textwidth}
		\centering
		\includegraphics[width=1.2\textwidth]{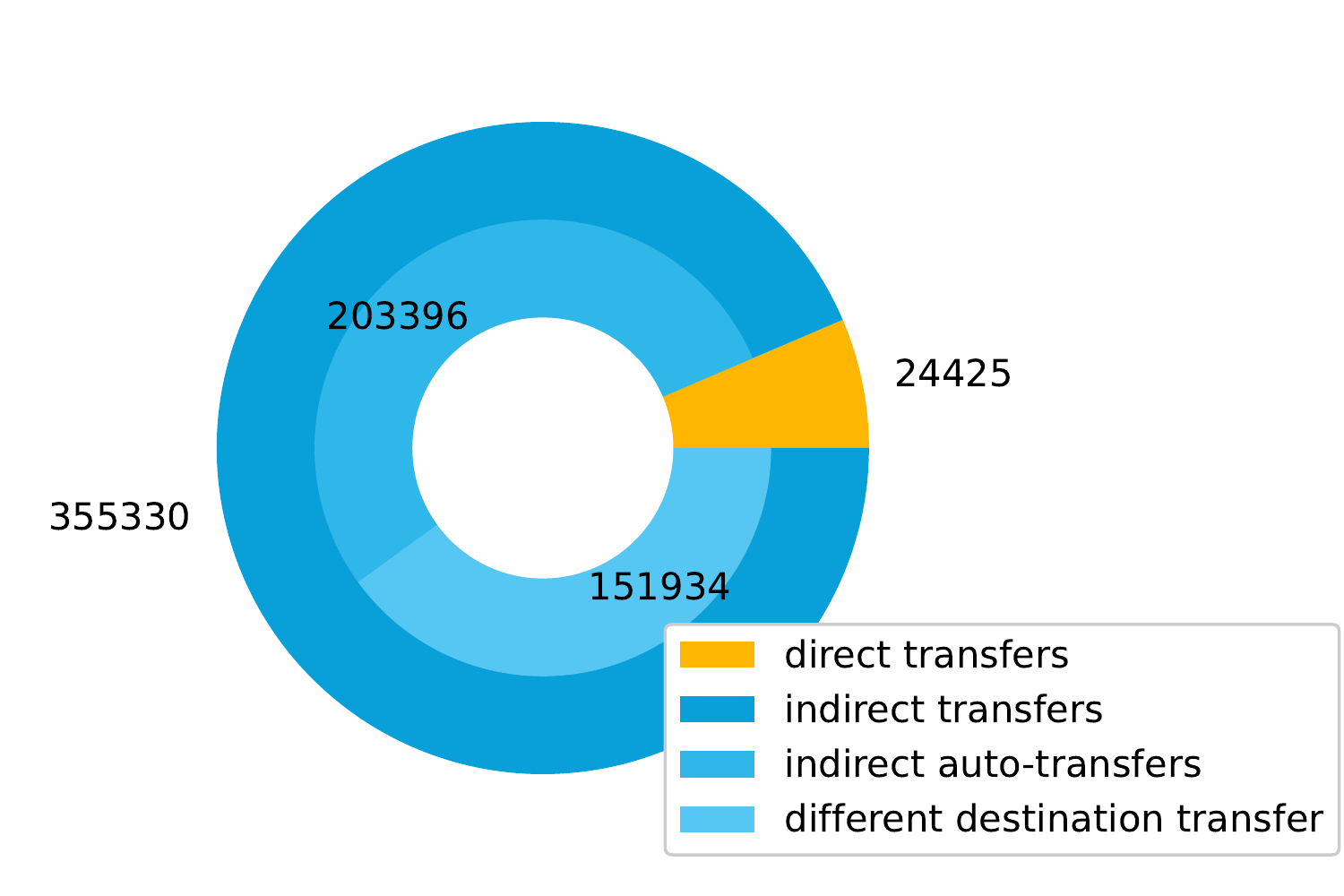}
		\caption{Hesse}
	\end{subfigure}
	\hfill
	\begin{subfigure}[b]{0.45\textwidth}
		\centering
		\includegraphics[width=1.2\textwidth]{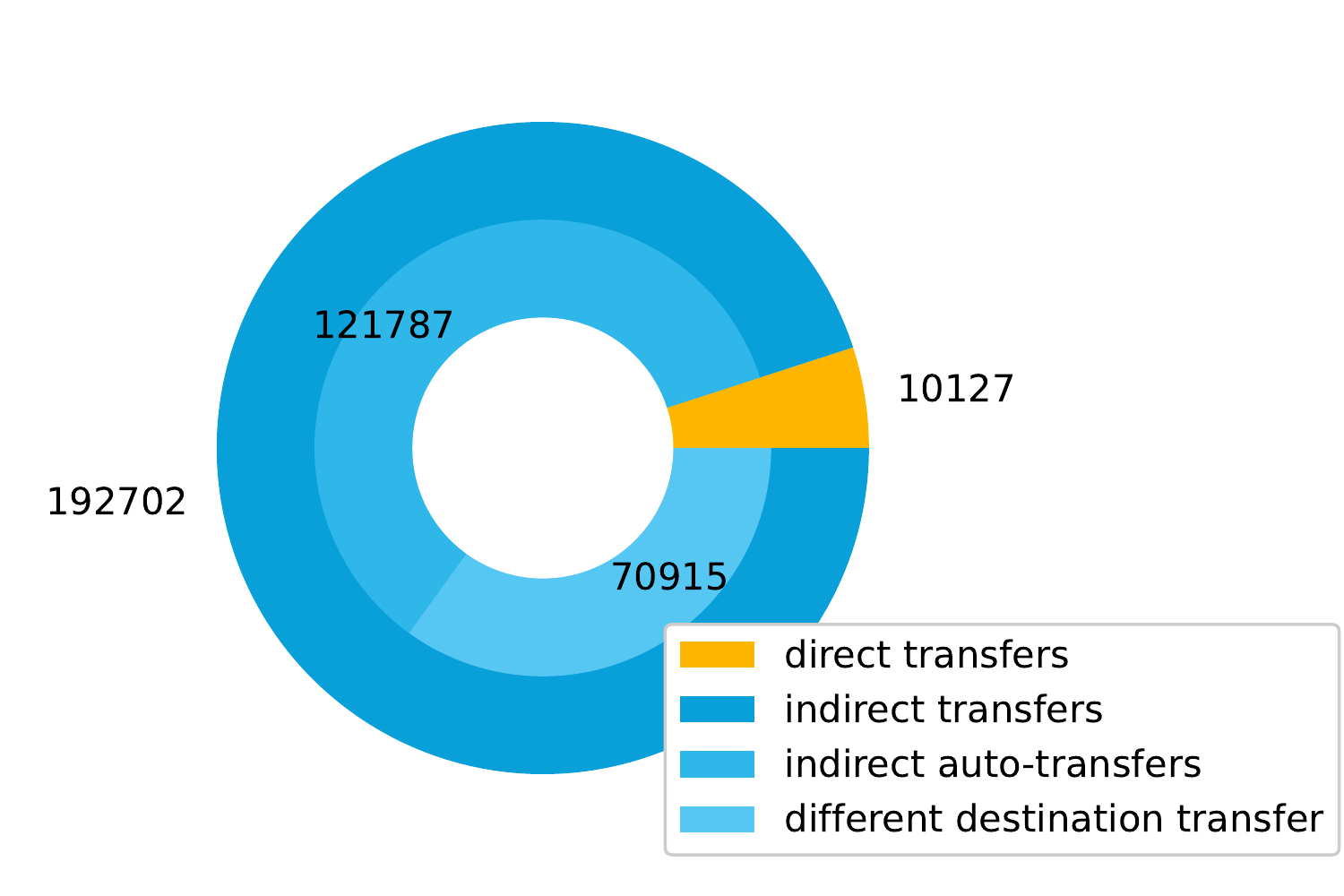}
		\caption{Rhineland-Palatinate}
	\end{subfigure}
\end{figure}
\clearpage   
\begin{figure}[tb]\ContinuedFloat         
	\begin{subfigure}[b]{0.45\textwidth}
		\centering
		\includegraphics[width=1.2\textwidth]{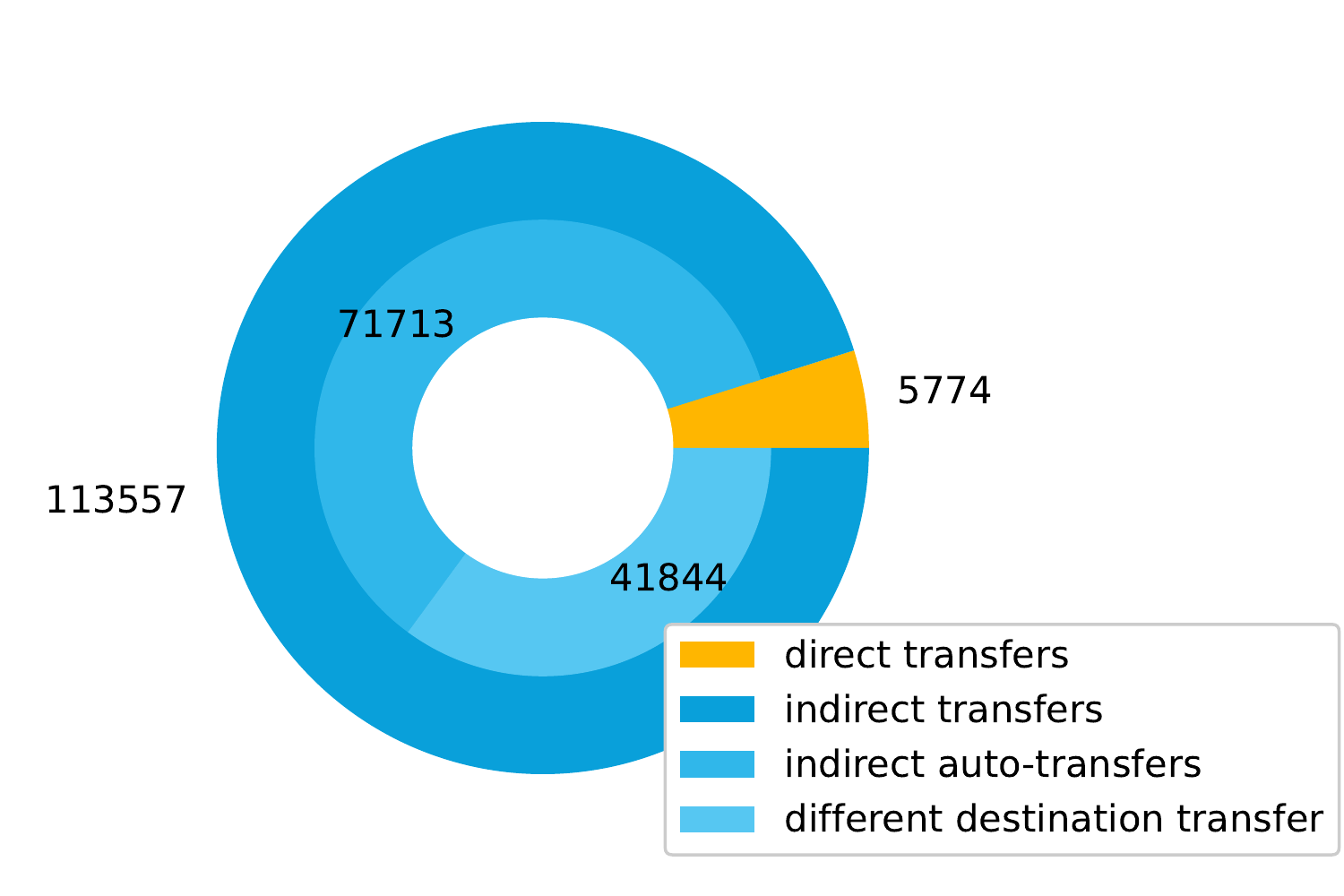}
		\caption{Saxony}
	\end{subfigure}
	\hfill
	\begin{subfigure}[b]{0.45\textwidth}
		\centering
		\includegraphics[width=1.2\textwidth]{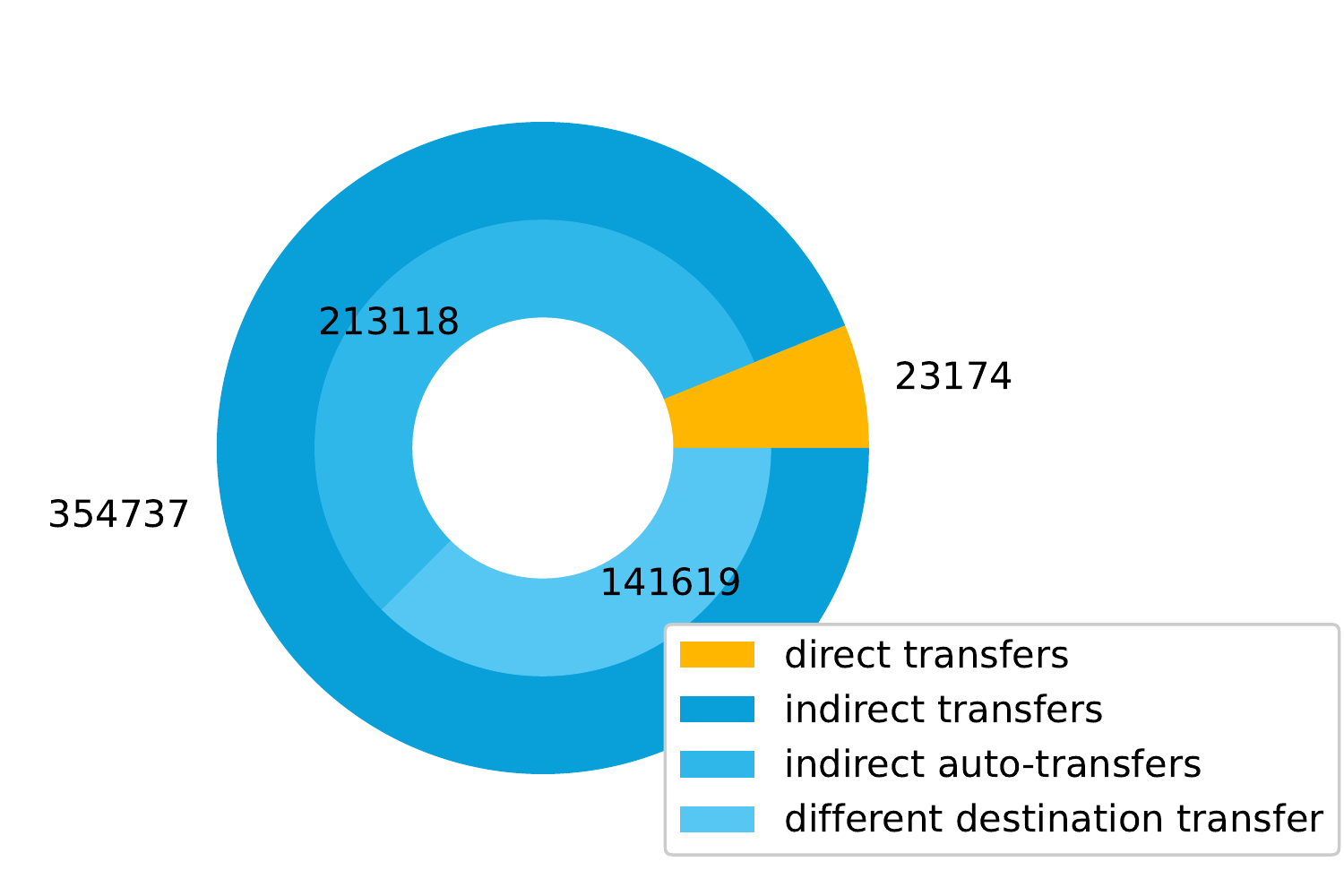}
		\caption{Berlin}
	\end{subfigure}
	
	\begin{subfigure}[b]{0.45\textwidth}
		\centering
		\includegraphics[width=1.2\textwidth]{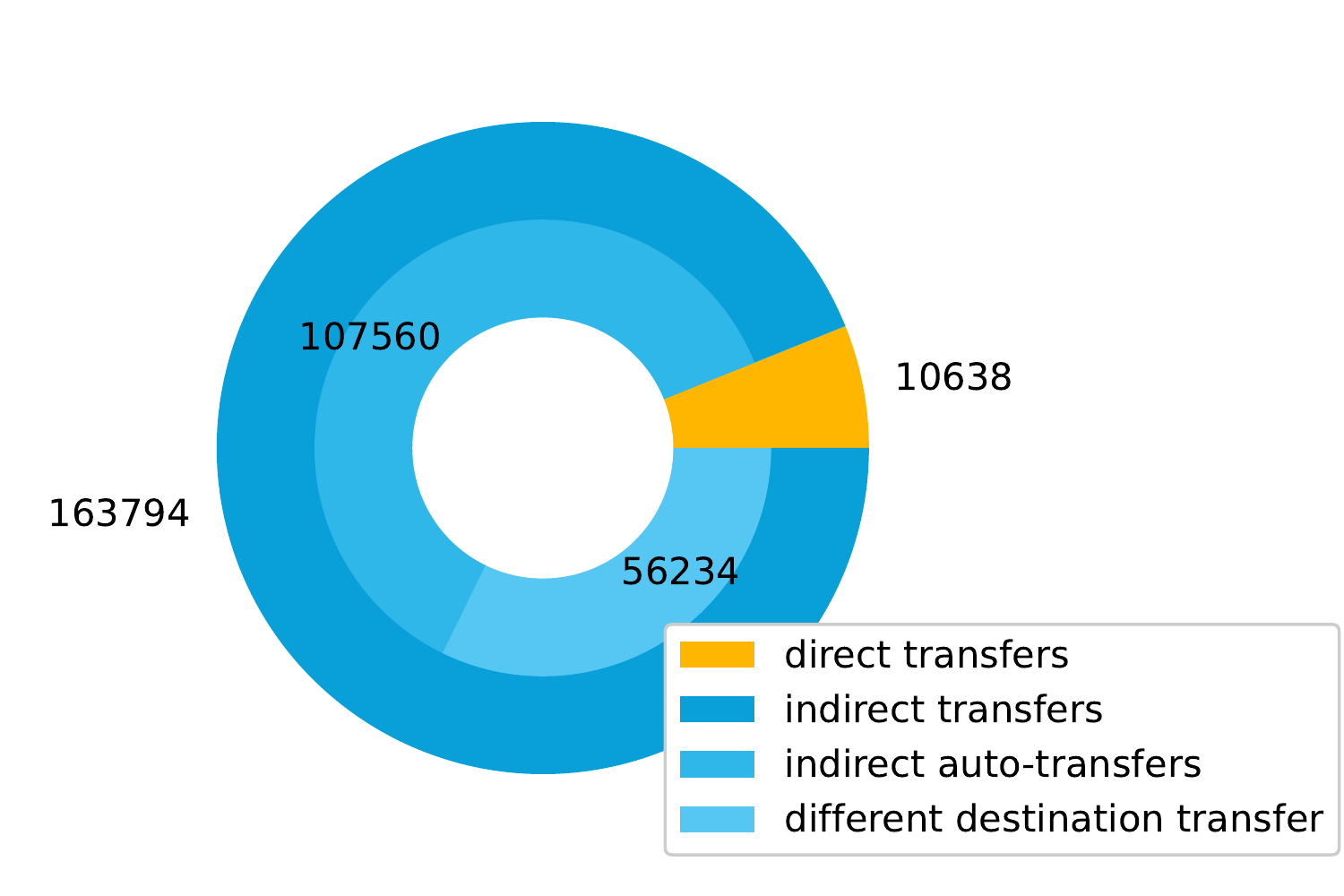}
		\caption{Schleswig-Holstein}
	\end{subfigure}
	\hfill
	\begin{subfigure}[b]{0.45\textwidth}
		\centering
		\includegraphics[width=1.2\textwidth]{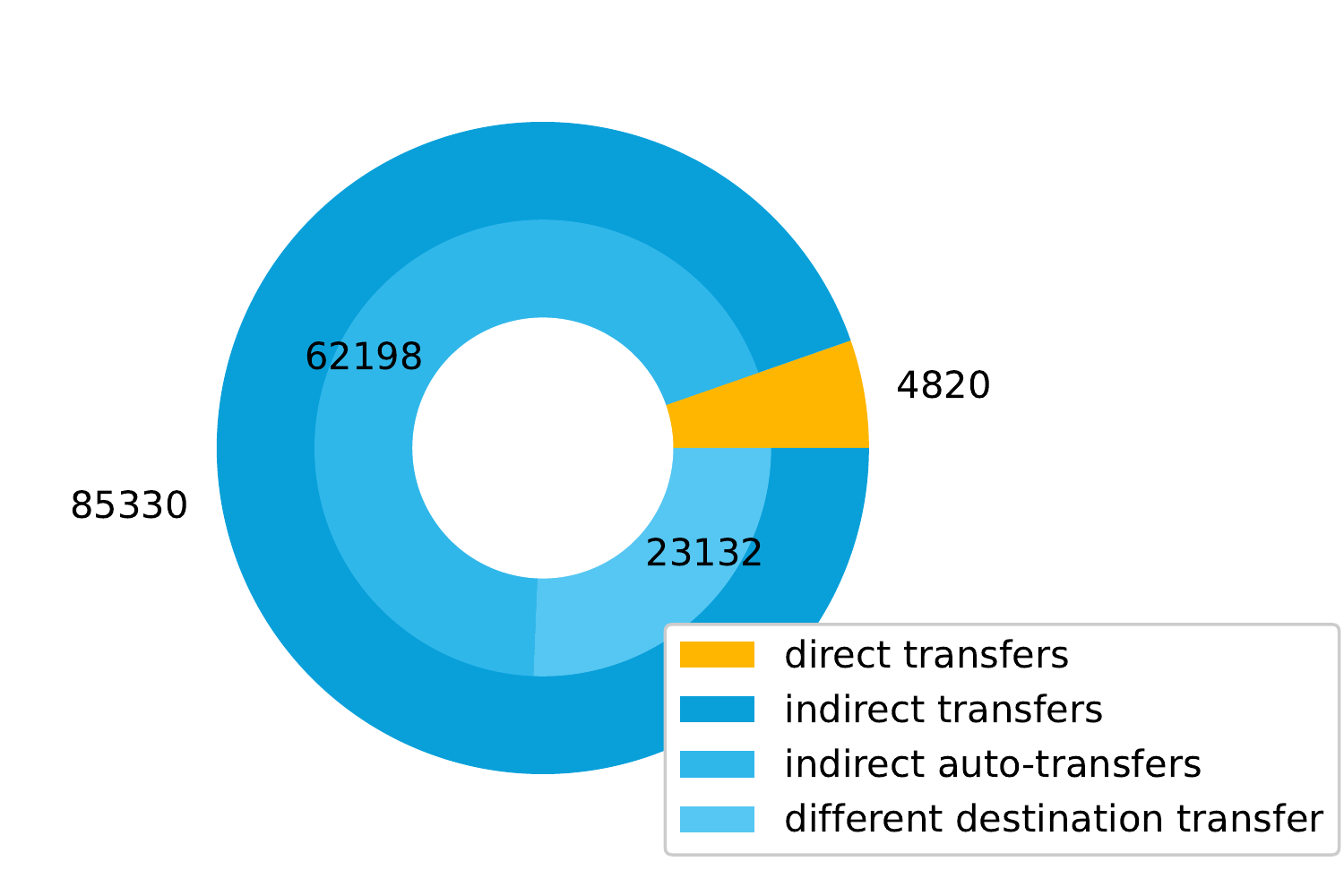}
		\caption{Brandenburg}
	\end{subfigure}   
	
	\begin{subfigure}[b]{0.45\textwidth}
		\centering
		\includegraphics[width=1.2\textwidth]{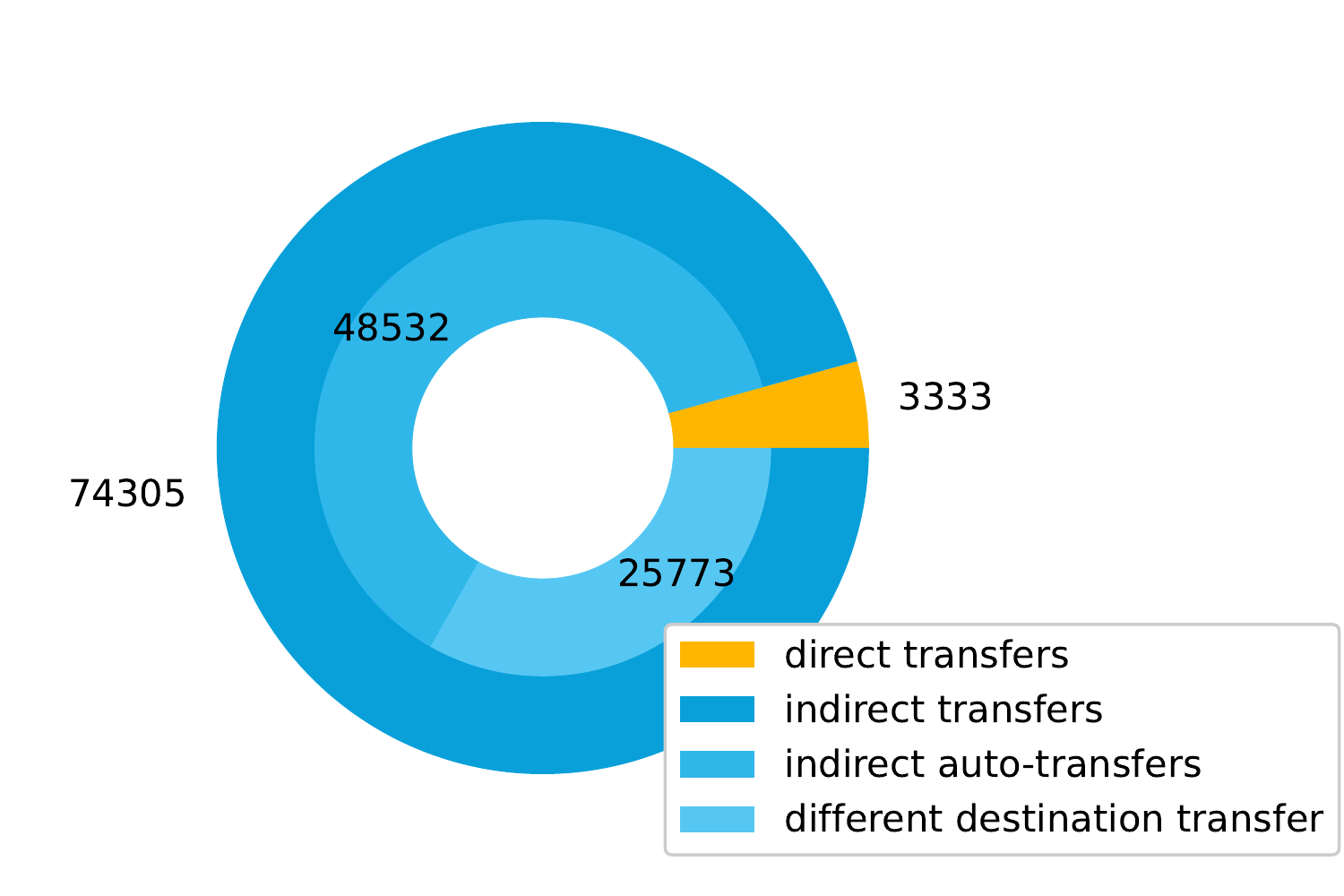}
		\caption{Saxony-Anhalt}
	\end{subfigure}
	\hfill
	\begin{subfigure}[b]{0.45\textwidth}
		\centering
		\includegraphics[width=1.2\textwidth]{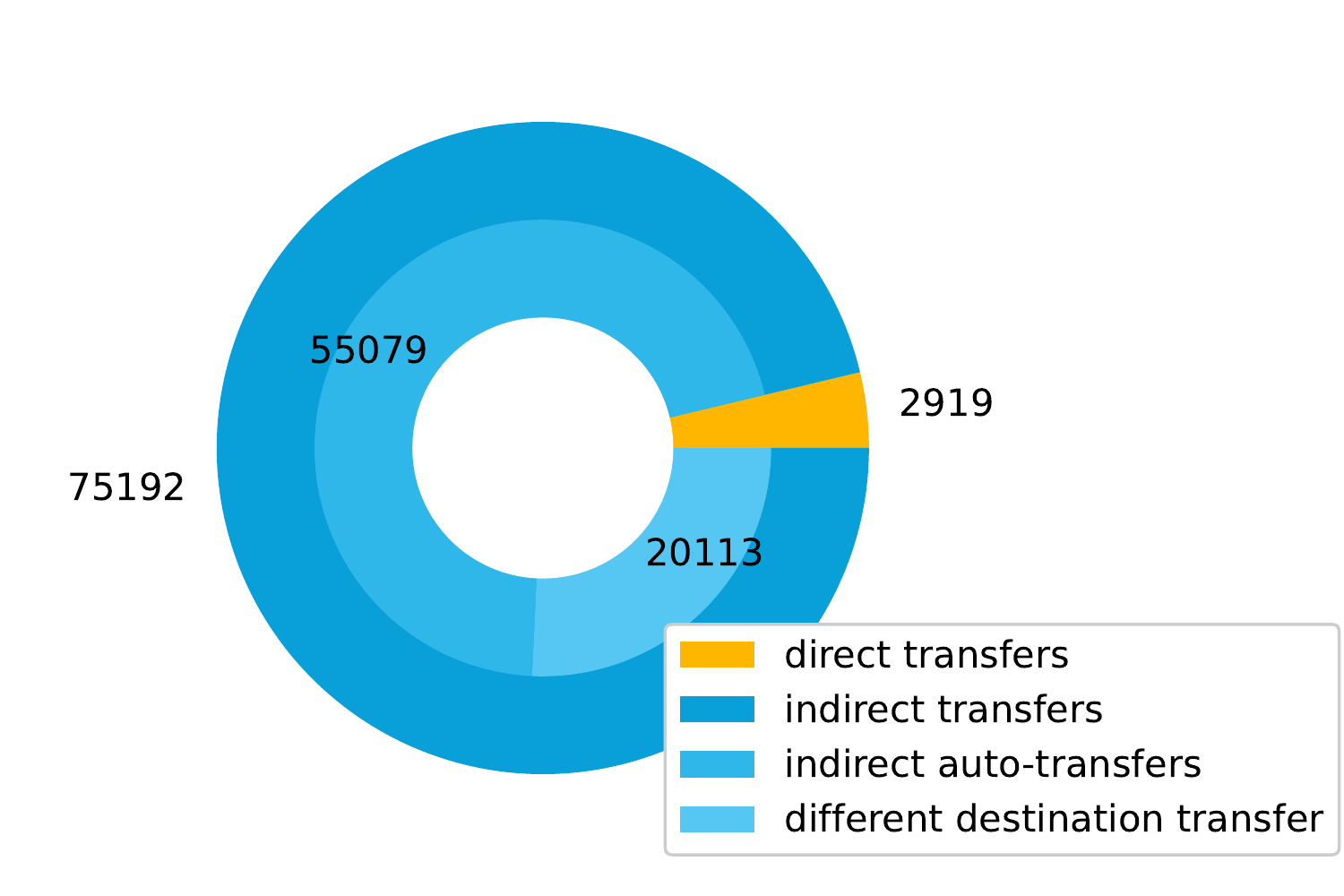}
		\caption{Thuringia}
	\end{subfigure}  
\end{figure}
\clearpage   
\begin{figure}[tbh!]\ContinuedFloat 
	\begin{subfigure}[b]{0.45\textwidth}
		\centering
		\includegraphics[width=1.2\textwidth]{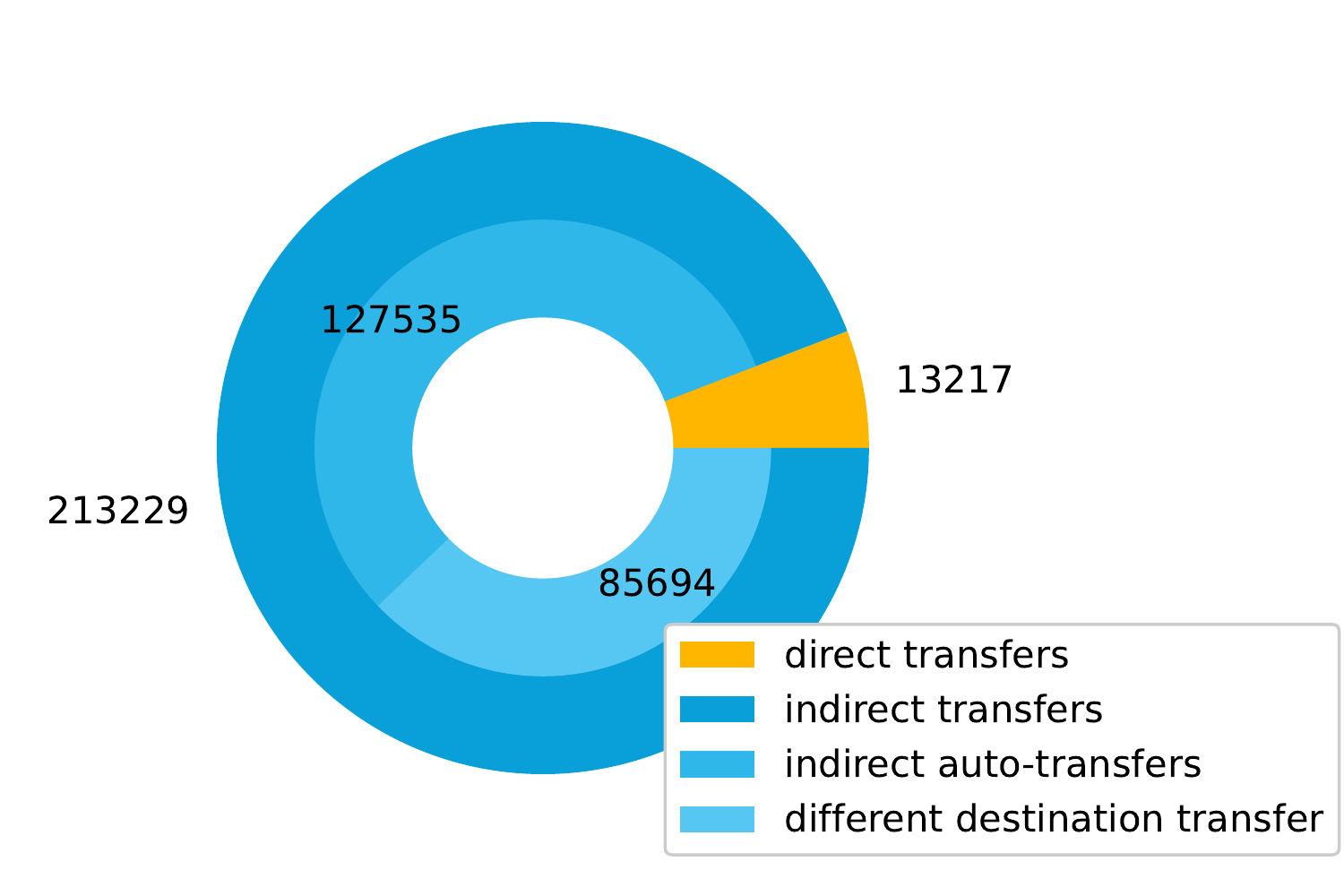}
		\caption{Hamburg}
	\end{subfigure}
	\hfill
	\begin{subfigure}[b]{0.45\textwidth}
		\centering
		\includegraphics[width=1.2\textwidth]{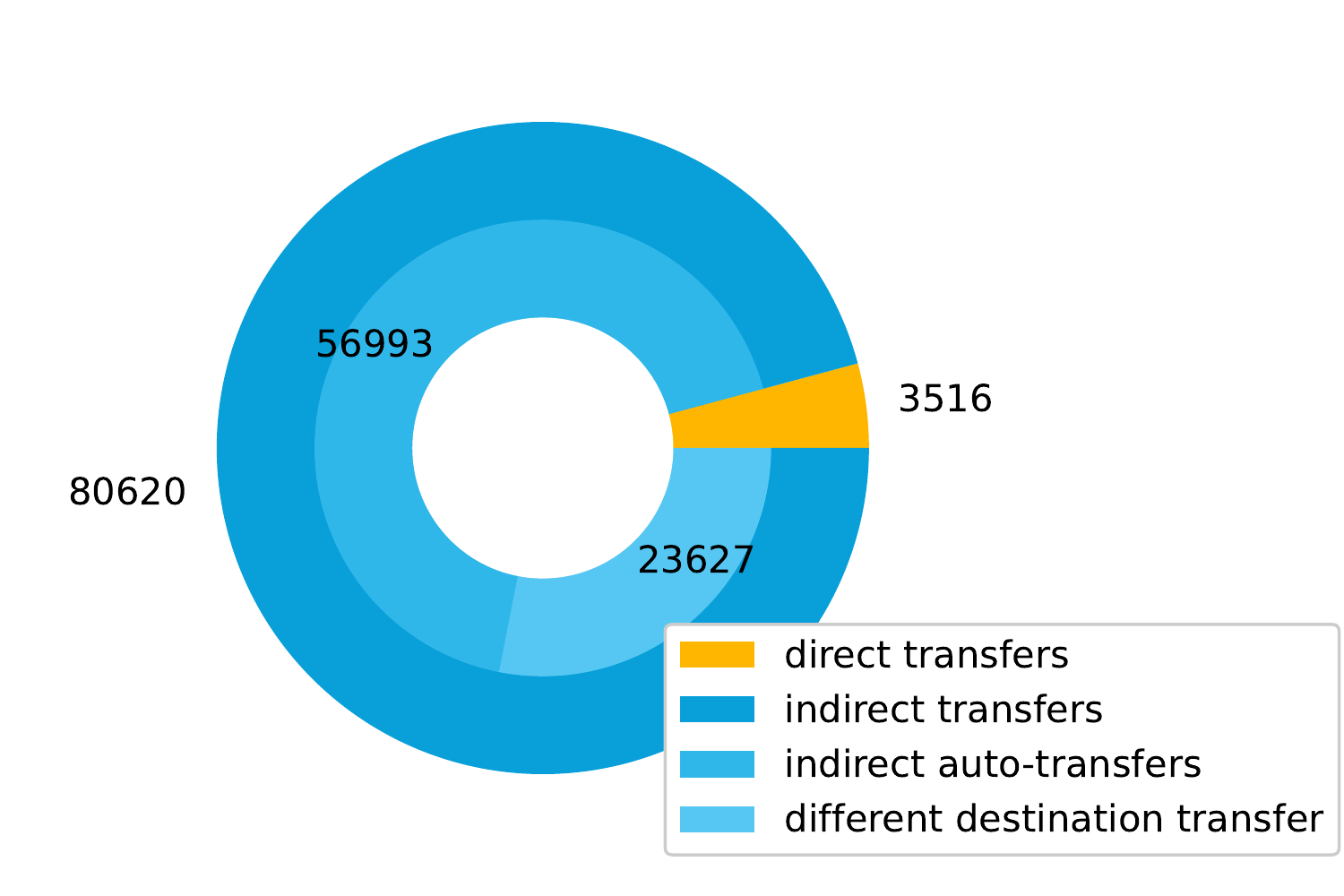}
		\caption{Mecklenburg-West Pomerania}
	\end{subfigure} 
	
	\begin{subfigure}[b]{0.45\textwidth}
		\centering
		\includegraphics[width=1.2\textwidth]{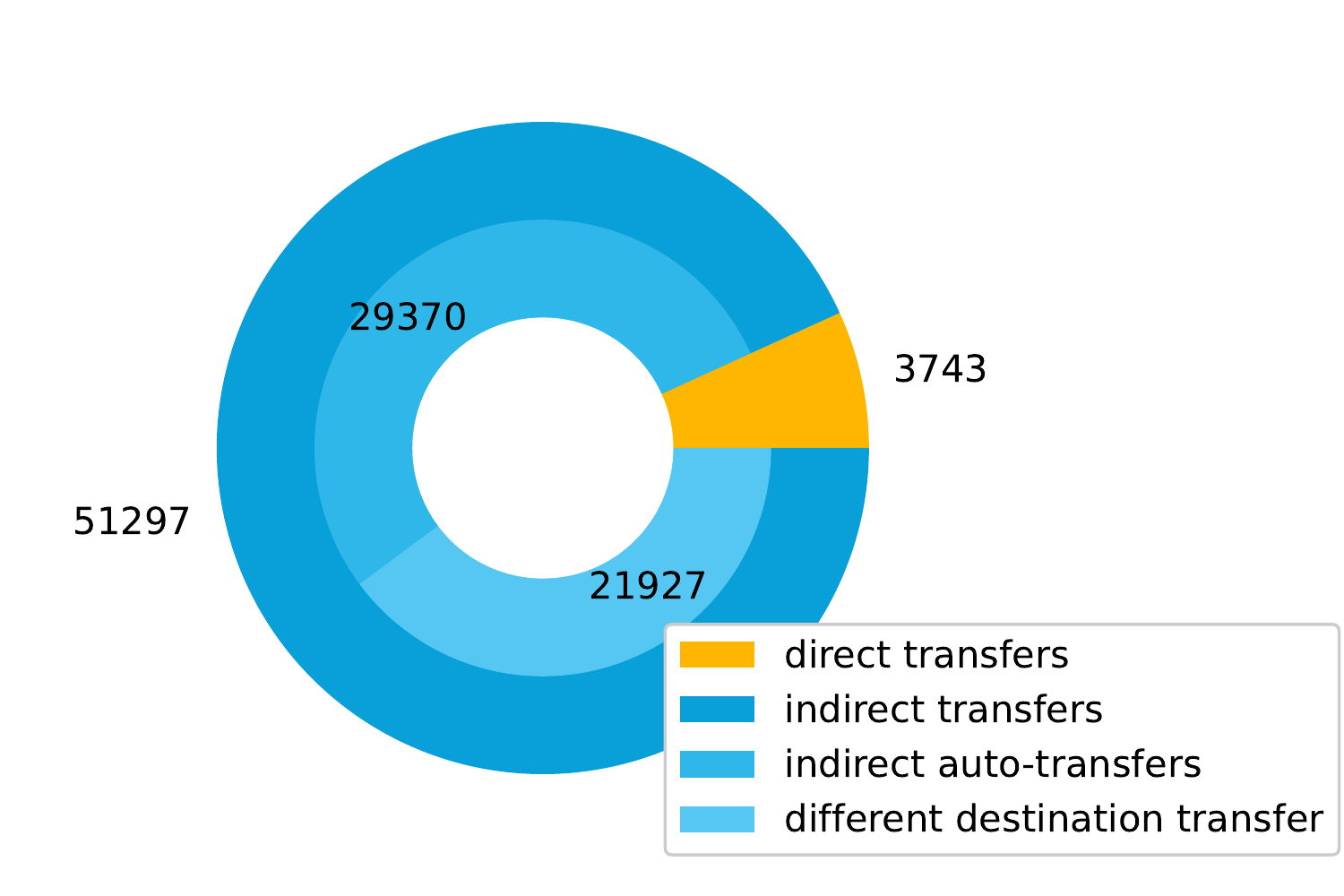}
		\caption{Saarland}
	\end{subfigure}
	\hfill
	\begin{subfigure}[b]{0.45\textwidth}
		\centering
		\includegraphics[width=1.2\textwidth]{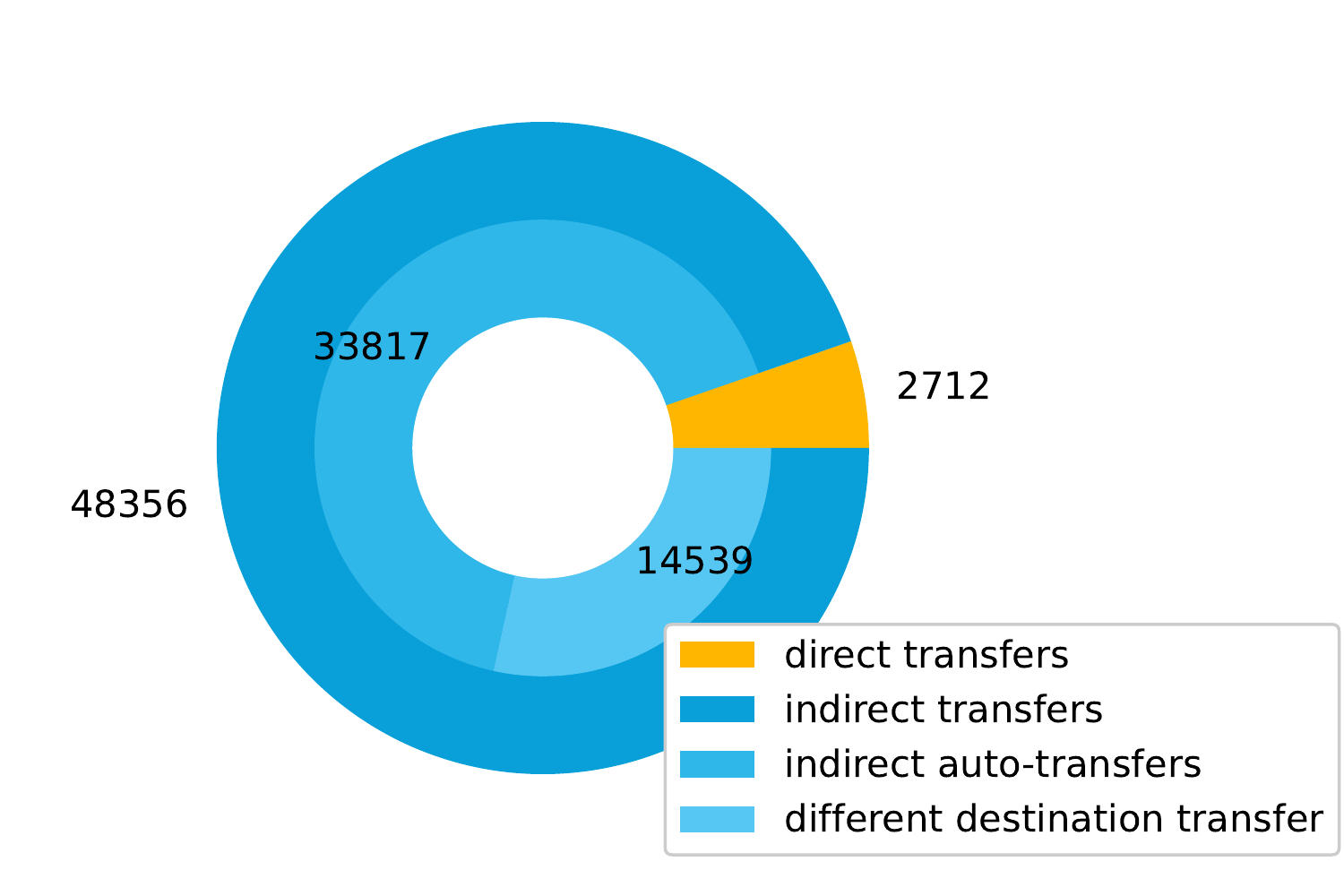}
		\caption{Bremen}
	\end{subfigure} 
	\caption{Pie chart of patient transfers for the considered state.}
	\label{fig_trans}
\end{figure}	

	\subsubsection{Transfers between states} 
	Almost all of the transfers between states were indirect (91.8\%) (for more details see Table~\ref{tab:transfer}, last row). For all the states the percentage of transfers between states was smaller or equal to 20\%. The state with the highest percentage of transfers to a different state is Brandenburg (20\%) and with the lowest is North Rhine-Westphalia (4\%), see Table \ref{tab:procent}.
	
\begin{figure}[h!]
		\centering
		\includegraphics[height = 9cm]{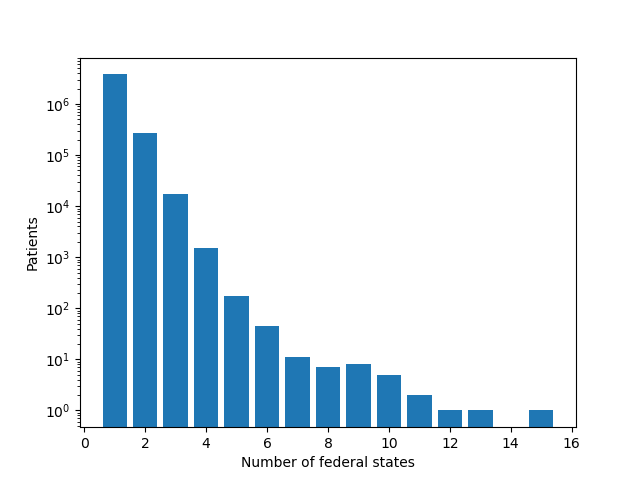}
		\caption{Number of patients having record in given number of states.}
		\label{patinlands}
	\end{figure}

	To analyse transfers between states more deeply, we checked number of states with records of hospital stays per patient.
	From Figure~\ref{patinlands} we see that most of the patients had records from only one state. Considerable amount of patients had visited healthcare facilities in two or three states and only three patients had records from more then 10 states.
	
	\begin{table}[h!]
		\centering
		\caption{Percentage of transfers between states (originating in given state and ending in a different one). States are ordered by their population according to~\cite{Destatis}.}
		\label{tab:procent}
		\begin{tabular}{|c|c|}
			\hline
			State & Percent\\\hline
			North Rhine-Westphalia & 4\%\\\hline
			Bavaria & 6\%\\\hline
			Baden-Württemberg & 7\%\\\hline
			Lower Saxony & 12\%\\\hline
			Hesse & 10\%\\\hline
			Rhineland-Palatinate & 16\%\\\hline
			Saxony & 8\%\\\hline
			Berlin & 8\%\\\hline
			Schleswig-Holstein & 16\%\\\hline
			Brandenburg & 20\%\\\hline
			Saxony-Anhalt & 10\%\\\hline
			Thuringia & 10\%\\\hline
			Hamburg & 15\%\\\hline
			Mecklenburg-West Pomerania & 10\%\\\hline
			Saarland & 10\%\\\hline
			Bremen & 18\%\\\hline
		\end{tabular}
	\end{table}

\begin{figure}[h!]
		\centering
		\includegraphics[height=10cm]{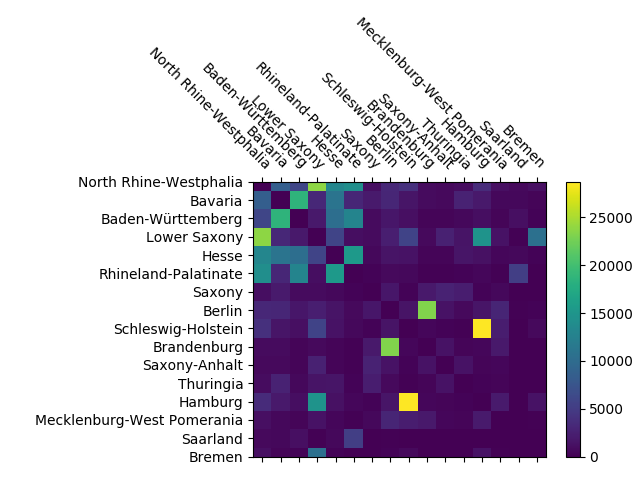}
		\caption{Number of patients for which records were found in two considered states.
		}
		\label{matrix:pat}
	\end{figure}

	Furthermore we explored the common parts of lists of patients between two given states. From Figure~\ref{matrix:pat} we see that states with biggest populations (according to~\cite{Destatis}) had many patients that also visited healthcare facilities in different states. It is also clear that many patients had records from both Hamburg and Schleswig-Holstein, Branderburg and Berlin, Lower Saxony and North Rhine-Westphalia, Baden-Württemberg and Bavaria while for some states exchange of patients was minor.

\begin{figure}[h!]
		\centering
		\includegraphics[height = 10cm]{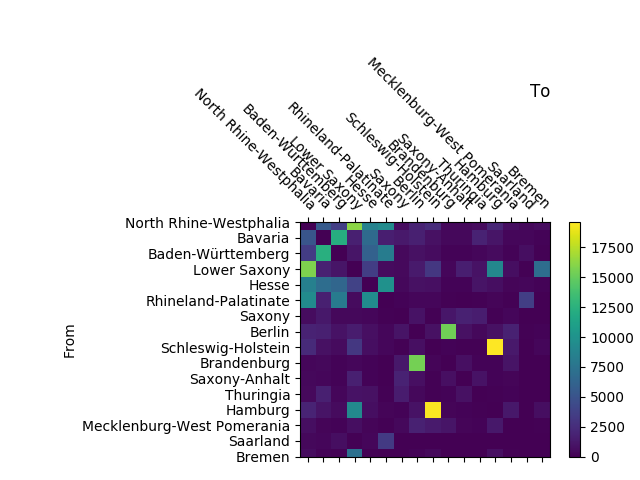}
		\caption{Number of indirect transfers between two considered states.}
		\label{matrix:indirect}
	\end{figure}

	Next, we analysed in detail indirect transfers between each two states. From Figure~\ref{matrix:indirect} we see that even thought the numbers of indirect transfers from one state to another and the other way around was not equal, the differences were small. Again the states that had the biggest number of transfers between each other were:  Hamburg and Schleswig-Holstein, Brandenburg and Berlin, Lower Saxony and North Rhine-Westphalia, Baden-Württemberg and Bavaria. 
	
\begin{figure}[h!]
		\centering
		\includegraphics[height = 10cm]{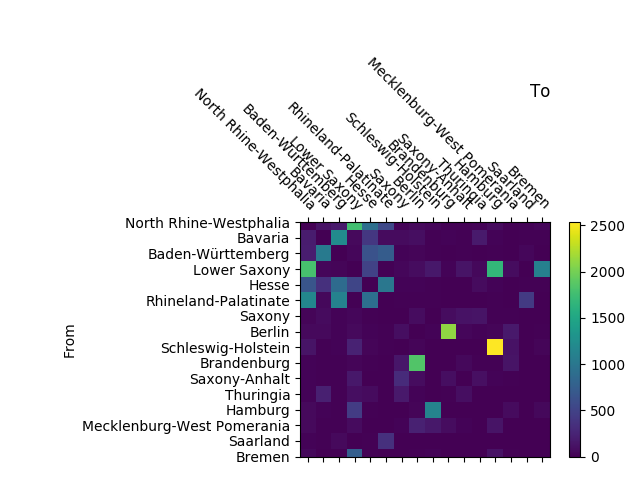}
		\caption{Number of direct transfers between two considered states.}
		\label{matrix:direct}
	\end{figure}

	On the other hand, considering only direct transfers, we see that they were not as symmetric as in the case of indirect transfers Figure \ref{matrix:direct}. There were more transfers from Schleswig-Holstein to Hamburg and from Berlin to Brandenburg than in opposite direction. 

	\clearpage
	\subsection{Overlaps}
	Among all data we found 330 093 overlapping entries. By overlapping records we understand distinct sets of two or more records for a given patient, with non-empty intersection of stay periods, either within the same facility or in other facilities. For the classification of overlaps we followed~\cite{Piotrowska2019arxiv, Lonc2019arxiv} and used publicly available code~\cite{emergenetpackage}.
	Most of the overlaps were between one or two healthcare facilities. Only 2 611 overlaps were between more than two healthcare facilities (2 577 between three, 34 between four), they correspond to 0.79\% of all overlaps and they were excluded from further analysis.
	
	In Table~\ref{tab:overlap_stat} we present results of our analysis of detected overlaps according to their location. We see that states with highest population had highest number of overlaps. The exception were Berlin and Hamburg.

	In all the cases most of the overlaps were classified as standard transfer or two entries in a single institution, while usually first type was more abundant. Other types of overlaps were responsible for around 20\% of detected overlaps. Except for Berlin and Hamburg, where it was 12.40\% and 11.50\%, respectively. In Bremen, percent of standard transfer and two entries in single institution was similar. In provided dataset we also found 29 661 overlaps for stays in hospitals located in different states. Considering overlaps for which hospitals were located in different states almost 80\% of them were standard transfers.

	In fact, for all states, the most common were typical transfers --- meaning that both stay periods were covered only by one day and the stays were reported for different institutions — standard, first day and last day transfers, see Figure~\ref{fig:overlap_hist}. Moreover, almost none overlaps were longer than a month.  
	
	\begin{landscape}
	\begin{table}
		\thispagestyle{empty}
		\centering
		\caption{Identified types of overlaps for given location. States are ordered by their population according to~\cite{Destatis}.
			Abbreviations for transfers:
			t. e. s. i. --- two entries in a single institution;
			temp. transfer --- temporary transfer;
			sim. e. s. i. ---  simultaneous two entries in a single institution;
			sim. e. t. i. --- simultaneous two entries in two institutions;
			u. t. e. i. --- unknown two entries in two institutions;
			u. m. e. --- unknown multiple entries. 
			Abbreviations for states:
			NRW --- North Rhine-Westphalia;
			BW --- Baden-Württemberg;
			RP --- Rhineland-Palatinate;
			SH --- Schleswig-Holstein;
			ST --- Saxony-Anhalt;
			MV --- Mecklenburg-West Pomerania.}
		\label{tab:overlap_stat}
		\begin{filecontents*}{tabels/overlap_stats.csv}
State,All,standard transfer,t. e. s. i.,temp. transfer,first day transfer,last day transfer,sim. e. s. i.,sim. e. t. i.,u. t. e. i,u. m. e.
NRW,85260,63.0\%,16.3\%,10.1\%,7.5\%,0.4\%,0.1\%,0.1\%,2.5\%,0.1\%
Bavaria,36057,57.0\%,20.5\%,8.7\%,11.1\%,0.6\%,0.1\%,0.2\%,1.9\%,0.1\%
BW,28871,62.6\%,17.2\%,8.8\%,8.2\%,0.5\%,0.1\%,0.1\%,2.5\%,0.1\%
Lower Saxony,25071,62.3\%,17.2\%,6.5\%,11.5\%,0.7\%,0.1\%,0.3\%,1.4\%,0.1\%
Hesse,26524,61.8\%,21.8\%,7.5\%,6.8\%,0.4\%,0.1\%,0.1\%,1.5\%,0.1\%
RP,10871,55.8\%,22.8\%,8.2\%,10.1\%,0.5\%,0.1\%,0.2\%,2.3\%,0.1\%
Saxony,7022,53.6\%,30.7\%,7.4\%,6.2\%,0.3\%,0.1\%,0.1\%,1.6\%,0.1\%
Berlin,27522,61.3\%,26.3\%,6.1\%,4.3\%,0.4\%,0.2\%,0.1\%,1.3\%,0.1\%
SH,11011,61.9\%,17.8\%,8.1\%,9.4\%,0.4\%,0.2\%,0.2\%,2.0\%,0.1\%
Brandenburg,5608,57.2\%,25.3\%,6.9\%,9.1\%,0.4\%,0.0\%,0.1\%,1.0\%,0.0\%
ST,3799,59.4\%,24.2\%,5.8\%,8.9\%,0.4\%,0.1\%,0.1\%,1.1\%,0.0\%
Thuringia,3445,56.3\%,27.6\%,6.1\%,7.9\%,0.4\%,0.1\%,0.1\%,1.5\%,0.0\%
Hamburg,14695,66.9\%,21.6\%,5.3\%,4.8\%,0.4\%,0.1\%,0.1\%,0.9\%,0.0\%
MV,4485,52.0\%,31.9\%,4.4\%,10.5\%,0.3\%,0.1\%,0.1\%,0.7\%,0.0\%
Saarland,3821,58.5\%,19.3\%,9.6\%,9.2\%,0.4\%,0.1\%,0.1\%,2.9\%,0.1\%
Bremen,3759,40.7\%,40.8\%,6.9\%,9.0\%,0.4\%,0.1\%,0.3\%,1.9\%,0.1\%
Between,2 9661,78.8\%,0.0\%,9.4\%,9.5\%,0.4\%,0.0\%,0.1\%,1.9\%,0.1\%
\end{filecontents*}
\pgfplotstabletypeset[
		col sep=comma, 
		display columns/0/.style={column type={|p{2.5cm}|},string type}, 
		display columns/1/.style={column type={p{1cm}|},,int detect, 1000 sep={\;},precision=3},
		display columns/2/.style={column type={p{1.4cm}|},string type},
		display columns/3/.style={column type={p{1.6cm}|},string type},
		display columns/4/.style={column type={p{1.4cm}|},string type},
		display columns/5/.style={column type={p{1.5cm}|},string type},
		display columns/6/.style={column type={p{1.5cm}|},string type},
		display columns/7/.style={column type={p{2.2cm}|},string type},
		display columns/8/.style={column type={p{2.2cm}|},string type},
		display columns/9/.style={column type={p{1.7cm}|},string type},
		display columns/10/.style={column type={p{1.7cm}|},string type},
		every head row/.style={before row=\hline,after row=\hline},
		every nth row={1}{before row=\hline},
		every last row/.style={after row=\hline},
		]{tabels/overlap_stats.csv}
    \end{table}
	\end{landscape}
	
	\begin{figure}
		\centering
		\begin{subfigure}[b]{0.45\textwidth}
			\centering
			\includegraphics[width=\textwidth]{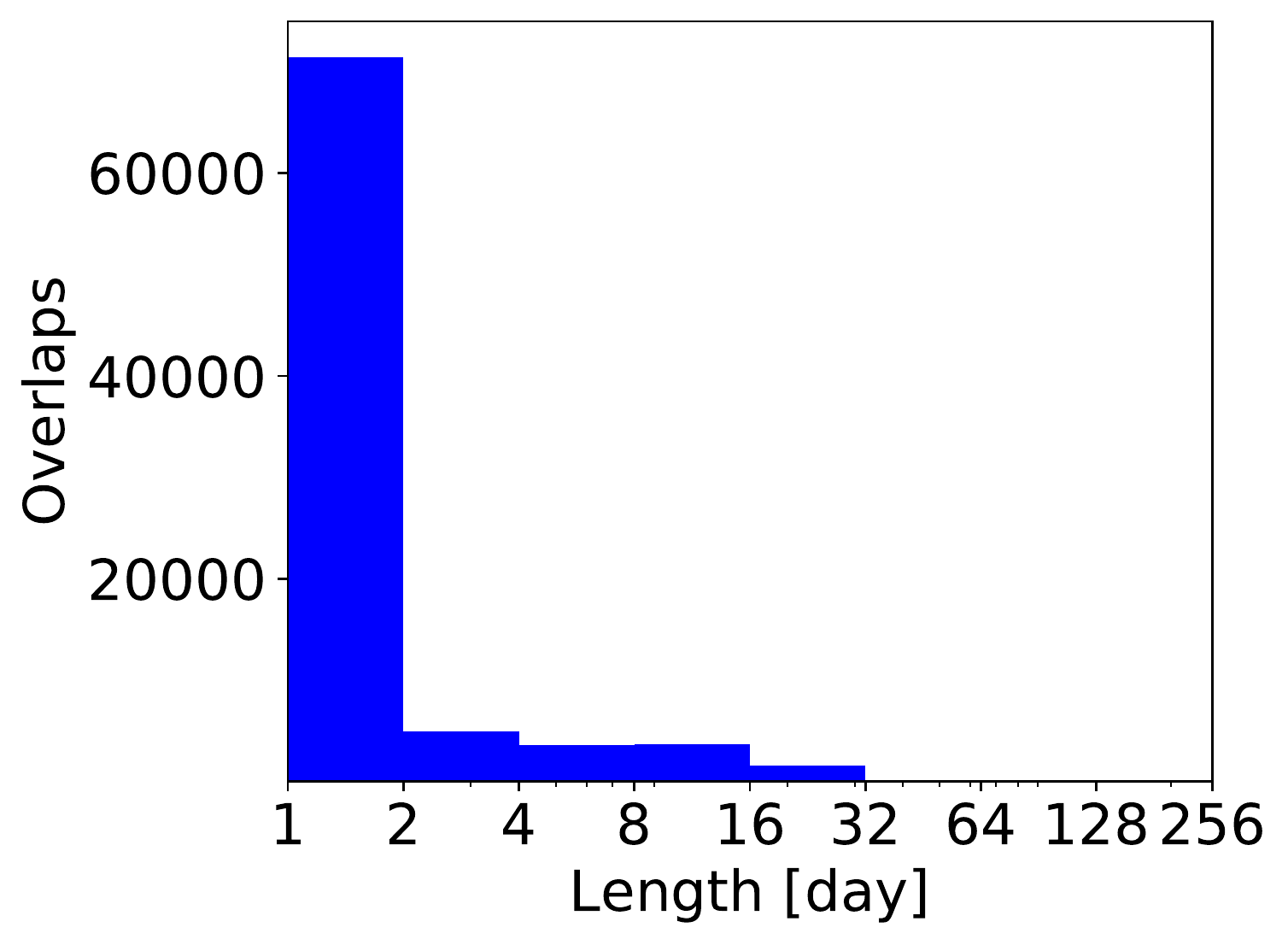}
			\caption{North Rhine-Westphalia}
		\end{subfigure}
		\hfill
		\begin{subfigure}[b]{0.45\textwidth}
			\centering
			\includegraphics[width=\textwidth]{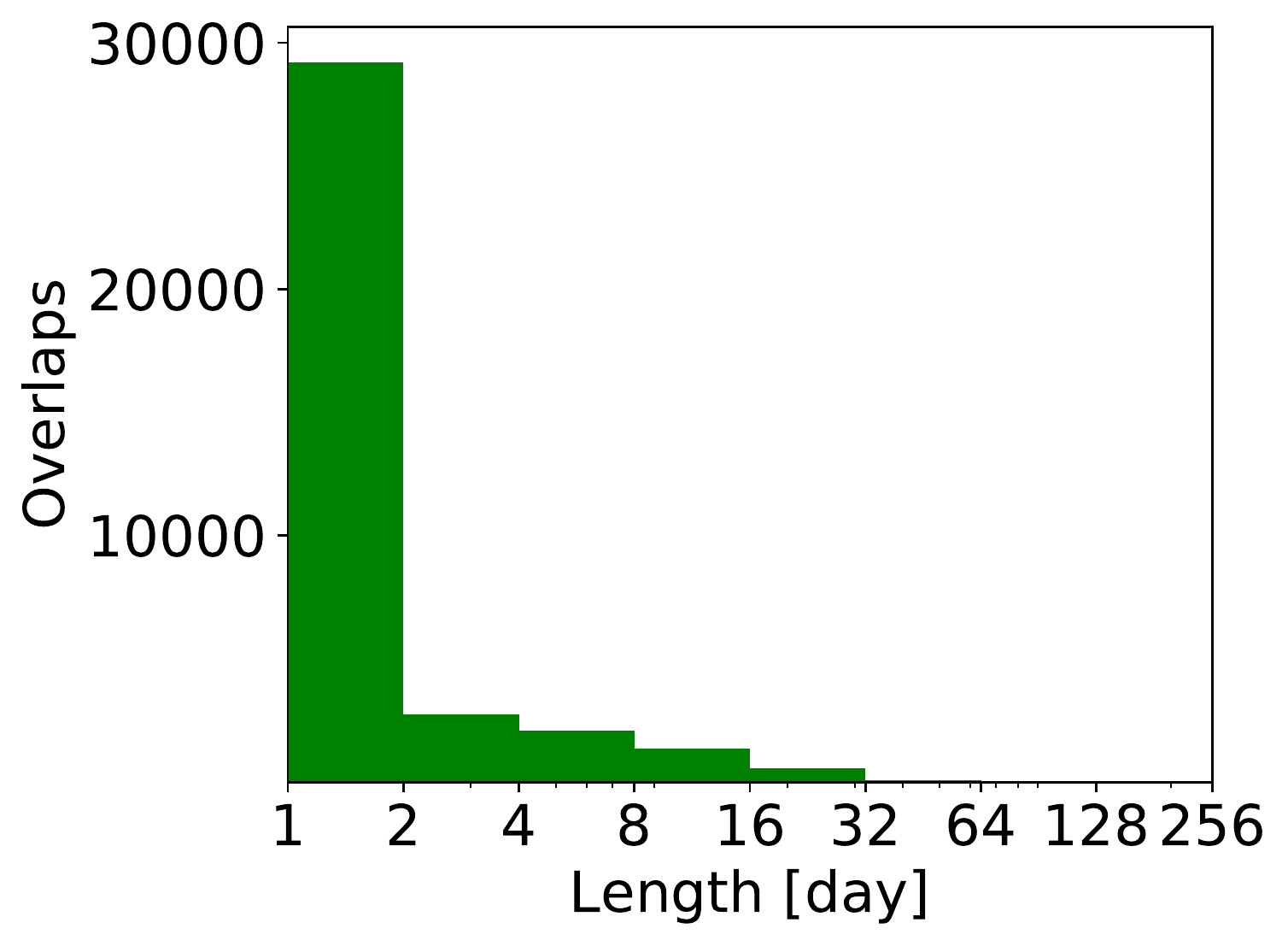}
			\caption{Bavaria}
		\end{subfigure}
		
		\begin{subfigure}[b]{0.45\textwidth}
			\centering
			\includegraphics[width=\textwidth]{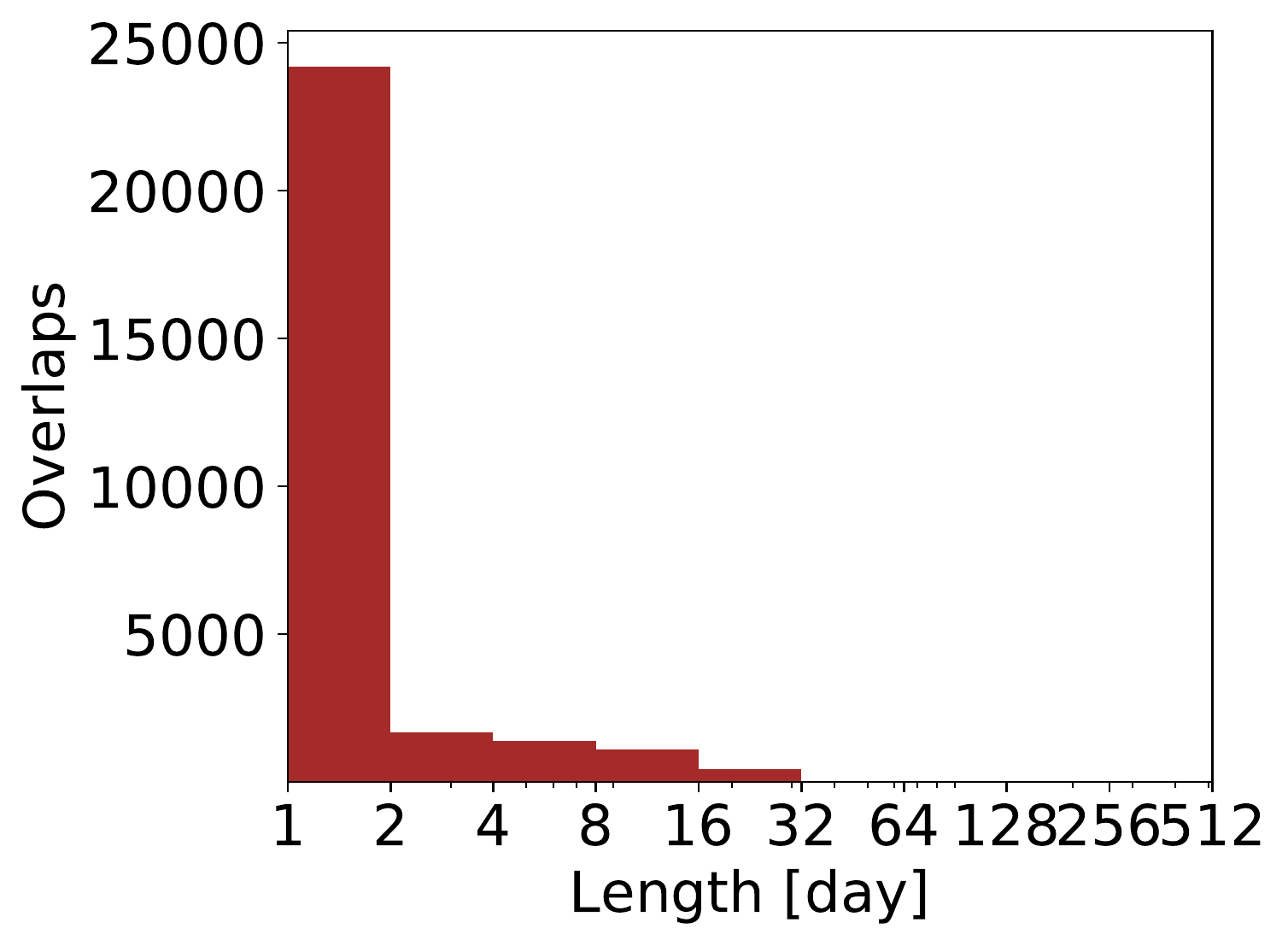}
			\caption{Baden-Württemberg}
		\end{subfigure}
		\hfill
		\begin{subfigure}[b]{0.45\textwidth}
			\centering
			\includegraphics[width=\textwidth]{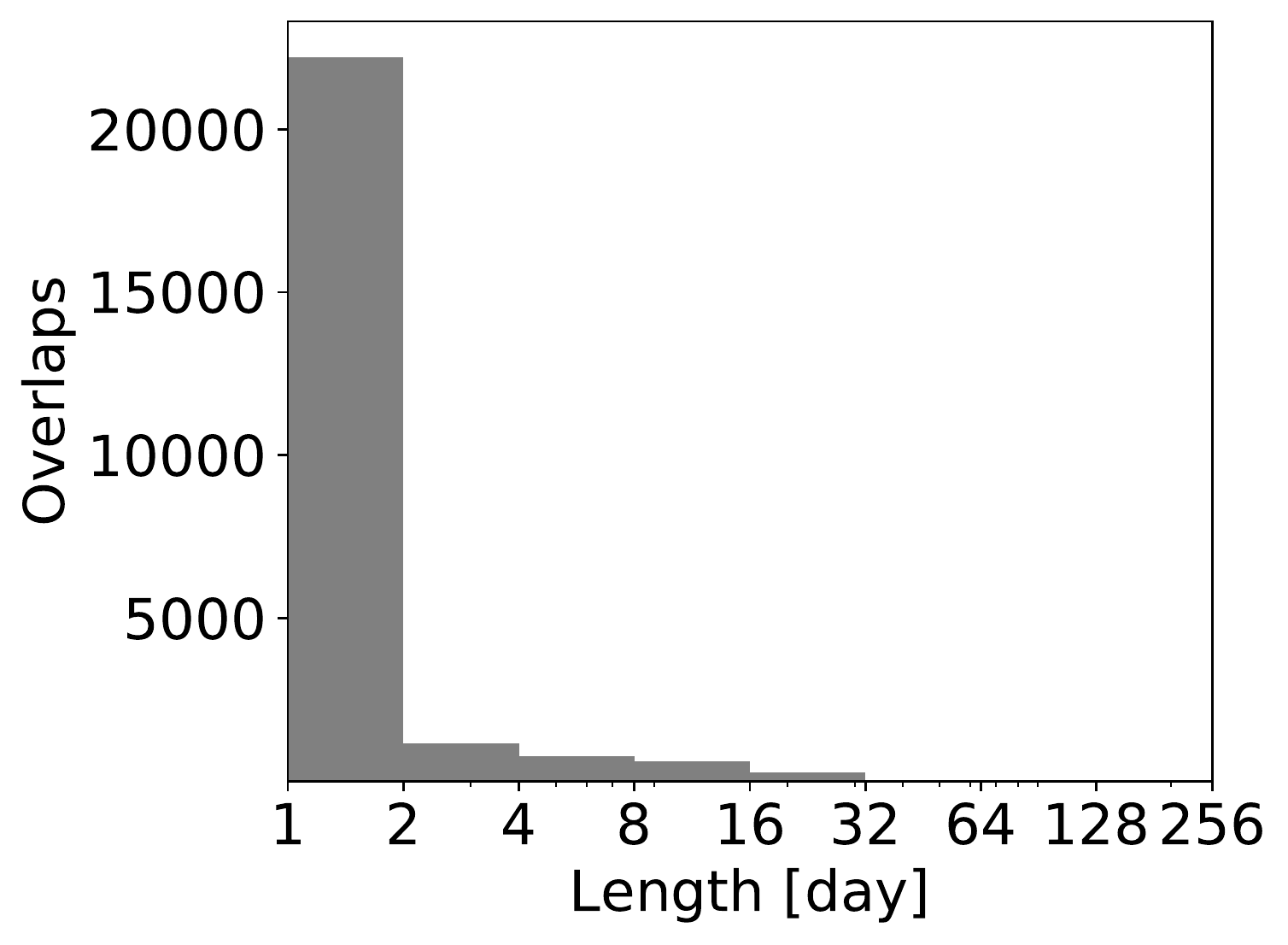}
			\caption{Lower Saxony}
		\end{subfigure}
		
		\begin{subfigure}[b]{0.45\textwidth}
			\centering
			\includegraphics[width=\textwidth]{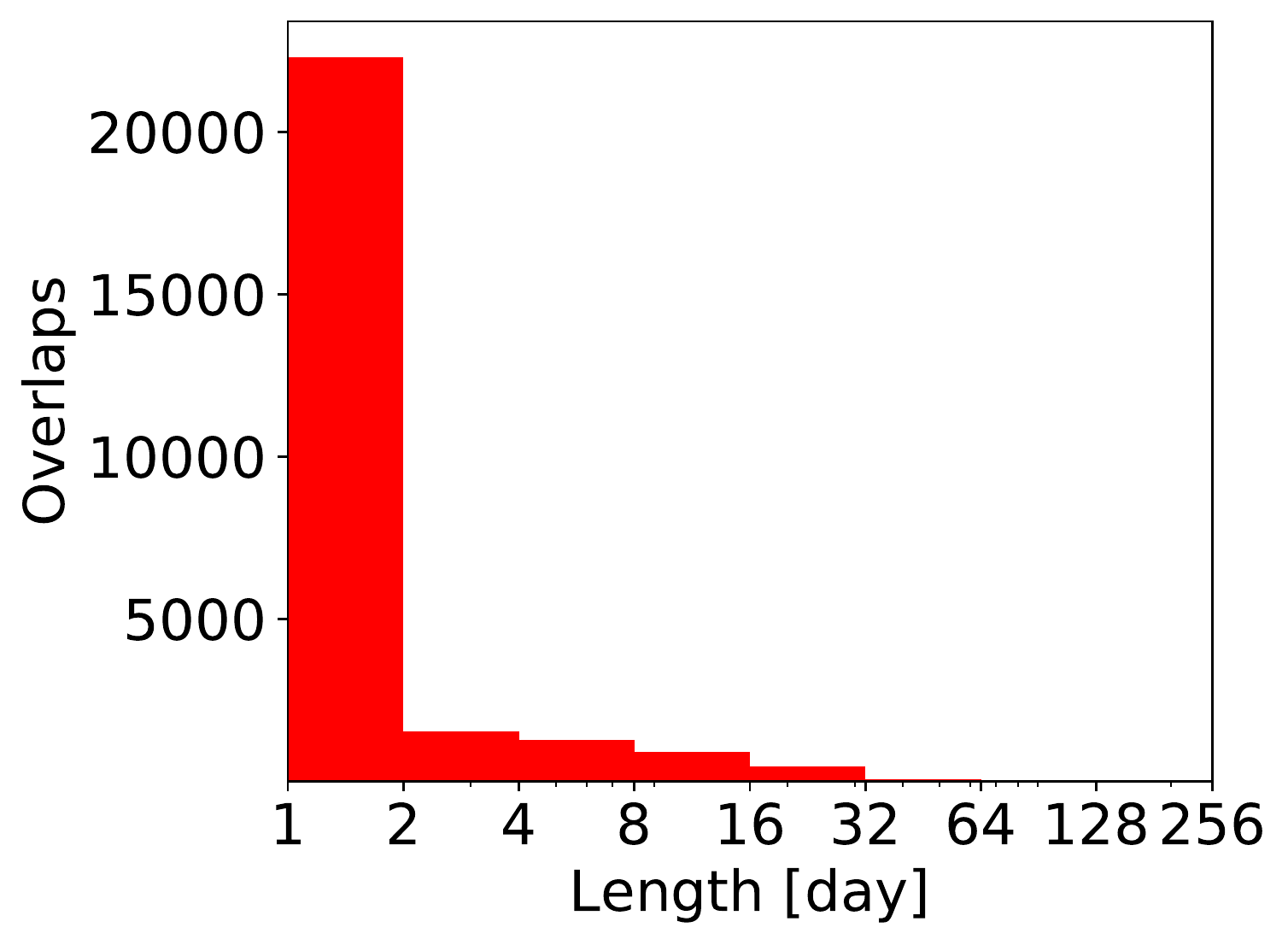}
			\caption{Hesse}
		\end{subfigure}
		\hfill
		\begin{subfigure}[b]{0.45\textwidth}
			\centering
			\includegraphics[width=\textwidth]{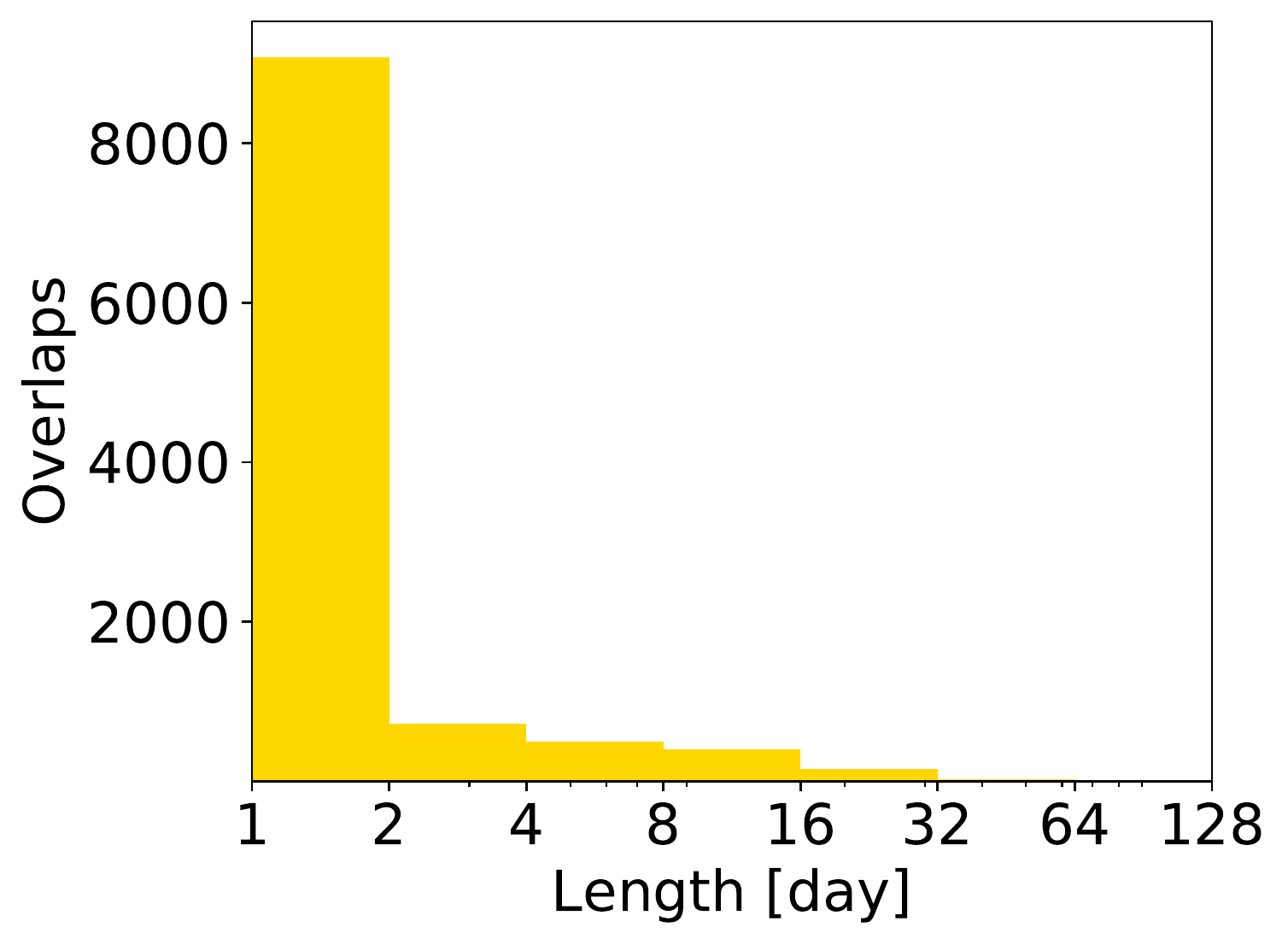}
			\caption{Rhineland-Palatinate}
		\end{subfigure}
	\end{figure}
	\clearpage   
	\begin{figure}[tb]\ContinuedFloat         
		\begin{subfigure}[b]{0.45\textwidth}
			\centering
			\includegraphics[width=\textwidth]{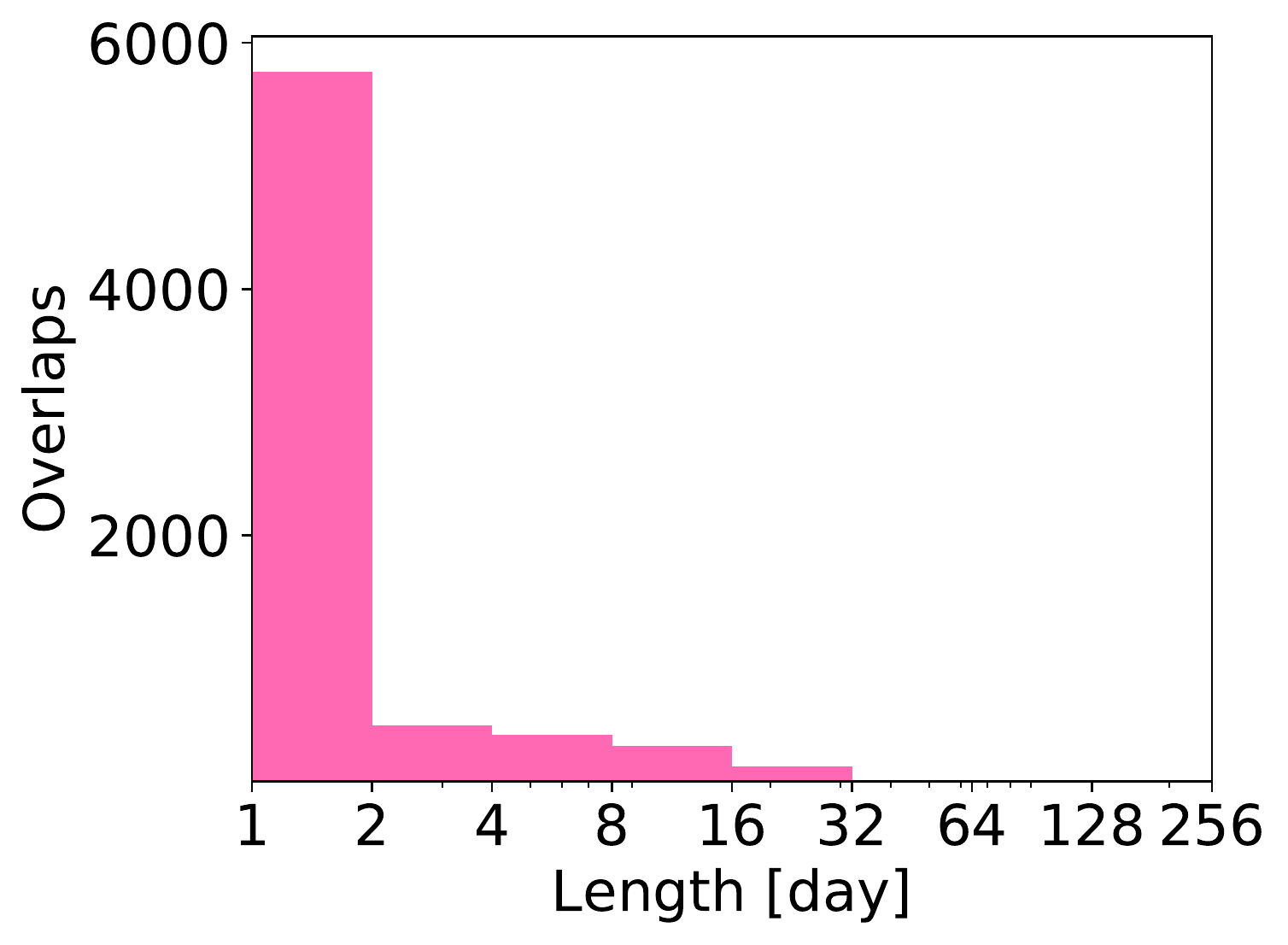}
			\caption{Saxony}
		\end{subfigure}
		\hfill
		\begin{subfigure}[b]{0.45\textwidth}
			\centering
			\includegraphics[width=\textwidth]{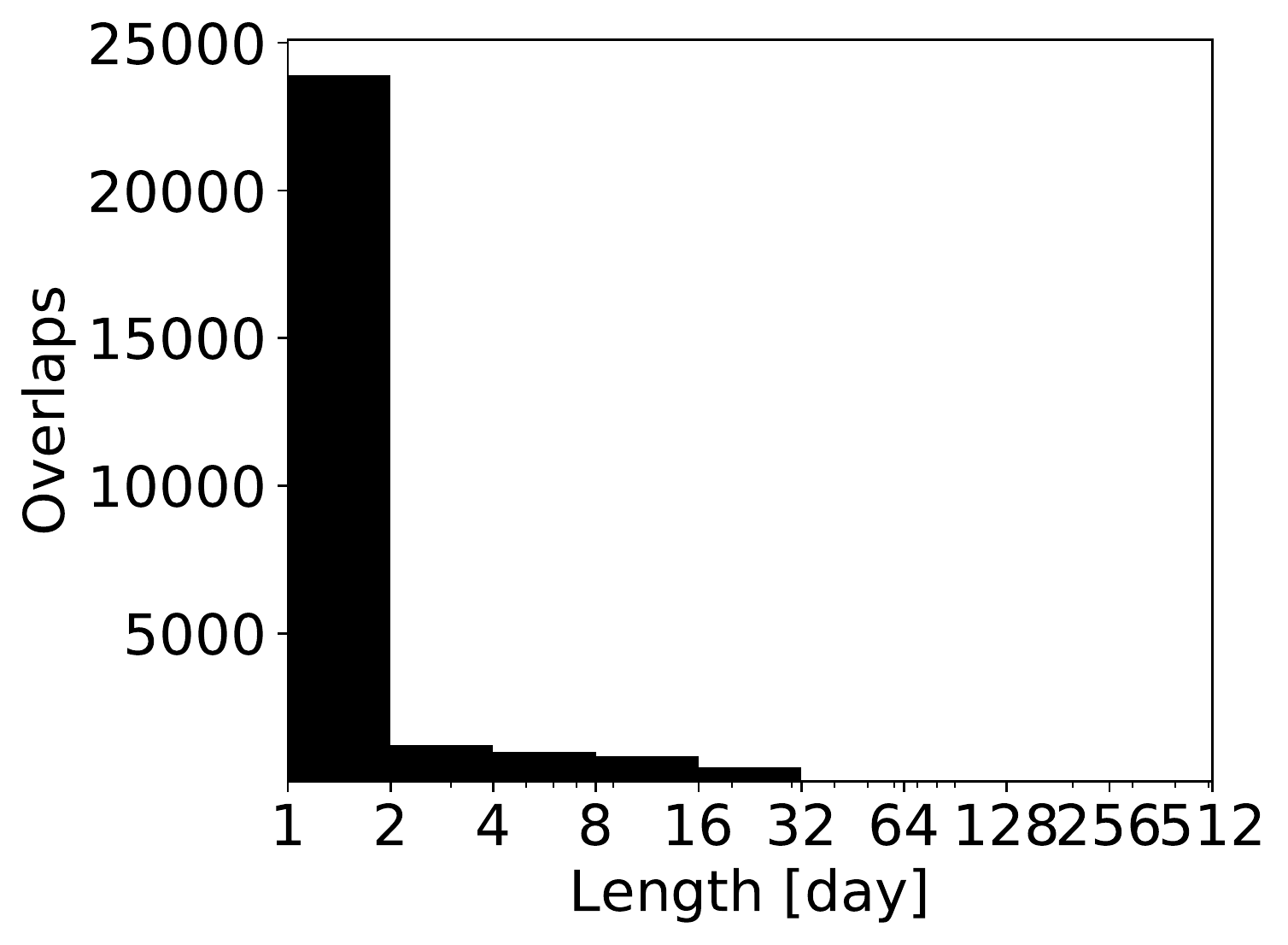}
			\caption{Berlin}
		\end{subfigure}
		
		\begin{subfigure}[b]{0.45\textwidth}
			\centering
			\includegraphics[width=\textwidth]{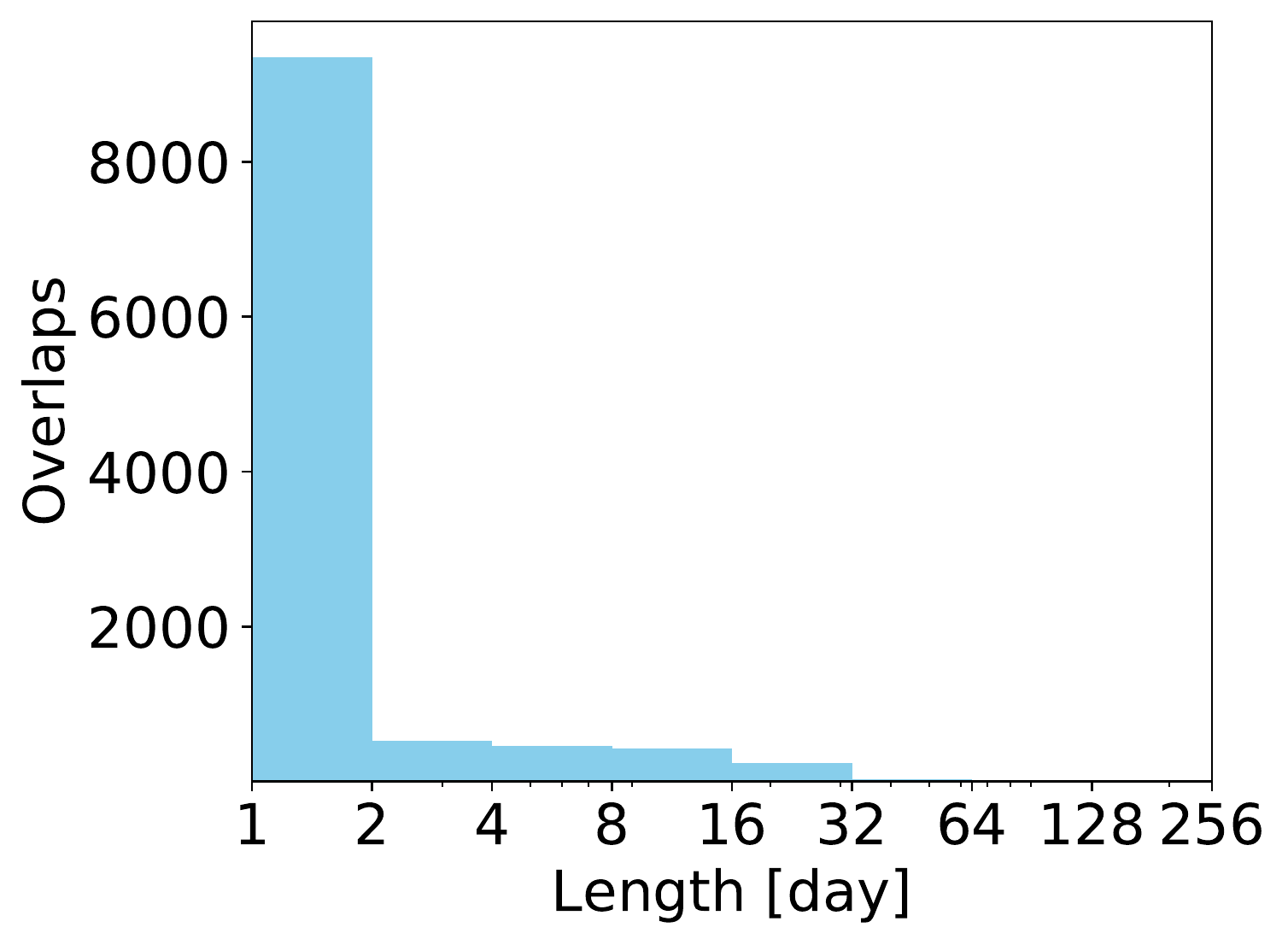}
			\caption{Schleswig-Holstein}
		\end{subfigure}
		\hfill
		\begin{subfigure}[b]{0.45\textwidth}
			\centering
			\includegraphics[width=\textwidth]{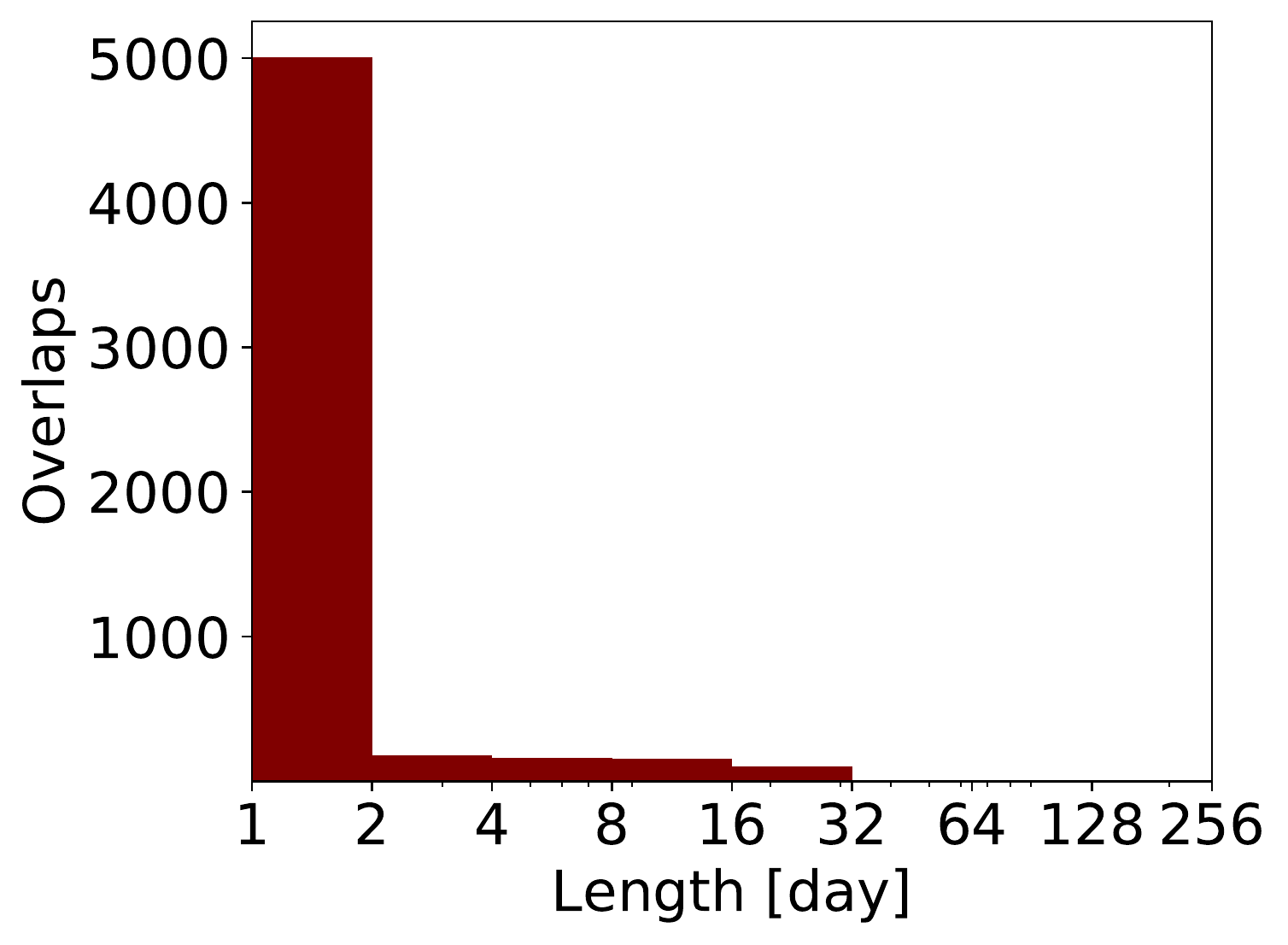}
			\caption{Brandenburg}
		\end{subfigure}   
		
		\begin{subfigure}[b]{0.45\textwidth}
			\centering
			\includegraphics[width=\textwidth]{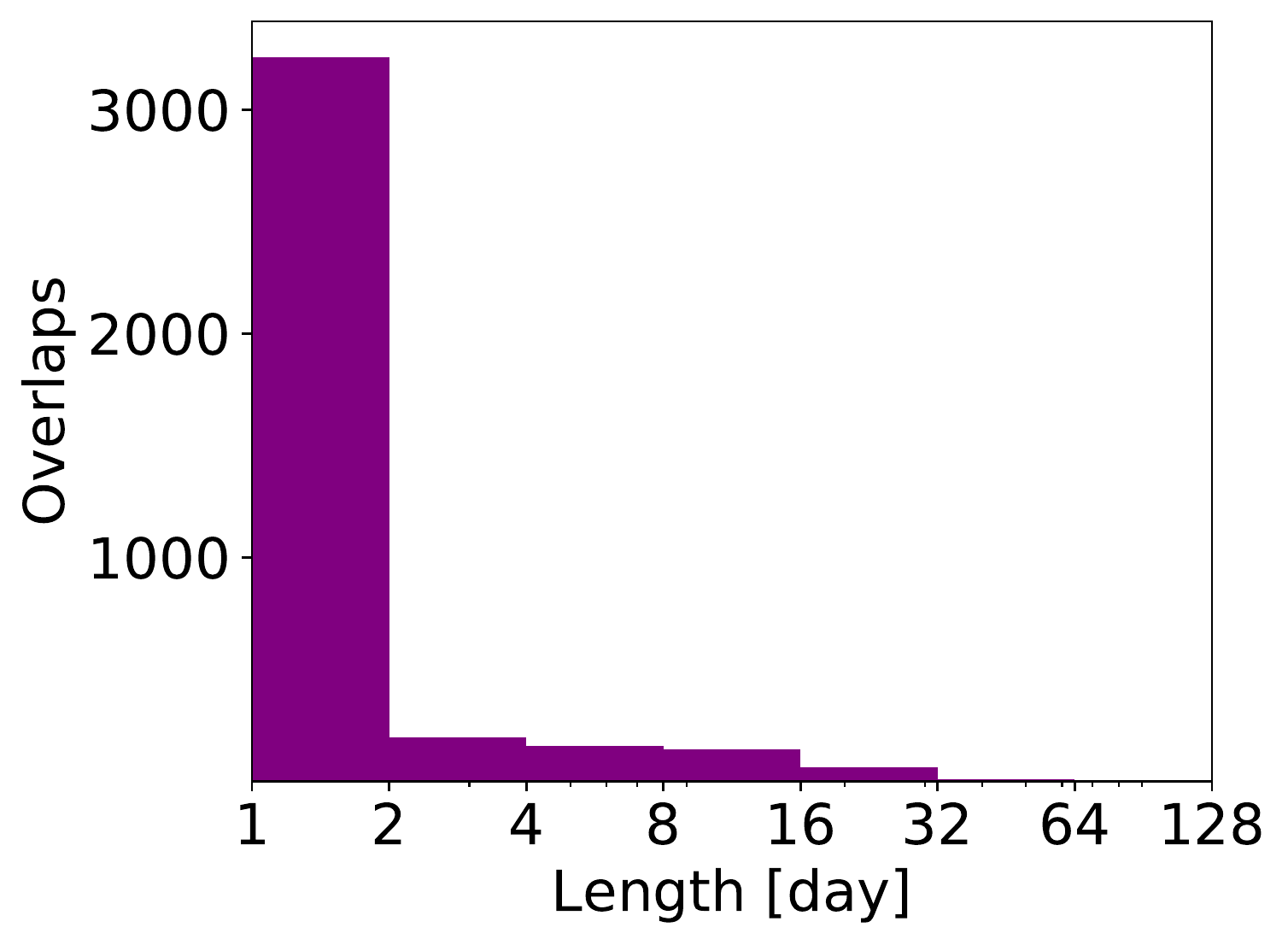}
			\caption{Saxony-Anhalt}
		\end{subfigure}
		\hfill
		\begin{subfigure}[b]{0.45\textwidth}
			\centering
			\includegraphics[width=\textwidth]{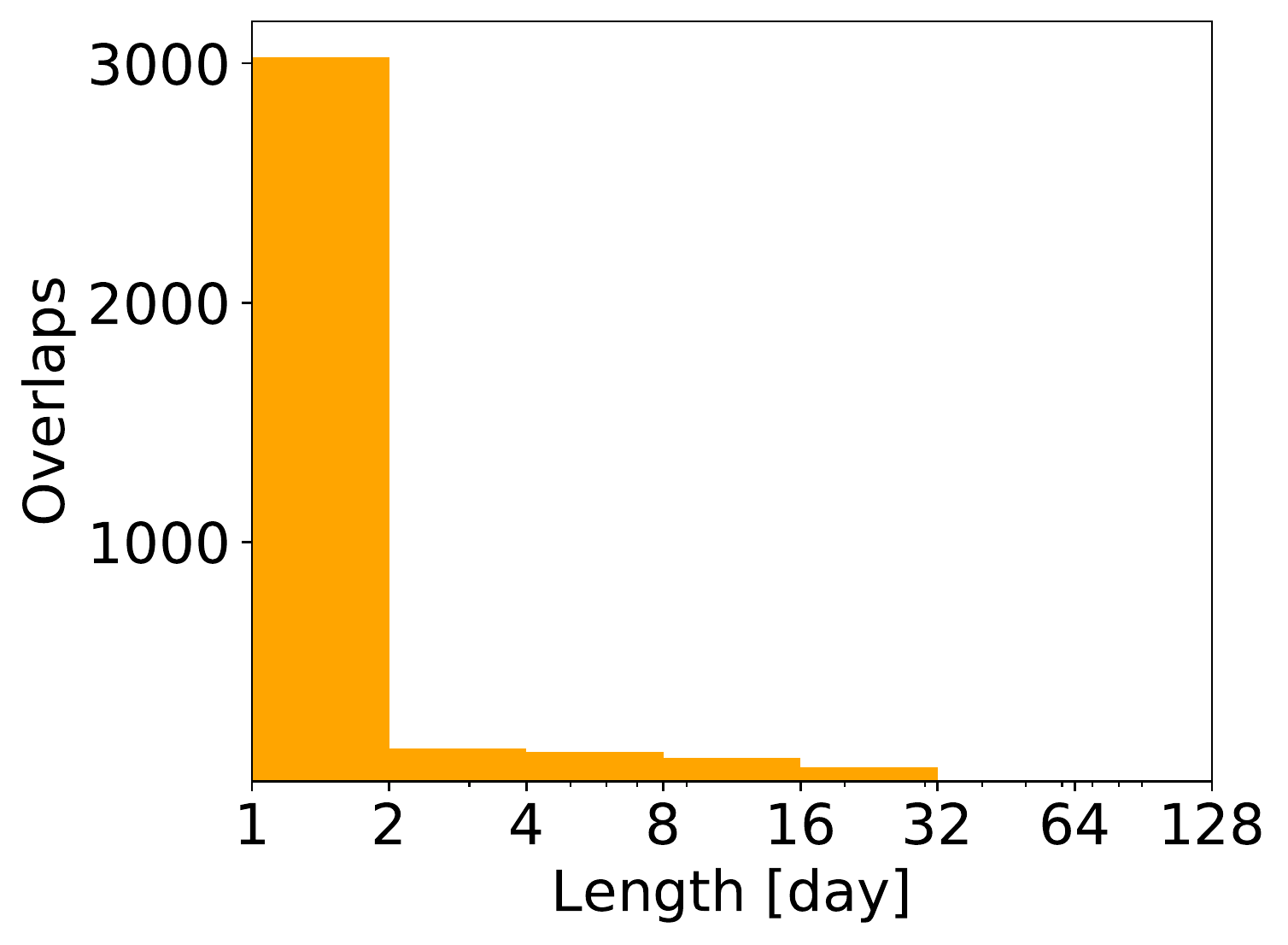}
			\caption{Thuringia}
		\end{subfigure}  
	\end{figure}
	\clearpage   
	\begin{figure}[tbh!]\ContinuedFloat 
		\begin{subfigure}[b]{0.45\textwidth}
			\centering
			\includegraphics[width=\textwidth]{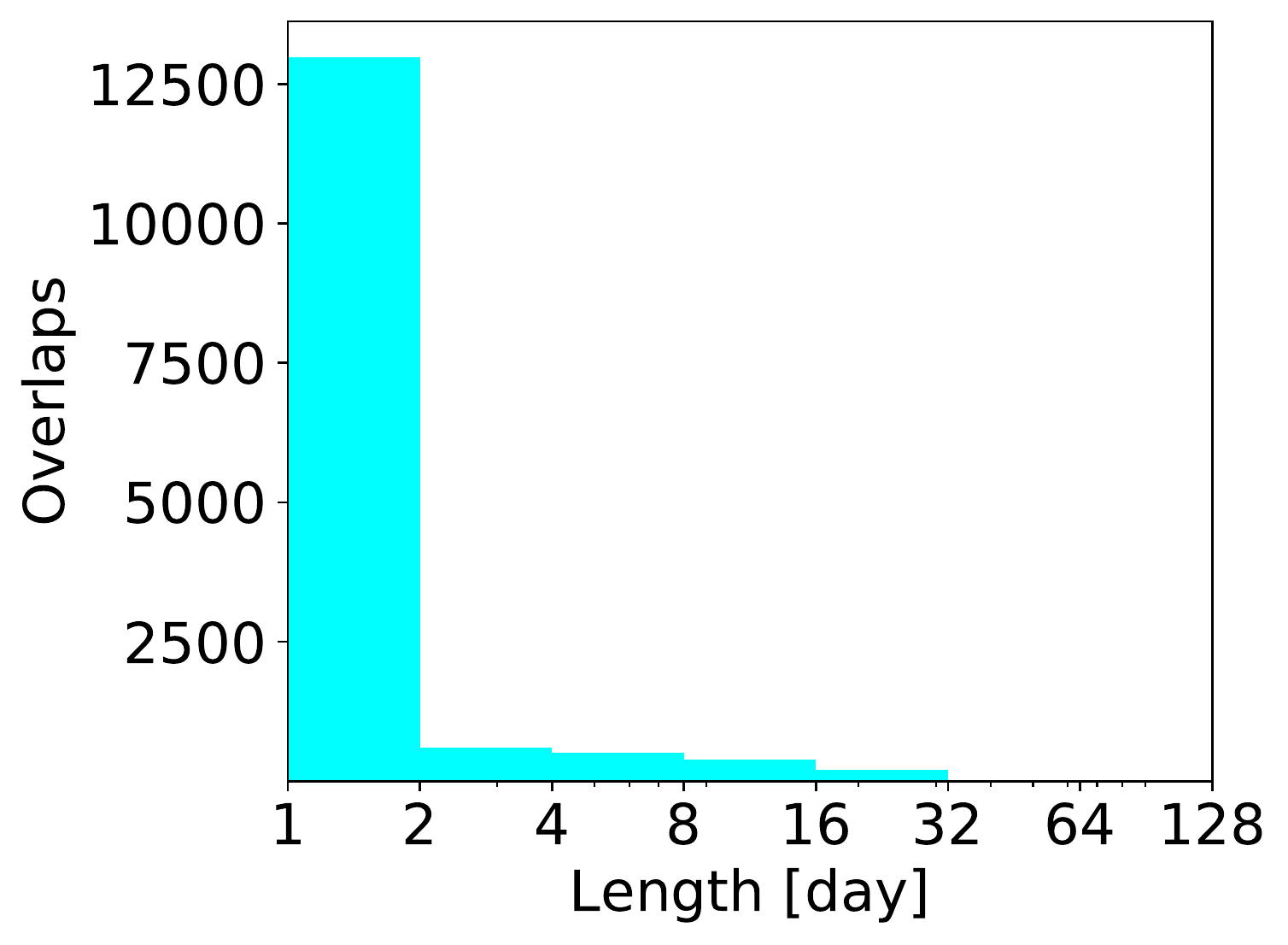}
			\caption{Hamburg}
		\end{subfigure}
		\hfill
		\begin{subfigure}[b]{0.45\textwidth}
			\centering
			\includegraphics[width=\textwidth]{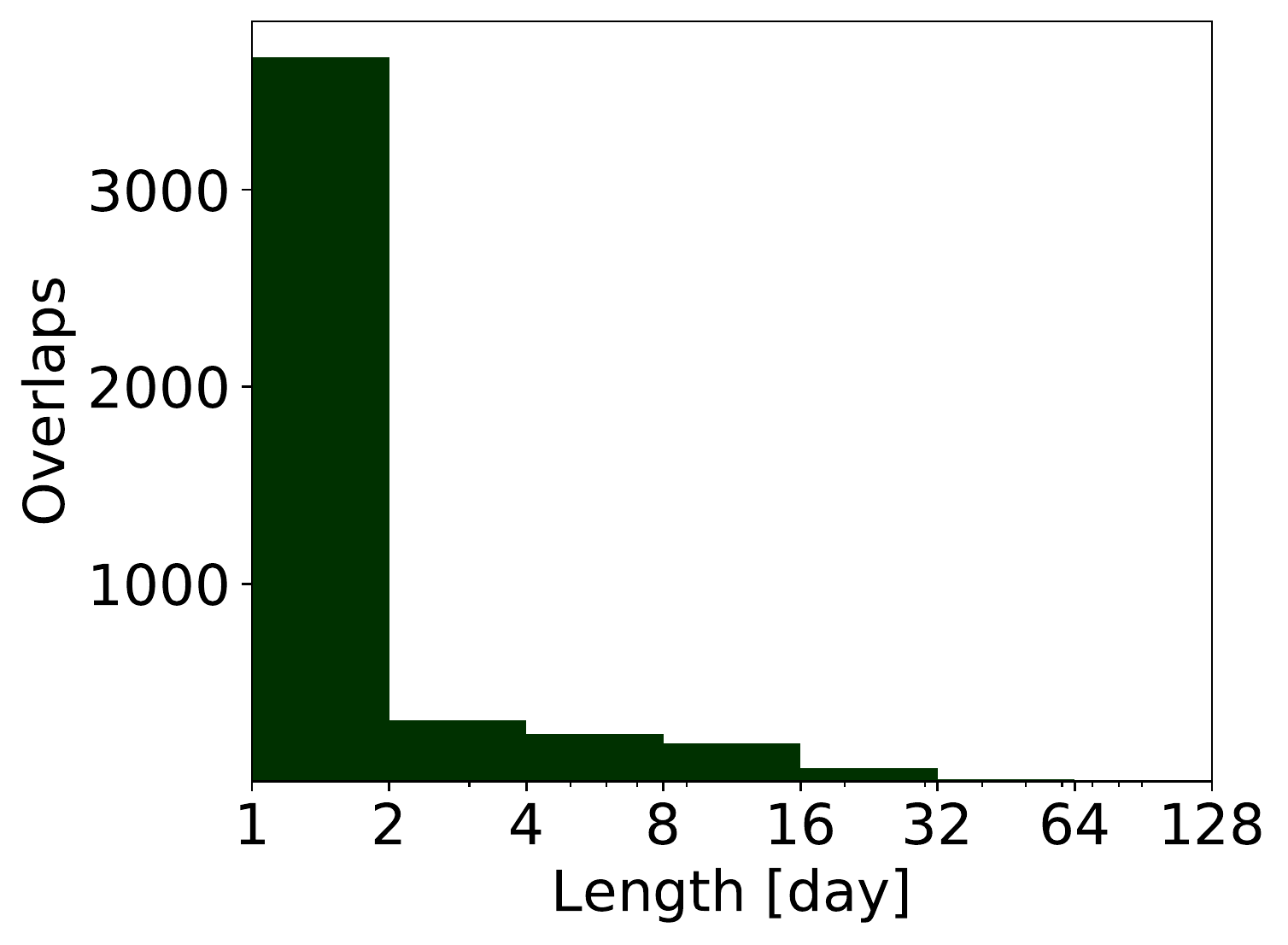}
			\caption{Mecklenburg-West Pomerania}
		\end{subfigure} 
		
		\begin{subfigure}[b]{0.45\textwidth}
			\centering
			\includegraphics[width=\textwidth]{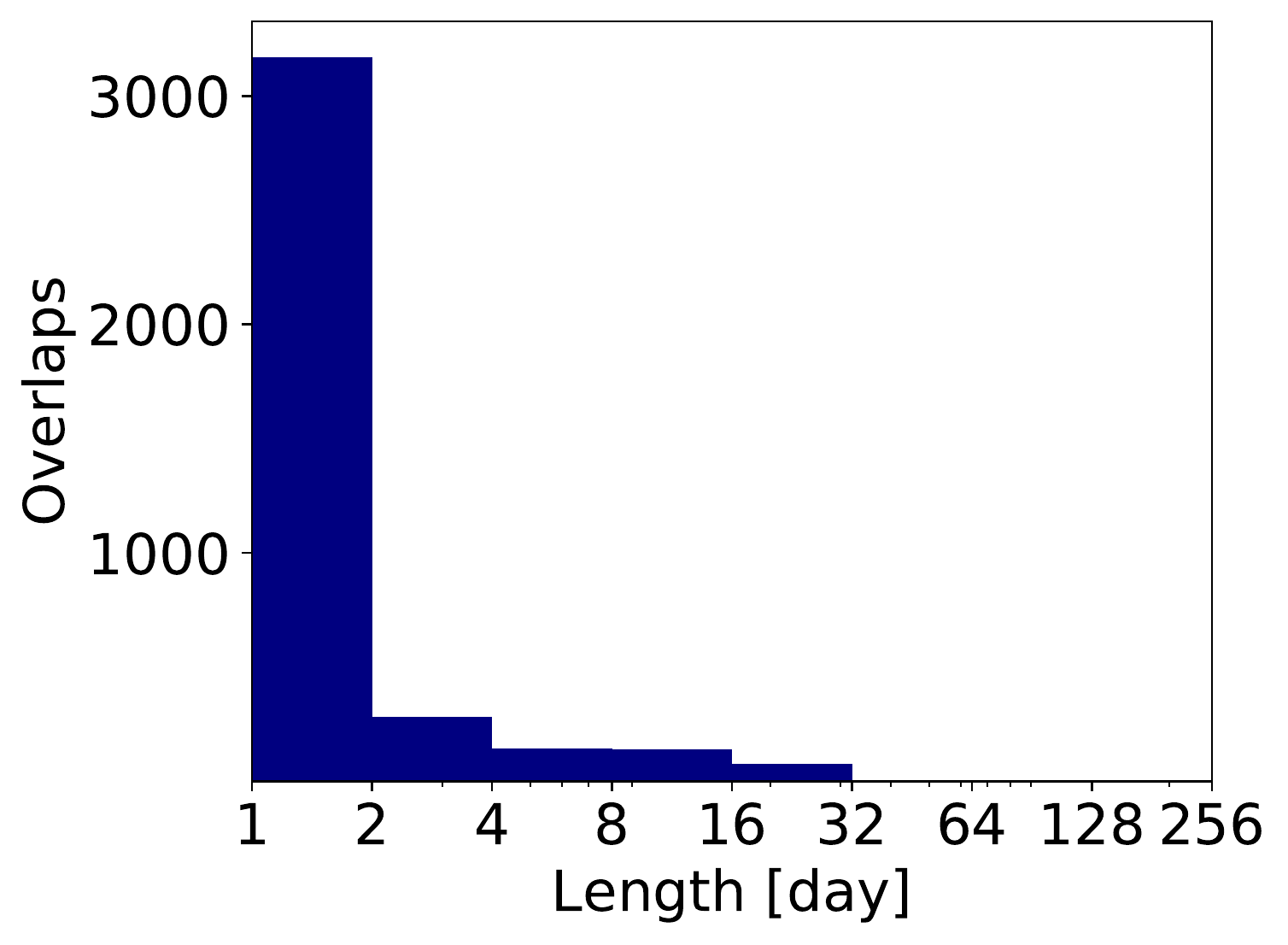}
			\caption{Saarland}
		\end{subfigure}
		\hfill
		\begin{subfigure}[b]{0.45\textwidth}
			\centering
			\includegraphics[width=\textwidth]{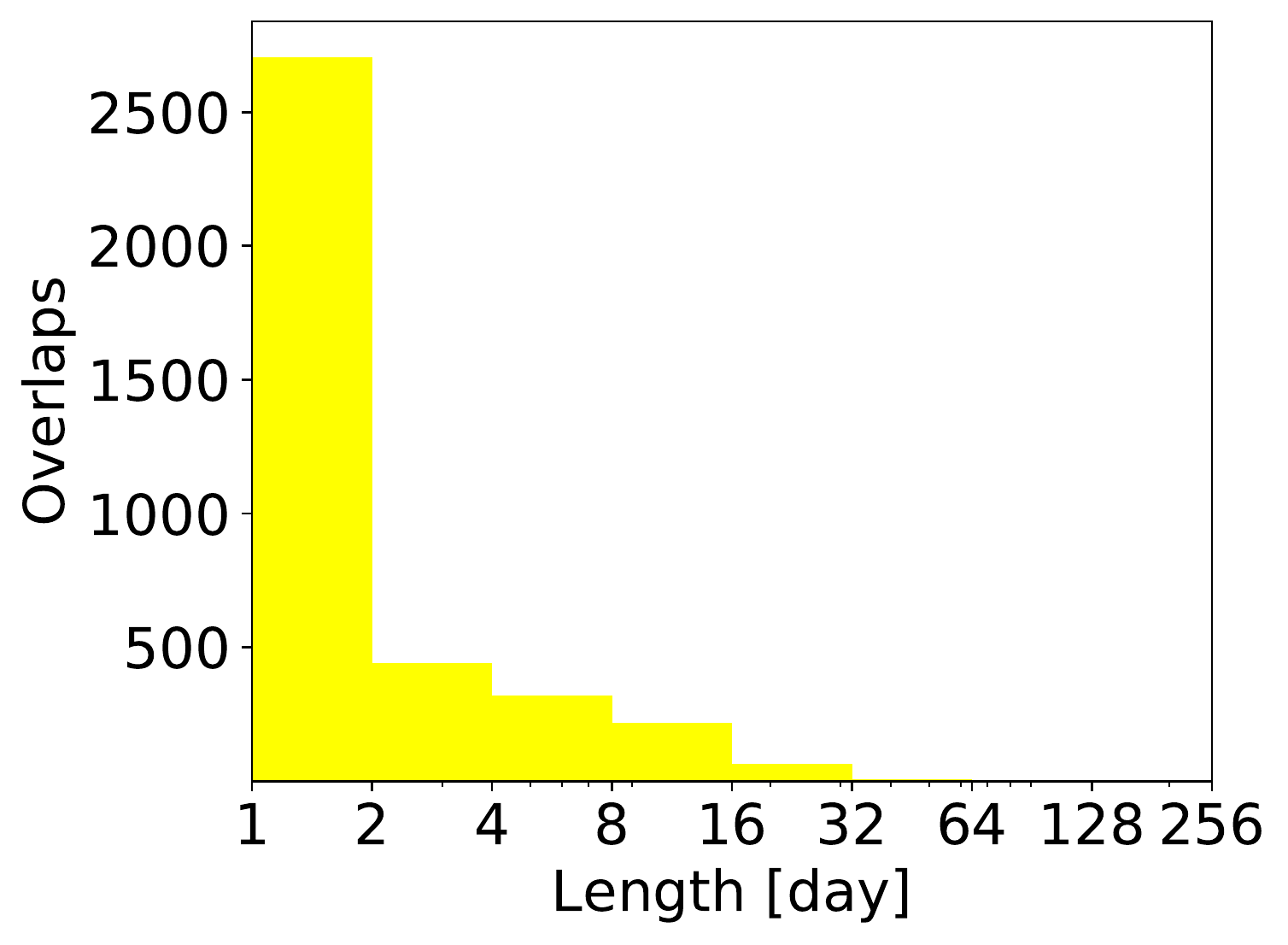}
			\caption{Bremen}
		\end{subfigure} 
		\begin{subfigure}[b]{0.45\textwidth}
			\centering
			\includegraphics[width=\textwidth]{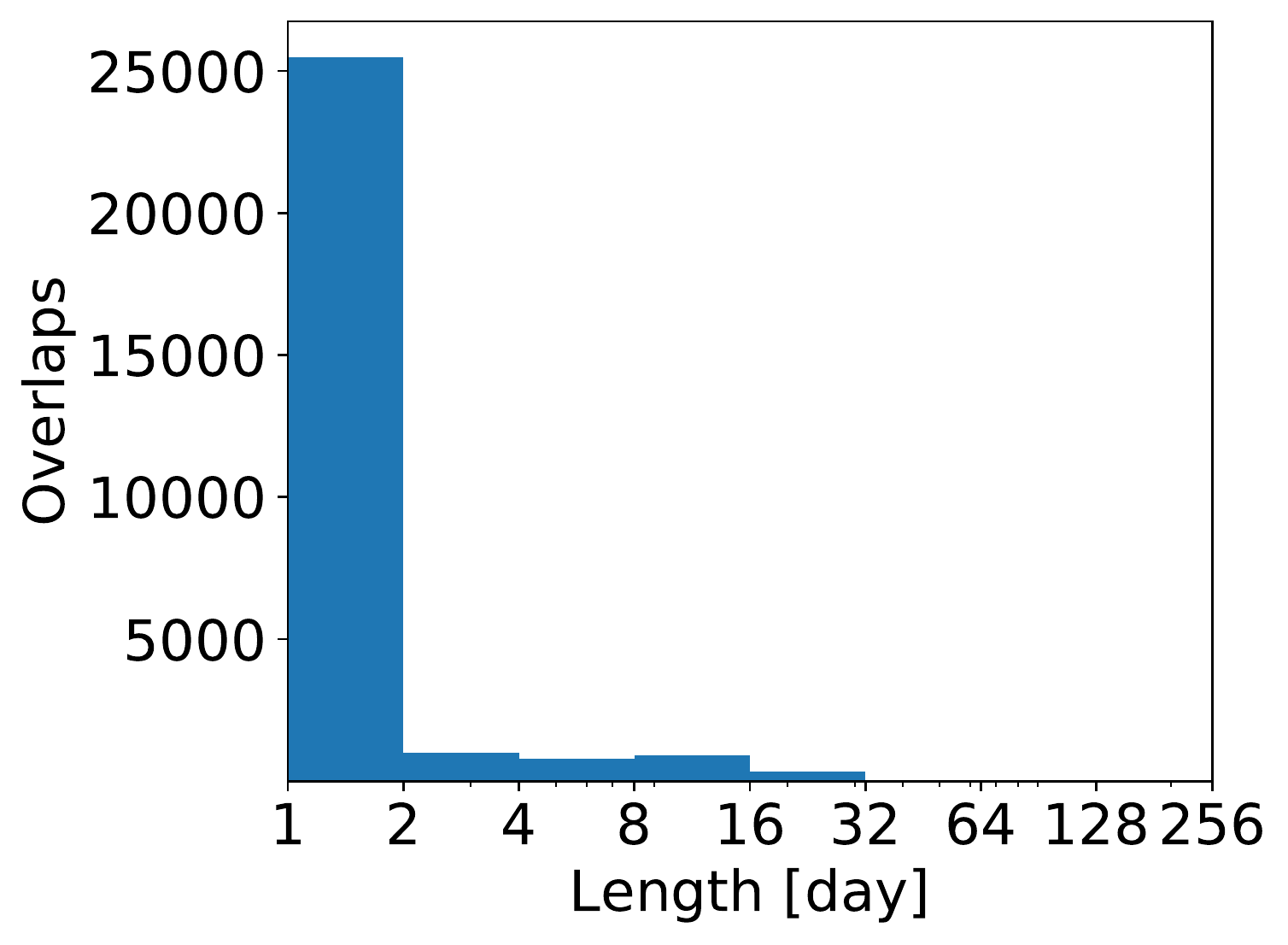}
			\caption{Between}
		\end{subfigure}
		\caption{Histograms of the duration of overlapping stays reported for in each state (a)-(p) and between states (q).}
		\label{fig:overlap_hist}
	\end{figure}

\begin{figure}[h!]
		\centering
		\includegraphics[width=\textwidth]{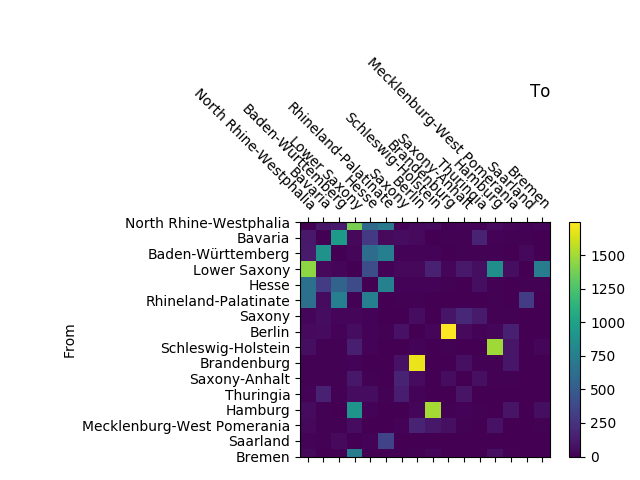}
		\caption{Number of overlaping records originating in one state and ending in a different one.}
		\label{fig:over_between}
	\end{figure}

	To further analyse overlaps between healthcare facilities from different states we generated a matrix indicating where overlapped transfers started and where ended, see Figure~\ref{fig:over_between}. The observed structure of transitions is similar to one showed for indirect transfers between states (cf. Figure~\ref{matrix:indirect}). There are many transfers from Brandenburg and Berlin, Hamburg and Schleswig-Holstein, Lower Saxony and North Rhine-Westphalia, Baden-Württemberg and Bavaria and another way around.
	
	To deeply characterize the types of overlaps in detail, as in previous reports \cite{Piotrowska2019arxiv, Lonc2019arxiv}, we used a four-digit classification. The truth is indicated by 1 while 0 means false. First digit indicate if two considered overlaps took place in the same healthcare facility, second digit: if overlaps have the same diagnoses, third: if two overlaps have the same admission dates and fourth if two overlaps have the same discharge dates. For example code 1100 simply means that two considered overlaps had been reported by the same healthcare facility, in both cases the diagnosis was the same, but there were different dates of admissions and discharges.
	
\begin{figure}[h!]
		\centering
		\includegraphics[height=15cm]{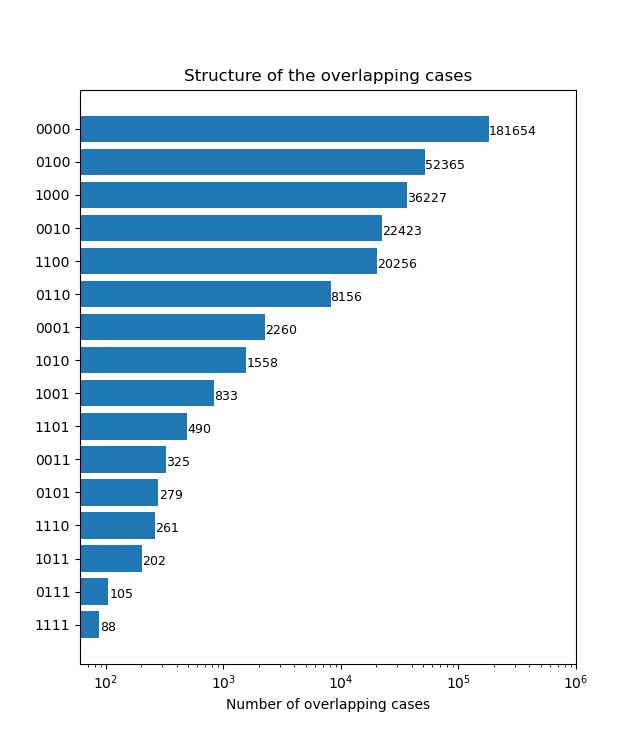}
		\caption{A structure of four-digit classification for whole dataset.}
		\label{fig:bin_str}
	\end{figure}

	The result of such classification is presented in Table~\ref{tab:bin_clas_stat} with a distinction to states and in~Figure \ref{fig:bin_str} without it. We see that in all the states most of the overlaps were classified as 0000 (different: diagnosis, healthcare facilities, admission date, discharge date). They were responsible for 181 654 overlaps out of 327 482. Other two significant types are 0100 and 1000 (respectively: same facility, same diagnose). 
	
	It is also worth noticing that the structure of overlaps was similar in all the locations. Four biggest states had the same order of four-digit classification according to the number of overlaps. The rest of the states had slight differences mainly in order of columns 0100, 1000 and columns 0010, 0010. 
	
	\begin{landscape}
		\thispagestyle{empty}
		\begin{table}[h!]
			\centering
			\caption{Effect of four-digit classification of the overlapping cases for given location.}
			\label{tab:bin_clas_stat}
			\begin{filecontents*}{tabels/overlap_bin_stats.csv}
Land,0000,0100,1000,0010,1100,0110,0001,1010,1001,1101,0011,0101,1110,1011,0111,1111
North Rhine-Westphalia,49186,13740,8254,5682,4856,2077,421,391,205,135,98,90,45,45,28,7
Bavaria,19341,4611,4365,3198,2565,1156,228,241,132,57,39,32,31,28,17,16
Baden-Württemberg,16542,4334,2867,2094,1707,716,145,263,63,34,22,24,30,16,11,3
Lower Saxony,13644,3740,2321,2318,1641,782,165,206,58,47,50,31,27,16,14,11
Hesse,14452,4123,3371,1480,2075,508,109,207,69,35,19,8,32,20,8,8
Rhineland-Palatinate,5750,1262,1603,970,700,312,64,107,37,18,12,9,10,10,6,1
Saxony,3502,830,1270,354,792,140,28,48,22,17,3,1,5,5,1,4
Berlin,13663,5129,4136,995,2676,300,96,263,97,55,14,18,25,37,2,16
Schleswig-Holstein,5983,1858,1262,902,466,220,48,150,37,23,15,7,17,9,2,12
Brandenburg,2871,757,961,329,391,200,21,47,9,7,3,5,4,1,2,0
Saxony-Anhalt,1927,567,611,249,255,113,15,33,10,4,3,1,5,3,2,1
Thuringia,1682,490,629,229,218,67,16,76,9,7,4,2,12,3,0,1
Hamburg,7825,2860,2242,504,752,245,53,94,47,30,6,19,7,5,2,4
Mecklenburg-West \mbox{Pomerania},2093,449,908,359,460,131,13,43,12,6,3,1,2,2,1,2
Saarland,2147,486,487,305,192,123,17,42,5,8,0,1,3,1,3,1
Bremen,1313,493,940,284,510,108,12,49,21,7,10,3,6,1,1,1
Between,19733,6636,0,2171,0,958,107,0,0,0,24,27,0,0,5,0
\end{filecontents*}
\pgfplotstabletypeset[
			col sep=comma, 
			display columns/0/.style={column name= State, column type={|p{2.5cm}|},string type}, 
			display columns/1/.style={column type={p{1cm}|},int detect, 1000 sep={\;},precision=3},
			display columns/2/.style={column type={p{0.9cm}|},int detect, 1000 sep={\;},precision=3},
			display columns/3/.style={column type={p{0.9cm}|},int detect, 1000 sep={\;},precision=3},
			display columns/4/.style={column type={p{0.9cm}|},int detect, 1000 sep={\;},precision=3},
			display columns/5/.style={column type={p{0.9cm}|},int detect, 1000 sep={\;},precision=3e},
			display columns/6/.style={column type={p{0.9cm}|},int detect, 1000 sep={\;},precision=3},
			display columns/7/.style={column type={p{0.8cm}|},string type},
			display columns/8/.style={column type={p{0.8cm}|},string type},
			display columns/9/.style={column type={p{0.8cm}|},string type},
			display columns/10/.style={column type={p{0.8cm}|},string type},
			display columns/11/.style={column type={p{0.7cm}|},string type},
			display columns/12/.style={column type={p{0.7cm}|},string type},
			display columns/13/.style={column type={p{0.7cm}|},string type},
			display columns/14/.style={column type={p{0.7cm}|},string type},
			display columns/15/.style={column type={p{0.7cm}|},string type},
			display columns/16/.style={column type={p{0.7cm}|},string type},
			every head row/.style={before row=\hline,after row=\hline},
			every nth row={1}{before row=\hline},
			every last row/.style={after row=\hline},
			]{tabels/overlap_bin_stats.csv}
		\end{table} 
		\vfill
		\raisebox{0ex}{\makebox[\linewidth]{\thepage}}
	\end{landscape}
	
	Each of four-digit set we analysed further by assigning each diagnose into groups indexed by numbers according to the rules presented in Table 3 in \cite{Piotrowska2019arxiv} or~\cite{Data}. In Table~\ref{tab:bin_icd} we summarised the most frequently appearing diagnosis within the particular types of overlaps for whole dataset. From presented result we see that most often diagnosis for two overlapping entries between different hospitals were diseases of the circulatory system (09,09). Other frequently present diagnosis were mental disorders (05, 05) and injuries (19, 19). Similarly, the most frequent diagnosis for two overlapping entries with in one healthcare facilities were also mental disorders (05, 05). Other frequent problems were neoplasms (02, 02) and related to pregnancy (15, x).
	
\pgfplotstableset{
    highlightrow/.style={
        postproc cell content/.append code={
           \count0=\pgfplotstablerow
            \advance\count0 by1
            \ifnum\count0=#1
            \pgfkeysalso{@cell content/.add={$\bf}{$}}
            \fi
        },
    },
}

	\begin{landscape}
		\thispagestyle{empty}
		\begin{table}[h!]
			\centering
			\caption{Number of cases for a given diagnosis for particular groups of overlaps. Two record
				overlaps are included in this table (vast majority among all overlaps).}
			\label{tab:bin_icd}
			\begin{filecontents*}{tabels/overlap_bin_icd_stat_format.csv}
0000	Over.	0001	Over.	0010	Over.	0011	Over.	0100	Over.	0101	Over.	0110	Over.	0111	Over.
['09,  09']	29116	['09,  09']	202	['09,  09']	5573	['09,  09']	90	['09,  09']	18168	['09,  09']	94	['09,  09']	3789	['09,  09']	44
['05,  05']	10512	['05,  05']	186	['05,  05']	1614	['05,  05']	51	['19,  19']	9808	['05,  05']	61	['19,  19']	1124	['05,  05']	26
['02,  02']	9651	['05,  19']	143	['19,  19']	1382	['05,  19']	34	['02,  02']	5009	['19,  19']	29	['16,  16']	664	['19,  19']	7
['19,  19']	8772	['05,  18']	62	['15,  15']	1177	['05,  18']	20	['05,  05']	4530	['02,  02']	23	['05,  05']	376	['18,  18']	6
['05,  19']	6848	['02,  02']	52	['16,  16']	1131	['09,  18']	18	['13,  13']	2288	['15,  15']	9	['15,  15']	372	['15,  15']	5
['10,  10']	5364	['05,  09']	44	['14,  14']	846	['19,  19']	8	['11,  11']	2044	['13,  13']	9	['14,  14']	336	['10,  10']	5
['13,  13']	4525	['09,  18']	42	['05,  19']	695	['06,  18']	7	['10,  10']	1702	['18,  18']	8	['11,  11']	319	['13,  13']	2
['11,  11']	4228	['19,  19']	31	['11,  11']	649	['09,  19']	6	['06,  06']	1620	['14,  14']	8	['18,  18']	198	['11,  11']	2
['05,  18']	3987	['05,  06']	31	['10,  10']	494	['18,  19']	5	['14,  14']	1527	['11,  11']	7	['06,  06']	168	['06,  06']	2
['09,  19']	3631	['10,  10']	29	['09,  19']	459	['14,  14']	5	['16,  16']	1363	['06,  06']	7	['10,  10']	163	['02,  02']	2
1000	Over.	1001	Over.	1010	Over.	1011	Over.	1100	Over.	1101	Over.	1110	Over.	1111	Over.
['05,  05']	7433	['05,  05']	111	['05,  05']	475	['15,  21']	62	['05,  05']	12622	['05,  05']	167	['05,  05']	169	['19,  19']	19
['05,  19']	2843	['05,  19']	57	['15,  15']	364	['14,  21']	61	['02,  02']	5362	['02,  02']	106	['02,  02']	30	['05,  05']	14
['02,  02']	2816	['09,  09']	43	['05,  19']	339	['15,  16']	20	['09,  09']	454	['09,  09']	48	['15,  15']	12	['15,  15']	13
['05,  18']	1674	['02,  02']	43	['14,  21']	277	['05,  19']	11	['19,  19']	401	['11,  11']	32	['09,  09']	11	['13,  13']	12
['05,  09']	1505	['05,  09']	34	['05,  18']	161	['05,  05']	9	['06,  06']	323	['19,  19']	30	['19,  19']	7	['21,  21']	7
['05,  06']	1411	['15,  21']	32	['05,  06']	67	['21,  21']	7	['13,  13']	190	['13,  13']	21	['06,  06']	7	['02,  02']	7
['05,  11']	1180	['05,  18']	32	['04,  05']	67	['15,  15']	3	['15,  15']	157	['14,  14']	16	['13,  13']	5	['03,  03']	3
['01,  02']	851	['02,  11']	20	['05,  09']	56	['05,  18']	3	['11,  11']	128	['10,  10']	16	['14,  14']	4	['18,  18']	2
['04,  05']	782	['19,  19']	19	['05,  11']	34	['02,  21']	3	['18,  18']	122	['06,  06']	14	['12,  12']	4	['14,  14']	2
['05,  10']	770	['11,  11']	16	['05,  10']	23	['17,  21']	2	['14,  14']	88	['18,  18']	8	['11,  11']	3	['11,  11']	2
\end{filecontents*}
\pgfplotstabletypeset[
			col sep=tab, 
			ignore chars={[,],'},
			display columns/0/.style={column type={|p{1.1cm}|},verb string type}, 
			display columns/1/.style={column name= Over.,column type={p{1cm}||},int detect, 1000 sep={\;},precision=3},
			display columns/2/.style={column type={p{1.1cm}|},verb string type},
			display columns/3/.style={column  name= Over.,column type={p{1cm}||},int detect, 1000 sep={\;},precision=3},
			display columns/4/.style={column type={p{1.1cm}|},verb string type},
			display columns/5/.style={column  name= Over.,column type={p{1cm}||},int detect, 1000 sep={\;},precision=3},
			display columns/6/.style={column type={p{1.1cm}|},verb string type},
			display columns/7/.style={column  name= Over.,column type={p{1cm}||},int detect, 1000 sep={\;},precision=3},
			display columns/8/.style={column type={p{1.1cm}|},verb string type},
			display columns/9/.style={column  name= Over.,column type={p{1cm}||},int detect, 1000 sep={\;},precision=3},
			display columns/10/.style={column type={p{1,11cm}|},verb string type},
			display columns/11/.style={column  name= Over.,column type={p{1cm}||},int detect, 1000 sep={\;},precision=3},
			display columns/12/.style={column type={p{1.1cm}|},verb string type},
			display columns/13/.style={column  name= Over.,column type={p{1cm}||},int detect, 1000 sep={\;},precision=3},
			display columns/14/.style={column type={p{1.1cm}|},verb string type},
			display columns/15/.style={column  name= Over.,column type={p{1cm}|},int detect, 1000 sep={\;},precision=3},
			every head row/.style={before row=\hline,after row=\hline},
			every nth row={1}{before row=\hline},
			every last row/.style={after row=\hline},
			highlightrow={11},
			assign column name/.style={/pgfplots/table/column name={\textbf{#1}}},
			every row 10 column 1/.style={verb string type},
			every row 10 column 3/.style={verb string type},
			every row 10 column 5/.style={verb string type},
			every row 10 column 7/.style={verb string type},
			every row 10 column 9/.style={verb string type},
			every row 10 column 11/.style={verb string type},
			every row 10 column 13/.style={verb string type},
			every row 10 column 15/.style={verb string type},
			]{tabels/overlap_bin_icd_stat_format.csv}
		\end{table}

		\vfill
		\raisebox{0ex}{\makebox[\linewidth]{\thepage}}
	\end{landscape}
	\medskip

\section{Derivation of inter-hospital networks and corresponding transfer matrices} \label{chapter: trensfer network}
	
We represented the whole healthcare system as a weighted directed graph, where the nodes represent healthcare facilities and corresponding community-nodes. In the case of direct transfer after the discharge patient goes to another hospital while in the case of indirect transfer to corresponding community-node. Patient from community-node can go either back to the same hospital or to a different one. Thus, in our approach the transfers between community-nodes are impossible. Following~\cite{Piotrowska2020PlosComp} and using the code~\cite{emergenetpackage}, based directly on the provided data (and on length of stay in nodes data in particular), we determined graph edge weights being probabilities of transfer between nodes. Next we used a matrix to code the graph structure, where $(i,j)$ entry is simply the probability of patient transfer from node $i$ to node $j$. Clearly, under that definition entry $(i,i)$ would describe the probability of a stay in node $i$.

\begin{figure}[h!]
	\centering
	\begin{subfigure}[b]{0.54\textwidth}
		\centering
		\includegraphics[width=\textwidth]{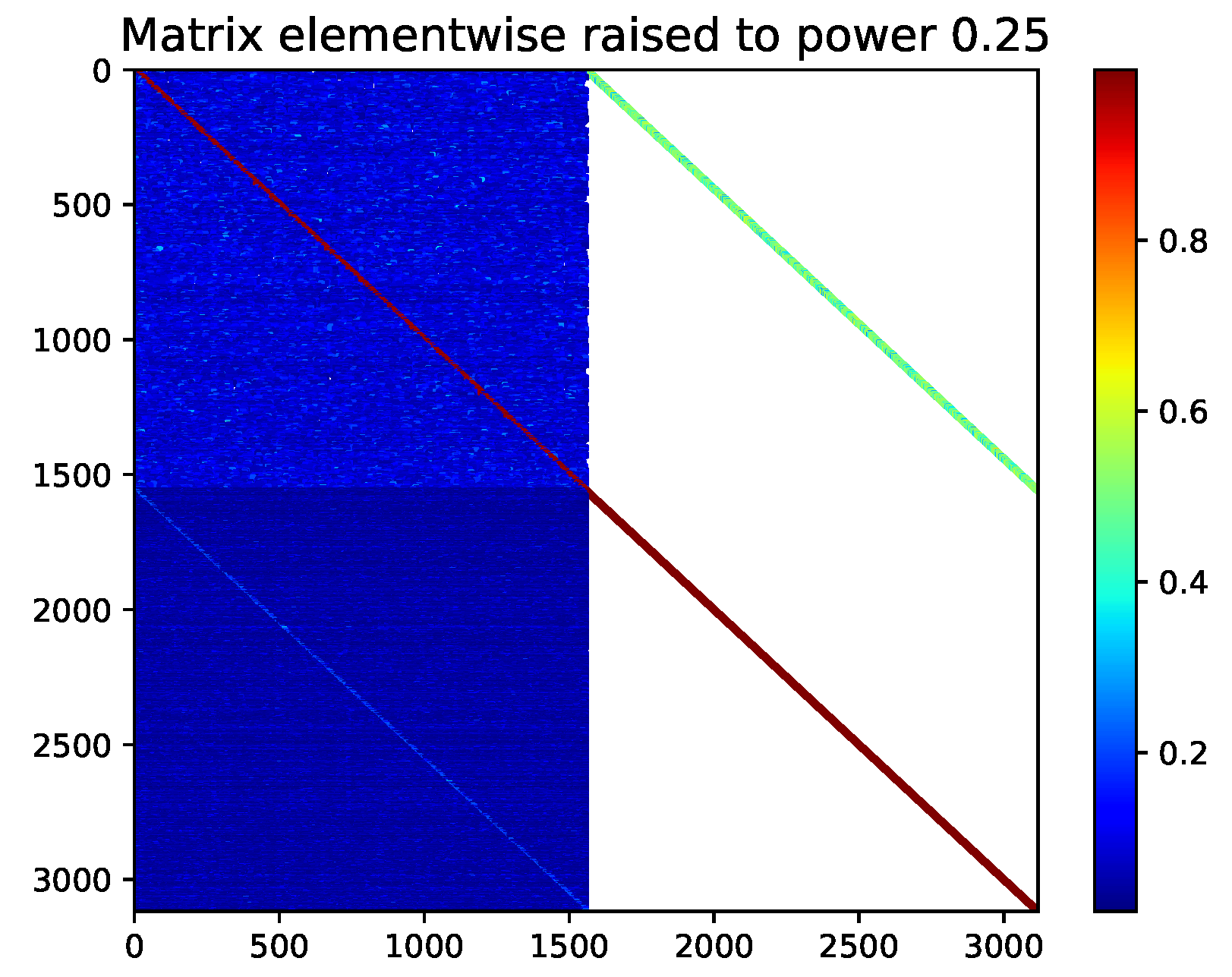}
		\caption{}
	\end{subfigure}
	\hfill
	\begin{subfigure}[b]{0.44\textwidth}
		\centering
		\includegraphics[width=\textwidth, height=\textwidth]{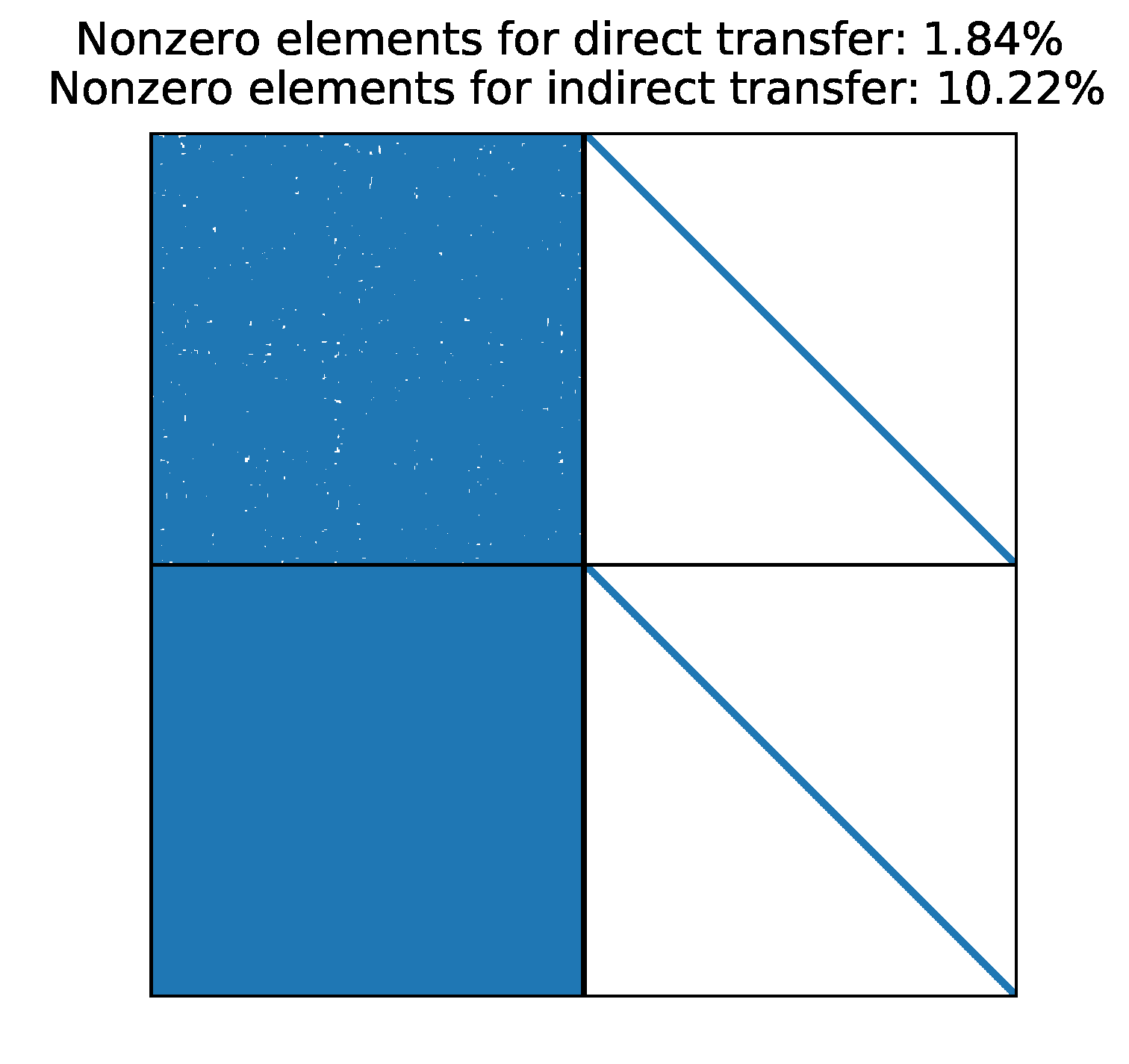}
		\caption{}
	\end{subfigure} 
	\caption{Visualization of the transfer probability matrix for the whole healthcare network created based on provided data. Healthcare facilities are numbered from 1 to $n$ and community-nodes numbered from $n + 1$ to $n + n$, $n = 1\,559$.(a) probability in matrix with all elements raised to power 0.25 (to emphasize the differences between elements, as they are mostly close to 0), (b) visualization of non-zero elements of matrix. The non-zero patterns of the left half suggest mostly non-zero elements, but this is only due to number of nodes --- actually number of non-zero elements is about 10\% for indirect transfers block (lower left) and 2\% for direct transfers block (upper left).
	}
	\label{fig:wiz_transfer_matrix}
\end{figure}

During the process of matrix derivation 19 healthcare facilities were omitted due to the fact that the corresponding community-node size was equal to 0. Moreover, we excluded facilities that were inactive for over 90 consecutive days. Thus finally, 1 559 facilities and 1 559 corresponding community-nodes were taken into account. The obtained transfer probability matrix for all states has a clear block structure, c.f. Figure~\ref{fig:wiz_transfer_matrix}.
The upper right diagonal block describes probabilities of discharges from healthcare facilities to corresponding community-nodes. The bottom right (diagonal due to the model assumptions) block corresponds to patient exchange between community-nodes. The left upper and bottom blocks represents probabilities of direct transfers between healthcare facilities (direct transfers) and admissions to healthcare facilities from society (indirect transfers), respectively. 
Transfer matrix has indirect transfer block clearly denser then direct transfer block underling the potential of indirect transfers as a pathogen transmission channel. Although it is not clear from the visualizations, due to matrix size, most of the elements in both direct and indirect blocks are zeros. For direct block only 1.84\% elements are not zeroes and for indirect it is 10.22\%.

\begin{figure}[h!]
	\centering
	\begin{subfigure}[b]{0.475\textwidth}
		\centering
		\includegraphics[width=\textwidth]{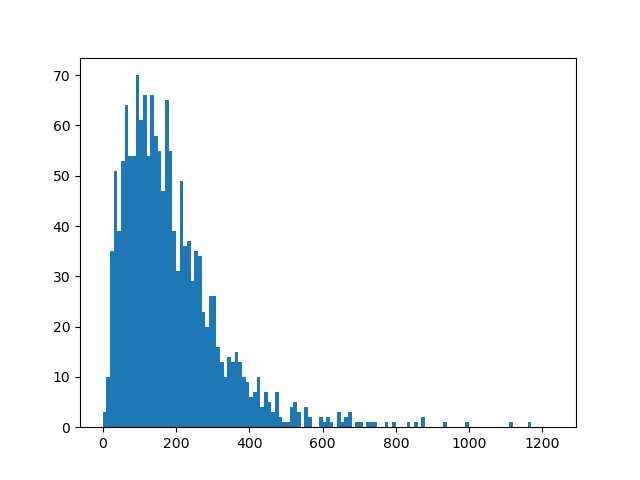}
		\caption{}
	\end{subfigure}
	\hfill
	\begin{subfigure}[b]{0.475\textwidth}
		\centering
		\includegraphics[width=\textwidth]{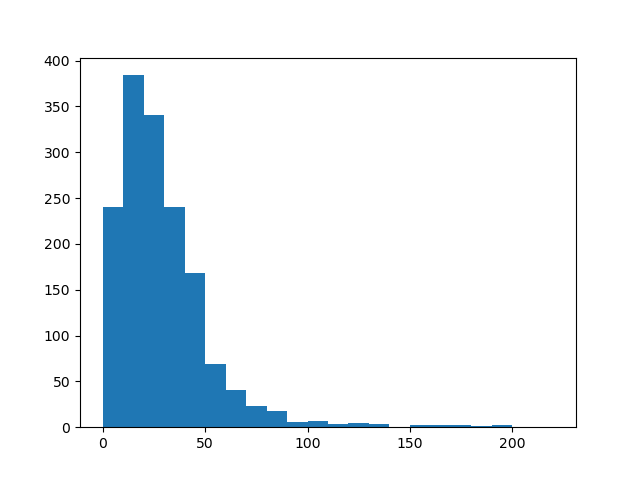}
		\caption{}
	\end{subfigure} 
	\caption{Histograms of in-degree (a) and out-degree (b) for the healthcare facility nodes.}
	\label{fig:hosp-in_out}
\end{figure}

\begin{figure}[h!]
	\centering
	\includegraphics[width = 0.475\textwidth]{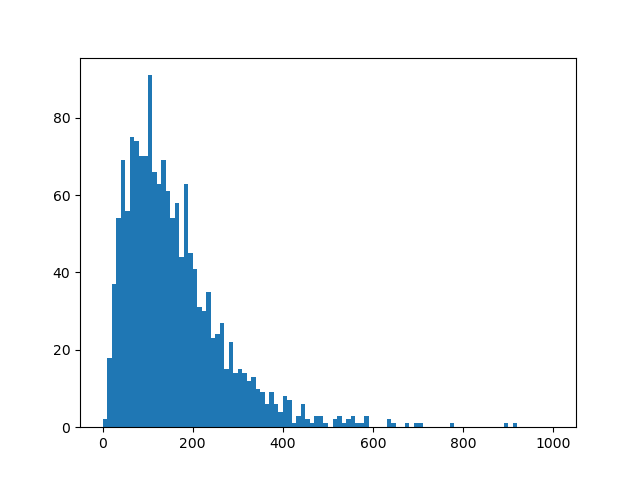}
	\caption{Histogram of out-degree for the community-nodes.}
	\label{fig:social_out}
\end{figure}

In Figure \ref{fig:hosp-in_out}, we present in- and out-degree histograms for the graph. We see that in-degrees were significantly higher then out degrees. That is related to the fact that to a hospital patient can be admitted from all other facilities and community-nodes, but can be discharged only to one corresponding community-node or other hospitals. Four hospitals had in-degrees higher than 900, two of them were located in Berlin, one in Baden-Württemberg and one in Hamburg. In Figure~\ref{fig:social_out} we present distribution of out degree for community-nodes. Analysing in-degree is pointless since its always 1.

The diameter of graph for all (not excluded) analysed data was 7 meaning that the distance between every two nodes was not greater then 7. The radius of whole graph was also 7 indicating that there was a node that is not further from any other node than 7. In addition, the density of the graph, understood as a ratio of all edges to $k(k-1)$  where $k$ is the number of all nodes, is equal to 0.0305.

If we neglect interstates transfers, the whole graph can be divided into disjoint sub-graphs corresponding to given states, cf. Figure~\ref{fig:hosp_graph}. Diameters of state sub-graphs ranged from 6 (for ones with largest population according~\cite{Destatis}) to 3 (for ones with smallest population), while the radius vary between states and it was either 4, 3 or 2. The average in-degrees were the same as out-degrees for all considered sub-graphs and they ranged from 77.17 for a state with biggest population to 12.35 for one with smallest one. Graphs for Berlin and Hamburg had relatively high in- and out-degrees, 32.48 and 24.23 respectively, compared to their populations. The densities of networks were between 0.0959 for Bavaria and 0.494 for Bremen. Information about all sub-graphs is summarized in Table~\ref{tab: trans_mat}.

\begin{figure}[h!]
	\centering
	\begin{subfigure}[b]{0.47\textwidth}
		\centering
		\includegraphics[width=\textwidth]{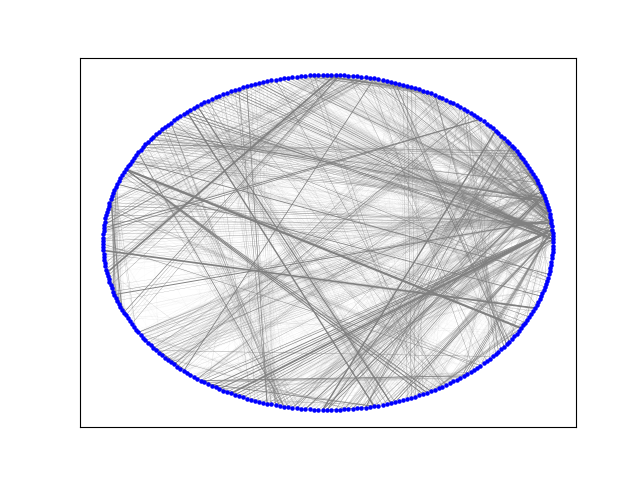}
		\caption{North Rhine-Westphalia}
	\end{subfigure}
	\hfill
	\begin{subfigure}[b]{0.47\textwidth}
		\centering
		\includegraphics[width=\textwidth]{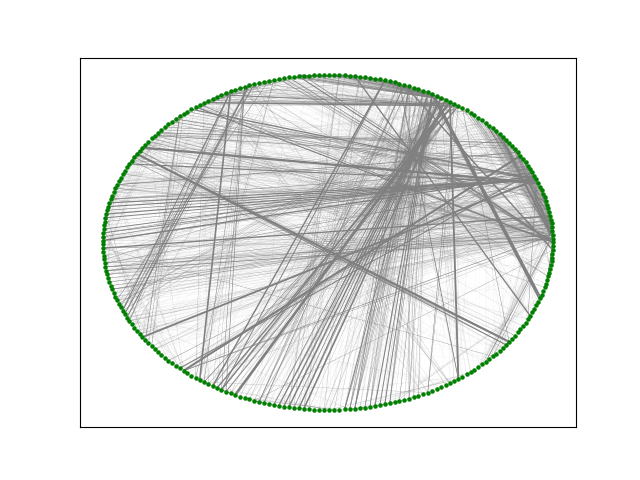}
		\caption{Bavaria}
	\end{subfigure}
	
	\begin{subfigure}[b]{0.47\textwidth}
		\centering
		\includegraphics[width=\textwidth]{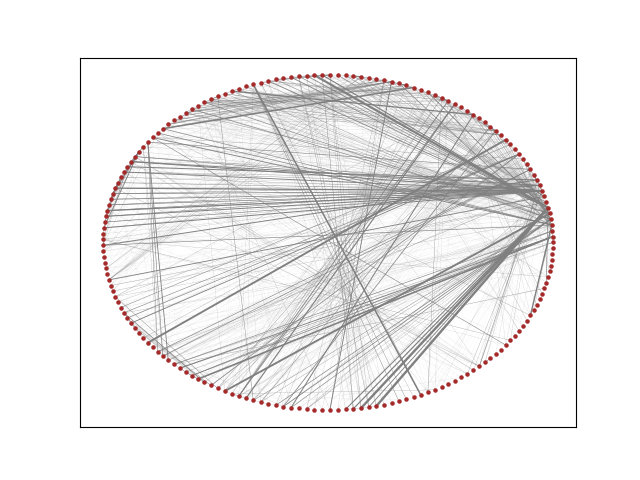}
		\caption{Baden-Württemberg}
	\end{subfigure}
	\hfill
	\begin{subfigure}[b]{0.47\textwidth}
		\centering
		\includegraphics[width=\textwidth]{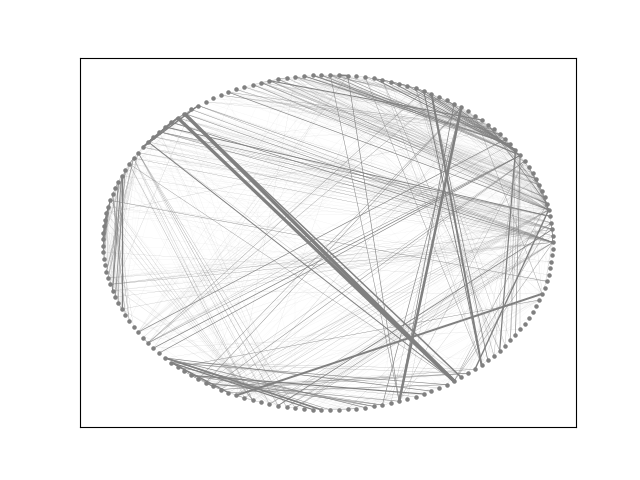}
		\caption{Lower Saxony}
	\end{subfigure}
\end{figure}
\clearpage   
\begin{figure}[tb]\ContinuedFloat  	
	\begin{subfigure}[b]{0.47\textwidth}
		\centering
		\includegraphics[width=\textwidth]{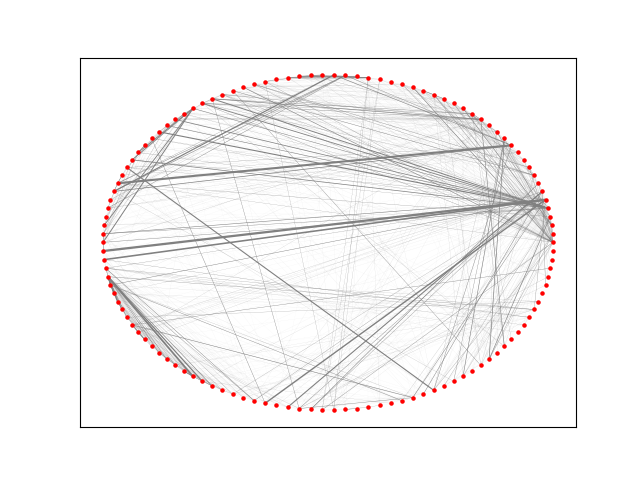}
		\caption{Hesse}
	\end{subfigure}
	\hfill
	\begin{subfigure}[b]{0.47\textwidth}
		\centering
		\includegraphics[width=\textwidth]{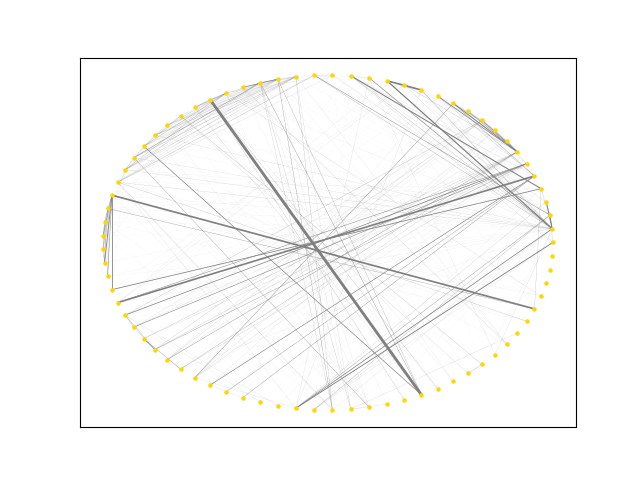}
		\caption{Rhineland-Palatinate}
	\end{subfigure}
       
	\begin{subfigure}[b]{0.47\textwidth}
		\centering
		\includegraphics[width=\textwidth]{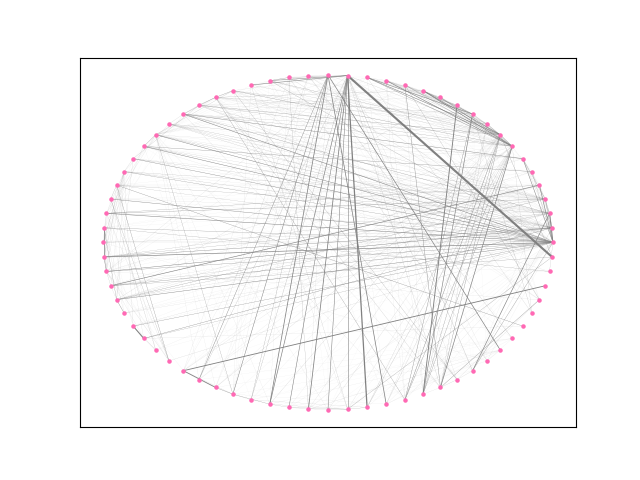}
		\caption{Saxony}
	\end{subfigure}
	\hfill
	\begin{subfigure}[b]{0.47\textwidth}
		\centering
		\includegraphics[width=\textwidth]{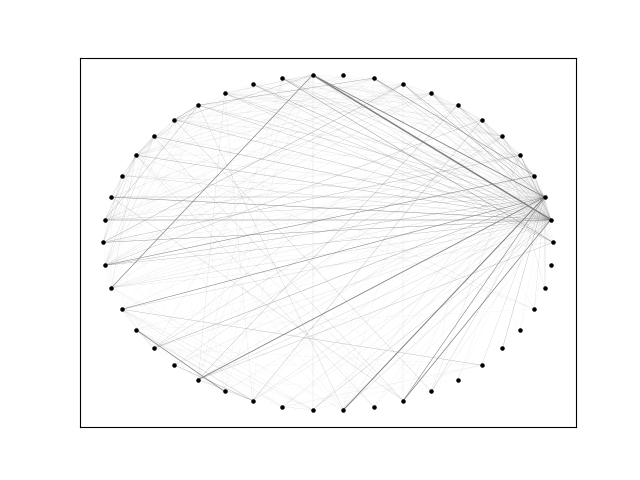}
		\caption{Berlin}
	\end{subfigure}
	
	\begin{subfigure}[b]{0.47\textwidth}
		\centering
		\includegraphics[width=\textwidth]{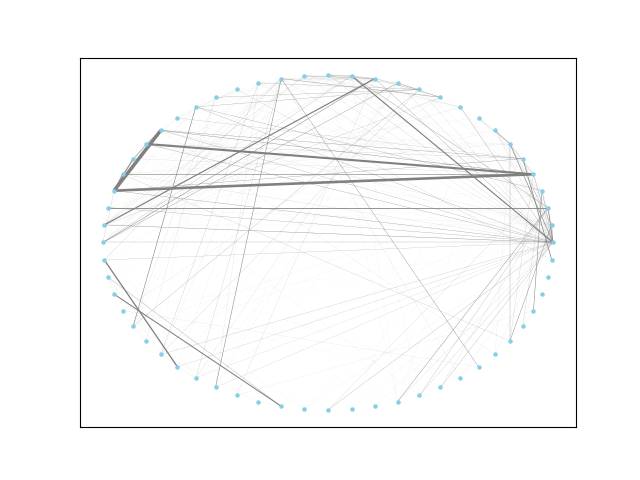}
		\caption{Schleswig-Holstein}
	\end{subfigure}
	\hfill
	\begin{subfigure}[b]{0.47\textwidth}
		\centering
		\includegraphics[width=\textwidth]{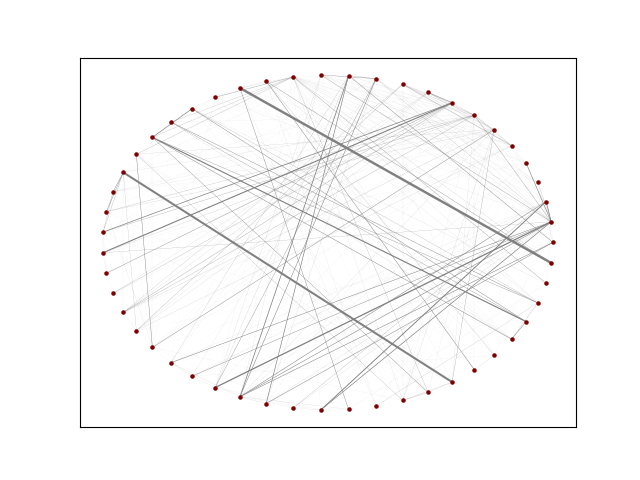}
		\caption{Brandenburg}
	\end{subfigure}   
\end{figure}
\clearpage   
\begin{figure}[!tb]\ContinuedFloat  	
	\begin{subfigure}[b]{0.47\textwidth}
		\centering
		\includegraphics[width=\textwidth]{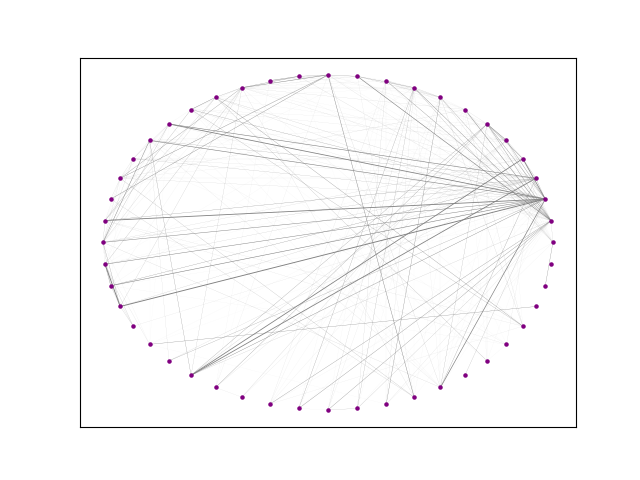}
		\caption{Saxony-Anhalt}
	\end{subfigure}
	\hfill
	\begin{subfigure}[b]{0.47\textwidth}
		\centering
		\includegraphics[width=\textwidth]{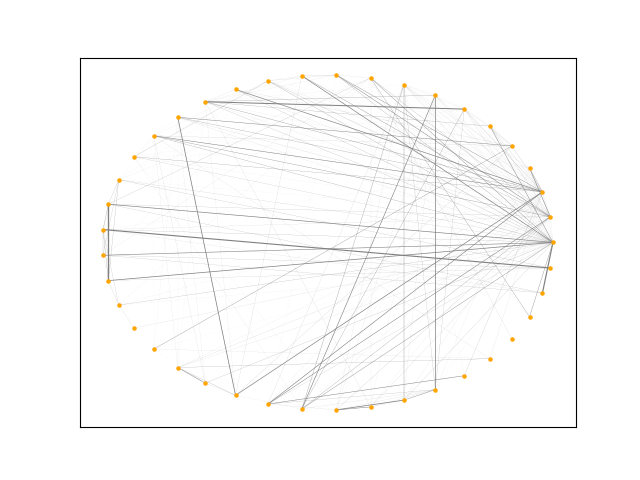}
		\caption{Thuringia}
	\end{subfigure}  

	\begin{subfigure}[b]{0.47\textwidth}
		\centering
		\includegraphics[width=\textwidth]{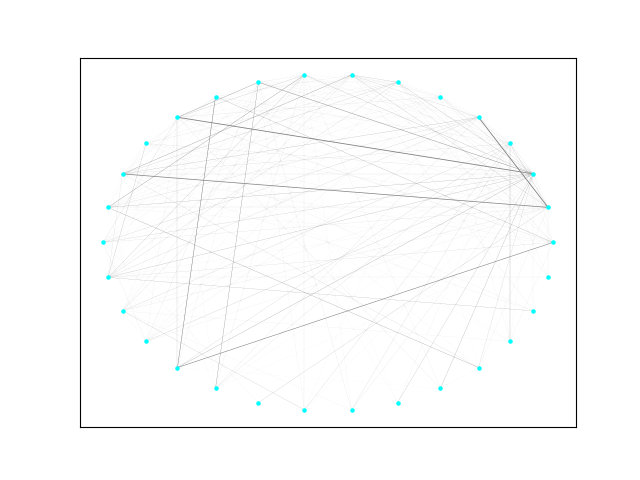}
		\caption{Hamburg}
	\end{subfigure}
	\hfill
	\begin{subfigure}[b]{0.47\textwidth}
		\centering
		\includegraphics[width=\textwidth]{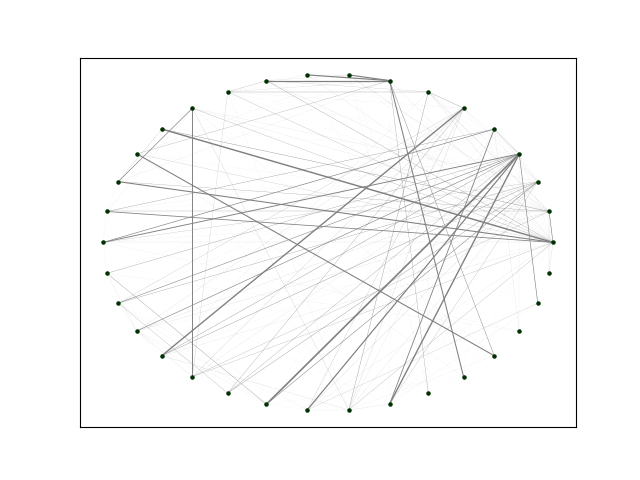}
		\caption{Mecklenburg-West Pomerania}
	\end{subfigure} 
	
	\begin{subfigure}[b]{0.47\textwidth}
		\centering
		\includegraphics[width=\textwidth]{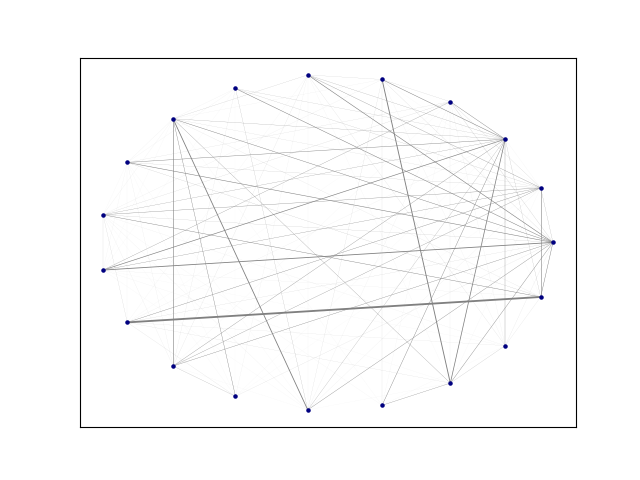}
		\caption{Saarland}
	\end{subfigure}
	\hfill
	\begin{subfigure}[b]{0.47\textwidth}
		\centering
		\includegraphics[width=\textwidth]{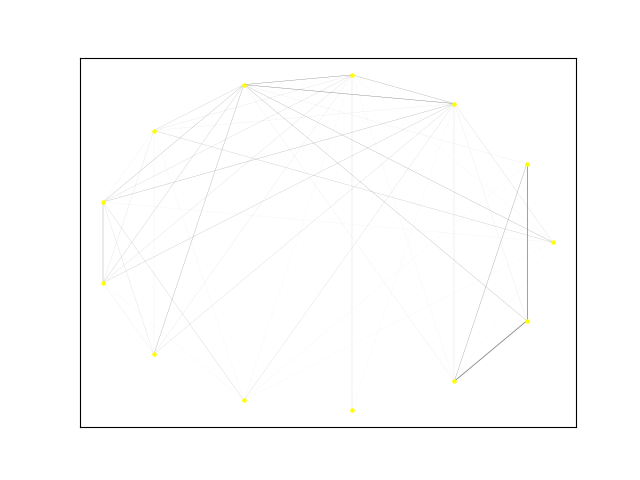}
		\caption{Bremen}
	\end{subfigure} 
	\caption{Visualization of hospital networks (community nodes are omitted) for given state.}
	\label{fig:hosp_graph}
\end{figure}

	\begin{table}[h!]
	\centering
	\caption{Transfer matrix statistics. States are sorted by their population according to~\cite{Destatis}.}
	\label{tab: trans_mat}
	\begin{filecontents*}{tabels/transfer_matrix_stat.csv}
State,No. nodes,No. edges,Avg. in degree,Avg. out degree,Density,Diameter,Radius
Whole,3118,296151,  94.9811,  94.9811,0.030471953036206298,7,4
North Rhine-Westphalia,654,50469,  77.1697,  77.1697,0.11817722016943676,6,4
Bavaria,554,29395,  53.0596,  53.0596,0.09594858370163402,6,4
Baden-Württemberg,362,15019,  41.4890,  41.4890,0.11492784010039639,6,4
Lower Saxony,322,10622,  32.9876,  32.9876,0.10276503937617307,6,4
Hesse,244,10046,  41.1721,  41.1721,0.16943263846724685,6,3
Rhineland-Palatinate,154,3978,  25.8312,  25.8312,0.16883116883116883,6,4
Saxony,144,3430,  23.8194,  23.8194,0.16656954156954157,5,3
Berlin,92,2988,  32.4783,  32.4783,0.3569039655996178,4,2
Schleswig-Holstein,120,2842,  23.6833,  23.6833,0.19901960784313724,5,3
Brandenburg,102,2065,  20.2451,  20.2451,0.20044651523975926,4,3
Saxony-Anhalt,96,1874,  19.5208,  19.5208,0.20548245614035088,5,4
Thuringia,82,1450,  17.6829,  17.6829,0.2183077386329419,5,3
Hamburg,60,1454,  24.2333,  24.2333,0.41073446327683616,4,2
Mecklenburg-West \mbox{Pomerania},68,1184,  17.4118,  17.4118,0.2598770851624232,4,3
Saarland,38,656,  17.2632,  17.2632,0.4665718349928876,3,2
Bremen,26,321,  12.3462,  12.3462,0.4938461538461538,3,2
\end{filecontents*}
\pgfplotstabletypeset[
	col sep=comma, 
	display columns/0/.style={column type={|p{2.3cm}|},string type}, 
	display columns/1/.style={column type={p{1.3cm}|}, int detect, 1000 sep={\;},precision=3},
	display columns/2/.style={column type={p{1.4cm}|},},
	display columns/3/.style={column type={p{1.4cm}|},},
	display columns/4/.style={column type={p{1.6cm}|},},
	display columns/5/.style={column type={p{1.7cm}|}, int detect, ,precision=2},
	display columns/6/.style={column type={p{1.5cm}|}, int detect, 1000 sep={\;},precision=3},
	display columns/7/.style={column type={p{1.3cm}|}, int detect, 1000 sep={\;},precision=3},
	every head row/.style={before row=\hline,after row=\hline},
	every nth row={1}{before row=\hline},
	every last row/.style={after row=\hline},
	]{tabels/transfer_matrix_stat.csv}
\end{table}

	\section{Summary}

	For the purpose of modelling the spread of antibiotic resistant bacteria or other infectious diseases, we analysed dataset provided for the EMerGE-Net project to the Martin Luther University Halle-Wittenberg. It allowed us to characterise both patients and healthcare facilities present in dataset. Due to high coverage of the dataset, which consists informations from all German states, we were able to compare data for individual regions. We discovered that most of them behave accordingly to their populations, i.e. number of healthcare facilities, patients, transfers and overlaps is proportional to the size of population reported in~\cite{Destatis}. The only exceptions are Berlin and Hamburg. That is reasonable due to nature of those states: these are large metropolitan areas. They have higher number of healthcare facilities, patients, transfers and overlaps than states with similar population size.  
	     
	From the dataset we were able to extract information about patients transfers. We characterised both transfers taking place within the same state and between states. Closer look into interstate transfer allowed us to indicate the states with the highest exchange of patients. Moreover, to better understand the movement of the patients in the network we analysed overlaps appearing in the records. In all the states around 80\% of them were classified as standard transfer or two entries in single institution, except for Berlin and Hamburg were it was almost 90\%.   
	   
	Our analysis allowed us to built transfer network for whole country as well as for individual states separately. The results indicate that for states with higher population the network has higher density and radius. Derived transfer matrices can be used further to model the spread of pathogen in details like e.g.~in~\cite{PiotrowskaSIVW2020} or~\cite{Piotrowska2020PlosComp}. 

	\section*{Acknowledgements} 

	This work was supported by grant no.~2016/22/Z/ST1/00690 "Effectiveness of infection control strategies against intra- and inter-hospital transmission of MultidruG-resistant Enterobacteriaceae --- insights from a multi-level mathematical NeTwork model" of the National Science Centre, Poland within the 3rd JPI ARM framework (Joint Programming Initiative on Antimicrobial Resistance) co-found grant no 681055 for the consortium EMerGE-Net (Effectiveness of infection control strategies against intra- and inter-hospital transmission of MultidruG-resistant Enterobacteriaceae).
	We thank the EMerGE-Net team at the Martin Luther University Halle-Wittenberg for providing access to the insurance data on their server in joint analyses.
	
	\section*{Supplementary materials}

		S1: Numerical data for probability transfer matrix.
	

	\bibliographystyle{ieeetr}
	\bibliography{emergenet_trojkat}

\begin{thebibliography}{10}

\bibitem{WHO}
``{Antibiotic resistance}.''
  \url{https://www.who.int/news-room/fact-sheets/detail/antibiotic-resistance}.

\bibitem{WHO2}
``{Antimicrobial resistance}.''
  \url{https://www.who.int/news-room/fact-sheets/detail/antimicrobial-resistance}.

\bibitem{ECDC}
E.~C. for Disease~Prevention and Control, ``{Antimicrobial resistance in the
  EU/EEA (EARS-Net) Annual Epidemiological Report for 2019},'' 2020.

\bibitem{Donker2010}
T.~Donker, J.~Wallinga, and H.~Grundmann, ``Patient referral patterns and the
  spread of hospital-acquired infections through national health care
  networks,'' {\em {PLoS} Computational Biology}, vol.~6, p.~e1000715, mar
  2010.

\bibitem{huang2010quantifying}
S.~S. Huang, T.~R. Avery, Y.~Song, K.~R. Elkins, C.~C. Nguyen, S.~K. Nutter,
  A.~A. Nafday, C.~J. Condon, M.~T. Chang, D.~Chrest, {\em et~al.},
  ``Quantifying interhospital patient sharing as a mechanism for infectious
  disease spread,'' {\em Infection Control \& Hospital Epidemiology}, vol.~31,
  no.~11, pp.~1160--1169, 2010.

\bibitem{Donker2012}
T.~Donker, J.~Wallinga, R.~Slack, and H.~Grundmann, ``Hospital networks and the
  dispersal of hospital-acquired pathogens by patient transfer,'' {\em {PLoS}
  {ONE}}, vol.~7, p.~e35002, apr 2012.

\bibitem{ciccolini2013infection}
M.~Ciccolini, T.~Donker, R.~K{\"o}ck, M.~Mielke, R.~Hendrix, A.~Jurke,
  J.~Rahamat-Langendoen, K.~Becker, H.~G. Niesters, H.~Grundmann, {\em et~al.},
  ``Infection prevention in a connected world: The case for a regional
  approach,'' {\em International Journal of Medical Microbiology}, vol.~303,
  no.~6-7, pp.~380--387, 2013.

\bibitem{Donker2014}
T.~Donker, J.~Wallinga, and H.~Grundmann, ``Dispersal of antibiotic-resistant
  high-risk clones by hospital networks: changing the patient direction can
  make all the difference,'' {\em Journal of Hospital Infection}, vol.~86,
  pp.~34--41, jan 2014.

\bibitem{Donker2017}
T.~Donker, T.~Smieszek, K.~L. Henderson, A.~P. Johnson, A.~S. Walker, and J.~V.
  Robotham, ``Measuring distance through dense weighted networks: The case of
  hospital-associated pathogens,'' {\em {PLOS} Computational Biology}, vol.~13,
  p.~e1005622, aug 2017.

\bibitem{Nekkab2017}
N.~Nekkab, P.~Astagneau, L.~Temime, and P.~Cr{\'{e}}pey, ``Spread of
  hospital-acquired infections: {A} comparison of healthcare networks,'' {\em
  {PLOS} Computational Biology}, vol.~13, p.~e1005666, aug 2017.

\bibitem{Lee2012}
B.~Y. Lee, S.~M. Bartsch, K.~F. Wong, S.~L. Yilmaz, T.~R. Avery, A.~Singh,
  Y.~Song, D.~S. Kim, S.~T. Brown, M.~A. Potter, R.~Platt, and S.~S. Huang,
  ``Simulation shows hospitals that cooperate on infection control obtain
  better results than hospitals acting alone,'' {\em Health Affairs}, vol.~31,
  pp.~2295--2303, oct 2012.

\bibitem{Belik2016}
V.~Belik, P.~H{\"o}vel, and R.~Mikolajczyk, ``Control of epidemics on hospital
  networks,'' in {\em Understanding Complex Systems}, pp.~431--440, Springer
  International Publishing, 2016.

\bibitem{Lonc2019}
A.~Lonc, M.~J. Piotrowska, and K.~Sakowski, ``Analysis of the {AOK} plus data
  and derived hospital network,'' {\em Mathematica Applicanda}, vol.~47, no.~1,
  2019.

\bibitem{Vilches2019}
T.~Vilches, M.~Bonesso, H.~Guerra, C.~Fortaleza, A.~Park, and C.~Ferreira,
  ``The role of intra and inter-hospital patient transfer in the dissemination
  of heathcare-associated multidrug-resistant pathogens,'' {\em Epidemics},
  vol.~26, pp.~104--115, mar 2019.

\bibitem{Nekkab2020}
N.~Nekkab, P.~Cr{\'{e}}pey, P.~Astagneau, L.~Opatowski, and L.~Temime,
  ``Assessing the role of inter-facility patient transfer in the spread of
  carbapenemase-producing enterobacteriaceae: the case of france between 2012
  and 2015,'' {\em Scientific Reports}, vol.~10, sep 2020.

\bibitem{PiotrowskaSIVW2020}
M.~Piotrowska, K.~Sakowski, A.~Lonc, H.~Tahir, and M.~Kretzschmar, ``Impact of
  inter-hospital transfers on the prevalence of resistant pathogens in a
  hospital{\textendash}community system,'' {\em Epidemics}, vol.~33, p.~100408,
  dec 2020.

\bibitem{Piotrowska2020PlosComp}
M.~J. Piotrowska, K.~Sakowski, A.~Karch, H.~Tahir, J.~Horn, M.~E. Kretzschmar,
  and R.~T. Mikolajczyk, ``Modelling pathogen spread in a healthcare network:
  Indirect patient movements,'' {\em {PLOS} Computational Biology}, vol.~16,
  p.~e1008442, nov 2020.

\bibitem{Xia2021}
H.~Xia, J.~Horn, M.~J. Piotrowska, K.~Sakowski, A.~Karch, H.~Tahir,
  M.~Kretzschmar, and R.~Mikolajczyk, ``Effects of incomplete inter-hospital
  network data on the assessment of transmission dynamics of hospital-acquired
  infections,'' {\em {PLOS} Computational Biology}, vol.~17, p.~e1008941, may
  2021.

\bibitem{Deeny2013}
S.~Deeny, B.~Cooper, B.~Cookson, S.~Hopkins, and J.~Robotham, ``Targeted versus
  universal screening and decolonization to reduce healthcare-associated
  meticillin-resistant staphylococcus aureus infection,'' {\em Journal of
  Hospital Infection}, vol.~85, pp.~33--44, sep 2013.

\bibitem{Sadsad2013}
R.~Sadsad, V.~Sintchenko, G.~D. McDonnell, and G.~L. Gilbert, ``Effectiveness
  of hospital-wide methicillin-resistant staphylococcus aureus ({MRSA})
  infection control policies differs by ward specialty,'' {\em {PLoS} {ONE}},
  vol.~8, p.~e83099, dec 2013.

\bibitem{Tahir2021}
H.~Tahir, L.~E. L{\'{o}}pez-Cort{\'{e}}s, A.~Kola, D.~Yahav, A.~Karch, H.~Xia,
  J.~Horn, K.~Sakowski, M.~J. Piotrowska, L.~Leibovici, R.~T. Mikolajczyk, and
  M.~E. Kretzschmar, ``Relevance of intra-hospital patient movements for the
  spread of healthcare-associated infections within hospitals - a mathematical
  modeling study,'' {\em {PLOS} Computational Biology}, vol.~17, p.~e1008600,
  feb 2021.

\bibitem{emergenetpackage}
``{EMerGE-NeT Package}.'' \url{https://www.mimuw.edu.pl/~monika/emergenet}.

\bibitem{Destatis}
``{Bevölkerung nach Nationalität und Bundesländern}.''
  \url{https://www.destatis.de/DE/Themen/Gesellschaft-Umwelt/Bevoelkerung/Bevoelkerungsstand/Tabellen/bevoelkerung-nichtdeutsch-laender.html}.

\bibitem{Piotrowska2019arxiv}
M.~J. Piotrowska and K.~Sakowski, ``Analysis of the {AOK} {L}ower {S}axony
  hospitalisation records data (years 2008 -- 2015),'' {\em arXiv},
  vol.~1903.04701v1, 2019.

\bibitem{Lonc2019arxiv}
A.~Lonc, M.~J. Piotrowska, and K.~Sakowski, ``Analysis of the hospital records
  from {AOK Plus},'' {\em arXiv}, vol.~1909.08169v1, 2019.

\bibitem{Data}
``{Internationale statistische Klassifikation der Krankheiten und verwandter
  Gesundheitsprobleme}.''
  \url{ttps://www.dimdi.de/static/de/klassifikationen/icd/icd-10-gm/kode-suche/htmlgm2019/}.

\end{thebibliography}

	
\end{document}